\documentclass[reprint,aps,nofootinbib]{revtex4-2}
\usepackage{amsmath}
\usepackage{amsfonts}
\usepackage{amssymb}
\usepackage{amsthm}
\usepackage{mathtools}
\usepackage{graphicx}
\usepackage[dvipsnames]{xcolor}
\usepackage[colorlinks=True,citecolor=Mahogany,linkcolor=myRed,urlcolor=Mahogany,hypertexnames=false]{hyperref}
\usepackage{hypcap}
\usepackage{yfonts}
\usepackage{bm}
\usepackage[normalem]{ulem}
\usepackage{dsfont}
\usepackage{braket}
\usepackage{marginnote}
\usepackage{adjustbox}
\usepackage{enumitem}
\usepackage{tikz}
\usepackage{quantikz}
\usetikzlibrary{shapes, arrows, arrows.meta}

\usepackage{kantlipsum}

\newcommand\eq[1]{\begin{align}#1\end{align}}

\DeclareMathAlphabet{\mathpzc}{OT1}{pzc}{m}{it}

\newcommand*{\NWarrow}{\rotatebox[origin=c]{135}{\(\leftrightarrow\)}}

\newcommand*{\NEarrow}{\rotatebox[origin=c]{45}{\(\leftrightarrow\)}}

\newcommand\mytitle{Dynamics of local quantum information in random unitary circuits}

\definecolor{myBlue}{RGB}{31,119,180}
\definecolor{myOrange}{RGB}{255,127,14}
\definecolor{myGreen}{RGB}{44,160,44}
\definecolor{myRed}{RGB}{214,39,40}
\definecolor{myPurple}{RGB}{148,103,189}

\begin{document}

\title{\mytitle}

\author{Ratul Thakur}
\email{ratul.thakur@icts.res.in}
\affiliation{International Centre for Theoretical Sciences, Tata Institute of Fundamental Research, Bengaluru 560089, India}

\author{Sthitadhi Roy}
\email{sthitadhi.roy@icts.res.in}
\affiliation{International Centre for Theoretical Sciences, Tata Institute of Fundamental Research, Bengaluru 560089, India}

\begin{abstract}
The physics of information scrambling in quantum many-body systems is intimately related to the fundamental questions of thermalisation and emergence of chaos. However, its standard characterisation through bipartite entanglement or operator growth remains inherently coarse-grained, obscuring the spatiotemporal anatomy of how quantum information flows between different regions of the system and across various length scales. In this work, we develop a theory for the dynamics of local information in locally interacting random unitary circuits which resolves the fine-grained microscopic structures of local information flow in such systems. Using the framework of information lattice which systematises the information content at each length scale within each subsystem, we identify a distribution of length scales at which information resides and obtain the dynamics of the distribution. For Haar-random circuits with infinite local Hilbert-space dimensions, we map this dynamics onto an exact classical stochastic process which reveals that this distribution is described by a Tracy-Widom form which moves ballistically in time ($ \propto t$) toward larger scales, accompanied by a $\sim t^{1/3}$ broadening. We also find the same qualitative behaviour for random Clifford circuits acting on qubits. The similar scaling behaviour in two rather different settings hints strongly towards the universality of our results. In the case of Clifford circuits, we develop a phenomenological Markov process for the dynamics of the stabiliser generators in a specific gauge which confirms the Tracy-Widom distribution and the $t^{1/3}$ scaling of the fluctuations. Ultimately, our results establish the universal properties of the dynamics of local quantum information in a length scale-resolved fashion and provide a possible route towards bridging exact microscopic theories for information dynamics with emergent hydrodynamic descriptions of information flow.
\end{abstract}

\maketitle

\tableofcontents

\section{Introduction\label{sec:intro}}
What are the universal properties of {\it local} spatiotemporal information flow in out-of-equilibrium quantum many-body systems? This is the central question with which we concern ourselves in this work. 
In the recent past, it has become increasingly evident that the dynamics of quantum information in isolated quantum systems lies at the core of one of the fundamental issues in statistical physics, namely, the mechanisms which underpin thermalisation~\cite{deutsch1991quantum,srednicki1994chaos,rigol2008thermalisation,dalessio2016from,deutsch2018eigenstate}. 
It is by now well understood that in isolated ergodic systems, for any local subsystem the rest of the system effectively acts as a bath which drives the former towards thermalisation under unitary dynamics of the entire system.
This naturally means that quantum information initially contained in a subsystem spreads and scrambles over the entire system as time goes on.
The resulting information loss from a local subsystem endows its reduced state with finite entropy which is eventually maximised at late times leading to universal thermodynamic descriptions, such as the Gibbs or generalised Gibbs ensembles, of the subsystems~\cite{jaynes1957information,*jaynes1957information2,popescu2006entanglement,rigol2007relaxation,rigol2008thermalisation}. 
The physics of information scrambling is therefore inextricably linked to that of thermalisation and naturally underscores the fundamental importance of understanding the spatiotemporal anatomy of quantum information in thermalising systems.

\subsection{Background and motivation}

Indeed, in the past several years there has been tremendous progress in studying information scrambling across a whole array of contexts and settings ranging from black holes placing bounds on scrambling~\cite{hayden2007black,sekino2008fast,shenker2014black,maldacena2016bound} to low-dimensional condensed-matter systems using diagnostics of scrambling for quantifying quantum supremacy~\cite{harrow2017nature,boixo2018characterising,arute2019quantum,mi2021information}.
In a vast majority of these works scrambling is characterised either by operator growth, often quantified via out-of-time-ordered correlators (OTOCs) and operator entanglement entropy~\cite{roberts2015diagnosing,maldacena2016bound,hosur2016chaos,aleiner2016microscopic,bohrdt2017scrambling,luitz2017information,zhou2017operator,nahum2018operator,keyserlingk2018operator,rakovszky2018diffusive,khemani2018operator,hahn2024eigenstate,pain2026imprints}
or via the dynamics of bipartite entanglement entropy of the time-evolving state~\cite{calabrese2005evolution,kim2013ballistic,liu2014entanglement,zhang2015thermalisation,nahum2017quantum,jonay2018coarse}.
These works have lead to several significant insights. 
For instance, in integrable 1D systems the entanglement growth following a quench can be understood in terms of ballistically propagating quasiparticles~\cite{calabrese2005evolution,calabrese2007entanglement}.
While such a picture ceases to exist for non-integrable systems, the entanglement growth can still be understood in terms of an effective classical statistical mechanics of interfaces referred to as the `entanglement membrane'~\cite{zhou2019emergent,zhou2020entanglement}.
Within this framework, the entanglement growth in one dimension maps onto the celebrated Kardar-Parisi-Zhang physics which leads to a linear in time growth of entanglement accompanied by fluctuations which scale as time\textsuperscript{1/3}.

However, despite these significant advancements, our understanding of scrambling remains fundamentally coarse-grained. Standard diagnostics such as bipartite entanglement or mutual information of macroscopic subregions, assign a single, integrated value to a given spatial block.
By construction, these measures wash out the fine-grained local structure of entanglement. 
Crucially, this lack of resolution means that such measures completely obscure the scale-by-scale anatomy of quantum information and their dynamics.

To make this precise, consider a contiguous subsystem $A$ of length $l$ in one dimension. While the bipartite entanglement of $A$ with the rest of the subsystem encodes how much quantum information is contained in $A$, it completely fails to capture how much of that information is contained exactly at length scale $l$ and is not present at any smaller length scale $l'<l$ within $A$.
Similarly, when the entanglement of $A$ changes, it only encodes the information flowing across the boundaries of $A$ and does not capture any feature of the overall reorganisation of the information across various length scales. This suggests that the dynamics of entanglement is insufficient to quantify the pattern of information flow across regions and length scales in the system.
This is precisely the lacunae in the theory of dynamics of quantum information in many-body systems that we address in this work.

Developing a theory for the dynamics of local information presents two fundamental challenges. 
First, information scrambling is inherently a non-local process; the entanglement entropy of any subsystem involves dynamical correlations of all the operators supported on the entire subsystem. 
We therefore need a formulation of local information which is sensitive to both the subsystem as well as the length scale at which the information resides.
Second, the total information content of a state is conserved under unitary dynamics but it cannot be written as a sum of local densities. 
It is therefore {\it a priori} not obvious that the dynamics of local information can be described in terms of equations which are also local; this just reflects the fact that changing the local information content at a certain location and length scale can, in principle, lead to a global reorganisation of the local information profile. 

Surmounting these challenges leads to a quantitative understanding of how local information is organised in different regions of the system and across different length scales. 
More importantly, this leads to an identification of distributions of length scales at which local information resides and how these distributions evolve in time. 
Developing a theory for the evolution of these distributions and the underlying patterns of information flow constitutes the central motivation of this work.

\subsection{Overview of the main results}

\begin{figure}
\input{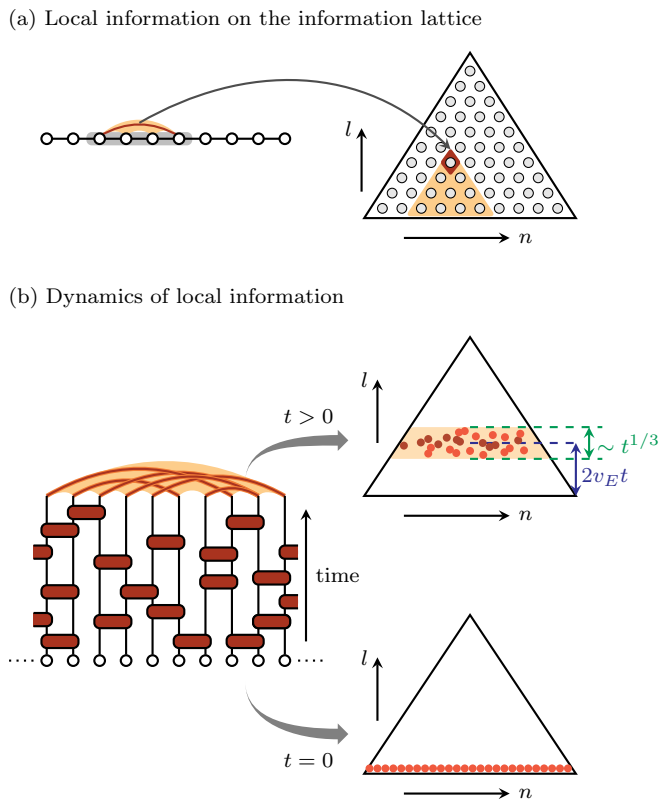}
\caption{Schematic summary of the main message of the paper. (a) The information lattice defined in the $(n,l)$ plane encodes the local information contained at length scale exactly $l$ within a subsystem centred at $n$. (b) Dynamics of  local information. At time $t=0$ the state is initialised in a product state and therefore all the information lives on scale $l=0$. Evolution of this state with a generic, chaotic circuit leads to entanglement growth and information scrambling. This is manifested on the information lattice at a later time $t$ by the local information residing at a length scale which grows ballistically, $2v_Et$, on an average accompanied by fluctuations which scale as $t^{1/3}$. While the local information in general is a positive real-valued quantity, for Haar-random circuits with infinite local Hilbert-space dimensions and Clifford circuits, they are integer valued which leads to a particle description of the information dynamics.}
\label{fig:summary}
\end{figure}

To the put the idea of local information on a concrete quantitative footing, we employ the recently introduced construction of the {\it information lattice}~\cite{kvorning2022time,artiaco2024efficient,artiaco2025universal,bauer2025local,flor2025higher}.
For a one-dimensional system, the information lattice is a lattice in two dimensions where the value at any point $(n,l)$ encodes the local information which resides only at length scale $l$ centred at $n$. This is the information contained within a contiguous subsystem of length $l$ centred at $n$, which is not contained in any of its smaller contiguous subsystems. A schematic of this is shown in Fig.~\ref{fig:summary}(a).
We find quite generally that under scrambling dynamics, there is a general tendency of the information to drift ballistically from lower length scales to higher. 
A comprehensive statistical understanding of this local information flow effected random unitary circuits comprises the main content of this paper; this is summarised schematically in Fig.~\ref{fig:summary}(b).

The rationale behind using random circuits as models of information scrambling is simply that they constitute the minimal models of quantum chaos and offer a degree of tractability that is significantly more than conventional non-integrable Hamiltonians.
More specifically, we employ two classes of circuits in our analyses where in both cases, the circuits are built from nearest-neighbour two-site unitary gates.
The first class corresponds to the family of circuits where the unitary gates are drawn from Haar measure. Such `Haar-random' circuits often lend themselves to solvability at the cost of obtaining results which are averaged over the family of the circuits by mapping the averaged dynamics onto classical statistical mechanics models~\cite{nahum2017quantum,nahum2018operator,keyserlingk2018operator,chan2018solution,zhou2019emergent,fisher2023random}.
The second class of circuits we study are  Clifford circuits as they are efficiently simulable classically using the stabiliser formalism again offering significant numerical tractability~\cite{gottesman1997stabilizer,gottesman1998heisenberg,aaronson2004improved}.

As we discuss in detail later, the characteristics properties of the dynamics of local information are identical in the two classes of circuits making a compelling case for the universality of the results.
In particular, we find that the average length scale at which information resides grows ballistically with time, $\braket{l(t)} = 2v_Et$ whereas the fluctuations in the length scale grow as $\sqrt{\braket{\Delta l^2(t)}}\sim t^{1/3}$.
In addition we find that the profile of local information over $l$ follows a Tracy-Widom law. 
For a finite-sized system of size $L$, at sufficiently late times, the information profile saturates to a profile which is peaked at $l=L/2$ and falls off exponentially away from it.

For circuits built from local unitary gates, we show that the locality and unitarity of the dynamics along with fundamental properties satisfied by the von Neumann entropy of entanglement impose non-local constraints on the dynamics on the information lattice. 
More specifically we show under a unitary operation acting on pair of sites real-space sites $(i,i+1)$, the information lattice values are constrained to change only along the two-site thick diagonals originating from the sites $(i,i+1)$ at $l=0$. 
Naturally, this also means that the action of such a gate generates information flows potentially over the entirety of these diagonals on the information lattice.
This makes the non-local effect on the information lattice of a unitary gate acting locally in space manifest.
These non-local constraints explicitly enter the dynamical rules for updating the local information content and in fact act as guiding principles for deriving effective theories for the dynamics.

For circuits composed of Haar-random unitaries in the limit of infinite local Hilbert-space dimension ($q \to \infty$), we leverage the exactness of the minimal cut bound~\cite{nahum2017quantum} to derive an exact classical stochastic process for the local information. 
Specifically, we show that the set of the minimal cuts for each bipartition of the system can be recast into a set of bigrams.
Each bigram is a classical extended object with two endpoints and contributes one bit of local information at its midpoint location and at a length scale equal to the length of the bigram.
The classical stochastic process that we derive exactly for updating the bigrams has the feature that a local unitary gate can only change the bigram endpoints on the sites it acts on.
However, given the extended nature of the bigrams, the stochastic process naturally incorporates the non-local change in the profile on the information lattice brought about by a local unitary gate.

In the case of Clifford circuits with $q=2$ we observe a very similar set of physical processes which underpin the dynamics of local information.
In this case, we can again define bigrams whose local information content is identical to that of the Haar-random circuits, except they now correspond to the stabiliser generators of the state written in a specific gauge. 
For Clifford circuits, the underlying non-locality of the dynamics is even more explicit as the dynamical update rules for the bigram endpoints also depend on the lengths of the bigrams. 

It is worth noting that defining the information lattice in one dimension higher than the physical dimension allows for a local description of the information content at different regions across various scales in the problem. In fact, for infinite-$q$ Haar-random circuits and Clifford circuits with any $q$, we show that the local information on the information lattice is integer valued. As such, this can be interpreted as a particle description of  entanglement and information dynamics, quite analogous to the quasiparticle description of entanglement in integrable systems~\cite{calabrese2005evolution}, but for strongly chaotic and non-integrable systems.
However, despite the local description being possible, the dynamics remains non-locally constrained on the information lattice.
This makes the continuum hydrodynamic description of the information flow extremely non-trivial. Nevertheless we do also make certain speculative but concrete remarks towards how such a description can be developed.

\subsection{Organisation of the paper}

The remainder of this paper is organised as follows. In Sec.~\ref{sec:info-latt}, we describe the framework of information lattice and the non-local constraints thereon, and formalise the definitions.  In Sec.~\ref{sec:res-general}, we formally define the information currents on the information lattice and describe their behaviour in all generality along with quantitative results at the leading order.
In Sec.~\ref{sec:res-haar}, we specialise to the case of Haar-random circuits where we first present the exact classical Markov process for the averaged local information dynamics in the $q\to\infty$ limit and the results from it, followed by direct numerical simulations for the $q=2$ case.
The analyses for the Clifford circuits constitute Sec.~\ref{sec:res-cliff}, where we first present exact numerical results obtained using the stabiliser formalism followed by a discussion of a phenomenological classical Markov process for the local information dynamics.
In Sec.~\ref{sec:hydro}, we discuss, somewhat speculatively, the first steps toward a continuum hydrodynamic description for the information lattice in systems without any conservation laws. 
Finally, we conclude with a summary and outlook in Sec.~\ref{sec:discussion}.
Technical details of various calculations are presented in the Appendices.

\begin{figure}
    \centering
    \begin{tikzpicture}[x=0.7cm, y=0.35cm]
\fill[fill=Apricot, rounded corners] (0.5,-2) rectangle (5.5,-1);
\foreach \l in {0,...,9}{
    \filldraw (\l, -1.5) circle (3pt);
    \node at (\l,-0.5) {$\l$};
}
\draw[-Stealth,thick] (0,0.5)--++(3,0);\node at (3.25,0.5) {$n$};
\node at (3,-4) {$l=4$};
\draw[-Stealth,thick] (4,-4)--++(1.5,0);
\draw[-Stealth,thick] (2,-4)--++(-1.5,0);
\draw[dashed, thick] (0.5,-1.5) -- ++(0.,-6);
\draw[dashed, thick] (5.5,-1.5) -- ++(0,-3.5);
\draw[dashed, thick,black!40] (1.5,-1.5) -- ++(0,-8.5);
\draw[dashed, thick,black!40] (4.5,-1.5) -- ++(0,-8.5);
\node at (3,-2.75) {{\color{Orange}{${\cal C}_{n}^l$}}};
\fill[rounded corners=2pt, black] (1.5,-3.75-1.5) rectangle (5.5,-3.25-1.5);
\node at (3,-6.1) {\small {\color{black}{${\cal C}_{n+1/2}^{l-1}$}}};
\fill[rounded corners=2pt, black!60] (0.5,-6.25-1.5) rectangle (4.5,-5.75-1.5);
\node at (3,-8.6) {\small {\color{black!60}{${\cal C}_{n-1/2}^{l-1}$}}};
\fill[rounded corners=2pt, black!35] (1.5,-8.75-1.5) rectangle (4.5,-8.25-1.5);
\node at (3,-10.95) {\small {\color{black!35}{${\cal C}_{n}^{l-2}$}}};

\draw[thick](0,-1.5) -- (9,-1.5);
\node at (-1.2,1) {(a)};
\end{tikzpicture}
    \begin{tikzpicture}[x=0.7cm, y=0.35cm]
\def\m{0.2} 
\fill[fill=Apricot, rounded corners=6pt] (6.25+0.125,-2) rectangle (8.75-0.125,-1);
\fill[fill=Salmon, rounded corners=6pt] (1.25+0.125,-2) rectangle (4.75-0.125,-1);
\fill[fill=Apricot, rounded corners=7pt] (7.5,2) -- (8.75,-0.5) -- (6.25,-0.5) --  cycle;
\fill[fill=Salmon, rounded corners=7pt] (3,3) -- ++(-1.75,-3.5) -- (4.75,-0.5) --  cycle;
\fill[fill=Gray!50,rounded corners =7pt] (5.5,2) -- ++(1,2) -- ++(-1.5,3) -- ++(-1,-2) --cycle;
\fill[rounded corners, OliveGreen!60] (3,3) --++ (0.5,1) -- ++(-0.5,1) -- ++(-0.5,-1) -- cycle;
\foreach \l in {0,...,9}{
    \edef\maxN{\numexpr9-\l\relax}
    \foreach \n in {0,...,\maxN}{
        \filldraw (\n+0.5*\l, \l) circle (1.4pt);
    }   
}
\foreach \l in {0,...,2}{
\node at (-1.2+0.5*\l,\l) {\small $l=\l$};
}
\foreach \l in {0,...,9}{
\filldraw (\l, -1.5) circle (3pt);
}
\draw[thick, dotted] (-0.5+0.8,3) -- (1.8,6);
\node at (3,-2.75) {{\color{red!70}{${\cal C}_{3}^2$}}};
\node at (7.5,-2.75) {{\color{Orange}{${\cal C}_{15/2}^1$}}};
\node at (7.5,7.5) {\small {\color{black!70}{$\mathfrak{I}({\cal C}_3^2:{\cal C}_{15/2}^1|{\cal C}_{11/2}^1)$}}};
\draw[Gray!80,-Stealth,thick] (5.5,6)--++(0.5,1);
\draw[OliveGreen!80,-Stealth,thick] (3,4.7)--++(-0.5,1);
\node at (2.3,6) {\small {\color{OliveGreen}{$I_3^4$}}};
\draw[thick](0,-1.5) -- (9,-1.5);
\node at (-1.2,9) {(b)};
\end{tikzpicture}
    \caption{(a) Regions $\mathcal C_n^l$, $\mathcal C_{n+1/2}^{l-1}$, $\mathcal C_{n-1/2}^{l-1}$ and $\mathcal C_n^{l-2}$, which appear in the equation for $I_n^l$ for $(n,l) = (3,4)$ are shown on a chain of $L = 10$ spins. (b) The information lattice for $L=10$ spins; The site $(n,l) = (3,4)$ is highlighted in green. Subsystem lattices for the subsystems $\mathcal C_3^2$ and $\mathcal C_{15/2}^1$ are highlighted (in red and orange respectively), where the sum of the local information content in these regions equals the respective von Neumann information. Local information in the gray region sums up to the conditional mutual information between he two subsystems, conditioned on the region which separates them.}
    \label{fig:info-lattice}
\end{figure}
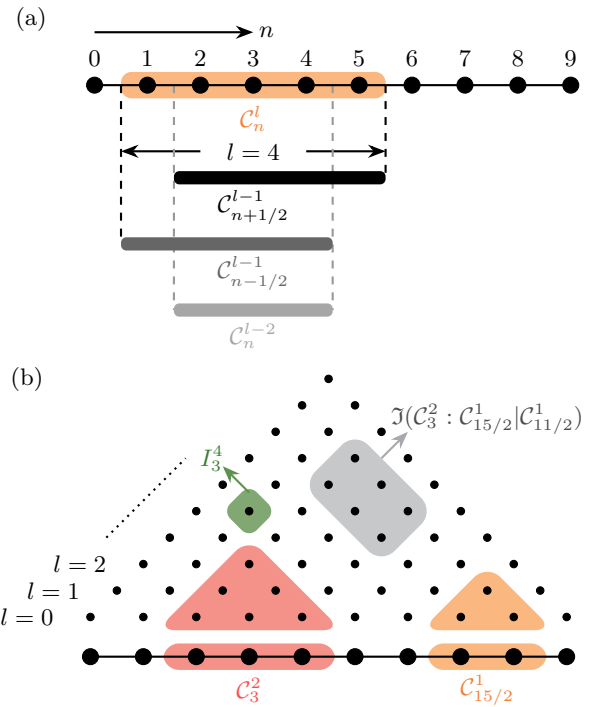

\section{Information lattice \label{sec:info-latt}}
In this section we discuss the theoretical backbone of this paper, namely the {\it information lattice}~\cite{kvorning2022time,artiaco2024efficient,artiaco2025universal,bauer2025local,flor2025higher},
which constitutes a concrete framework to organise systematically the quantum information contained at various length scales in a state. 
The information lattice, for a state in one spatial dimension, is a lattice in two dimensions.
The first dimension, denoted by $n$, corresponds to the physical dimension of the original lattice whereas the second dimension, denoted by $l$, corresponds to a length scale on the original lattice.
As such, the value at any point $(n,l)$ on this two-dimensional lattice encodes the information contained in a subsystem which is centred at point $n$ on the physical lattice and is of length $l$.

In Sec.~\ref{sec:info-latt-framework}, we set up the notations and provide the definitions of the main quantities required to describe the local information content on the information lattice.
In Sec.~\ref{sec:info-latt-constraints}, we describe how completely general properties of quantum mechanical states impose non-local constraints on the local information when viewed on the information lattice. 
Taking these constraints into account, in Sec.~\ref{sec:info-latt-homogeneous}, we show how the dynamics of information lattice may simplify over length scales over which the entanglement is homogeneously distributed.

\subsection{Framework and definitions \label{sec:info-latt-framework}}

Let us consider a chain of $L$ qudits --- degrees of freedom with a $q$-dimensional local Hilbert space --- placed at $i = 0,1,...,L-1$. We denote by $\mathcal C_n^l$, a contiguous subsystem centred at $n$ containing $l+1$ sites [see Fig.~\ref{fig:info-lattice}(a)]. 
For even and odd $l$, $n$ takes integer and half-integer values respectively. In both cases, ${\cal C}_n^l$ contains sites $n-l/2$ through $n+l/2$, both inclusive.
Note that by varying $n$ over all half-integers and $l$ over all integers, the collection $\{{\cal C}_n^l\}$ uniquely identifies all contiguous subsystems.

The information contained in the subsystem ${\cal C}_n^l$ is quantified by the deficit of the entropy of its reduced state from its maximal value. 
This is physically natural as if this deficit is zero, it implies that the reduced state of ${\cal C}_n^l$ is maximally mixed and therefore it contains no information.
In the other limit, if the entropy is itself zero, it implies that ${\cal C}_n^l$ is in a product state with the rest of the system and hence it contains the maximal information allowed by the purity of the state. 

To make this formal, consider a pure state $\rho\equiv \ket{\psi}\bra{\psi}$ defined on the entire system. The reduced density matrix of the subsystem ${\cal C}_n^l$ will be denoted by 
$\rho_n^l = {\rm Tr}_{\overline{\mathcal C_n^l}} \rho$, where $\overline{\mathcal C_n^l}$ denotes the complement of $\mathcal C_n^l$.
The von Neumann information of ${\cal C}_n^l$ is then defined as 
\begin{equation}
    E_n^l = S_{{\rm max},n}^l - S_n^l = (l+1) - S_n^l\,.
    \label{eq:Enl-def}
\end{equation}
where 
$S_n^l = -{\rm Tr}(\rho_n^l\log_q\rho_n^l)$ is the von Neumann entropy of entanglement~\footnote{Unless otherwise stated, all logarithms are with base $q$.} of ${\cal C}_n^l$ with the rest of the system in the state $\ket{\psi}$ and $S_{{\rm max},n}^l$ is the entanglement entropy when $\rho_n^l\propto \mathbb{I}$, which corresponds to the maximum it can be.

The von Neumann information in Eq.~\ref{eq:Enl-def} quantifies the total information contained in a contiguous subsystem, which also includes the information encoded 
in its smaller constituent subsystems. 
To quantify the {\it local information} contained in ${\cal C}_n^l$ at length scale $l$ (denoted by $I_n^l$) and therefore, not in any of its constituent, contiguous subsystems we need to subtract out the information encoded at smaller length scales within ${\cal C}_n^l$.
Formally, this is given by
\begin{equation}
    I_n^l = E_n^l - E_{n-1/2}^{l-1} - E_{n+1/2}^{l-1} + E_{n}^{l-2}\,.
    \label{eqn:local-info}
\end{equation}
The equation above can be understood physically as follows. 
The first term, $E_n^l$, denotes the total information in ${\cal C}_n^l$, for instance the segment shown in yellow in Fig.~\ref{fig:info-lattice}(a). To extract the information solely at length scale $l$, we subtract the total information contained at length scales up to $l-1$; there are two such segments, ${\cal C}^{l-1}_{n\pm 1/2}$ [shown in grey in Fig.~\ref{fig:info-lattice}(a)], which corresponds to the second and third terms in Eq.~\ref{eqn:local-info}.
However, this subtracts the total information contained in the segment ${\cal C}_n^{l-2}$ twice and hence we add it back in the last term in Eq.~\ref{eqn:local-info}.

From an information theoretic viewpoint, $I_n^l$ is simply the quantum conditional mutual information (QCMI)~\cite{fawzi2015quantum} between the leftmost and rightmost qudits of ${\cal C}_n^l$, conditioned on all the qudits in between them,
\eq{
\begin{split}
I_n^l &= \mathfrak{I}(\mathcal C_{n-l/2}^0: \mathcal C_{n+l/2}^0\vert\; \mathcal C_n^{l-2})\,,\\
&=S_{n-1/2}^{l-1} + S_{n+1/2}^{l-1} - S_n^l - S_n^{l-2}\,.
\end{split}
}
Since the state, $\ket{\psi}$, on the entire system is pure, the total von Neumann information in the state if trivially $E_{L/2}^{L-1} = L$.
The local information $I_n^l$ resolves this into length scales $l$ and spatial locations $n$; as such, the set $\{I_n^l\}$ over all $n$ and $l$ completely specifies the local information in each contiguous subsystem of the chain. 
This data, when arranged on the $n-l$ plane, furnishes the information lattice as shown in Fig.~\ref{fig:info-lattice}(b).

A remarkable property of this construction is additivity: the local information on all sites of the information lattice sums to the total von Neumann information of the state, $\sum_{n,l}I_n^l = E_{L/2}^{L-1} = L$.
However, this additivity can be generalised for a subsystem as well.
Consider a subsystem $A\equiv {\cal C}_n^l$, corresponding to which we define the {\it subsystem information lattice} as the set of all the sites on the information lattice lying inside the triangle which extends from the base consisting of the points $(n-l/2,0)$ through $(n+l/2,0)$ to its apex at $(n,l)$; we will refer to the set of these points as $\widetilde{\mathcal C_n^l}$. For instance, the red triangle in Fig.~\ref{fig:info-lattice}(b) denotes $\widetilde{\mathcal C_3^2}$.
The additivity implies the total information in the subsystem ${\cal C}_n^l$ is given by the sum of the information lattice values within $\widetilde{\mathcal C_n^l}$,
\eq{
E_n^l = \sum_{(n,l)\in \widetilde{\mathcal C_n^l}}I_n^l\,.
}
Similar ideas can be used to show that summing up the information lattice values within an arbitrary diamond gives the QCMI between the two subsystems onto which the two lower edges of the diamond project onto conditioned on the intervening subsystems. 
For instance, the grey diamond in Fig.~\ref{fig:info-lattice}(b) corresponds to $\mathfrak{I}({\cal C}_3^2:{\cal C}_{15/2}^1|{\cal C}_{11/2}^1)$.
When the bottom of the diamond includes a site at $l = 1$, there is no intervening subsystem and the above reduces to the quantum mutual information (QMI) between the adjacent subsystems.

In much of what follows, we will be interested in the length scales at which information lives in a quantum state and how does that evolve in time under scrambling dynamics. 
To quantify this, we sum the local information over the subsystem centres to define the information at length scale $l$ as 
\begin{equation}
    \mathcal I(l) = \sum_n I_n^l\,.
\end{equation}
However, to minimise the effects of the boundaries of the system,
it is natural to only work with local information within a subsystem $A$, sufficiently far from boundaries. To this end, we define the information at length scale $l$ within $A$ by summing the $I_n^l$ over sites contained in the associated subsystem information lattice $\widetilde A$,
\begin{equation}
    \mathcal I_A(l) = \sum_{n:\; (n,l) \in \widetilde A} I_n^l.
    \label{eqn:info-per-scale-A}
\end{equation}

\begin{figure}
    \centering
    \includegraphics[width=\linewidth]{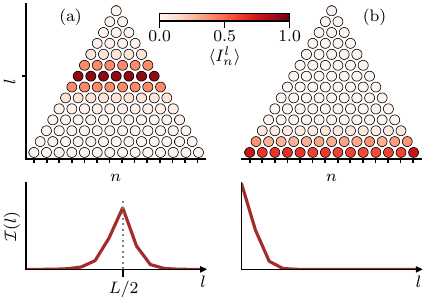}
    \caption{Examples of the anatomy of the information lattice and the local information per scale. The colour intensity denotes the value of $I_n^l$ at the information lattice point $(n,l)$ (a) For Haar-random states, the average $I_{n}^l$ is peaked at $l=L/2$ with an exponential decay away from it. (b) For area-law states, generated by by low-depth local random unitary circuits, the information is almost entirely concentrated on scales of $l\sim O(1)$.}
    \label{fig:lengthscale-info-latt}
\end{figure}

By appropriately normalising this quantity, it can be treated as a distribution over $l$.
This allows us to define a notion of the average length scale at which information resides within $A$ as 
\begin{equation}
     \zeta_A = \frac{1}{E_A}\sum_{l=0}^{L_A}\; l\mathcal I_A(l)\,,
     \label{eqn:expected-length}
\end{equation} 
where the normalisation $E_A = \sum_{l=0}^{L_A} {\cal I}_A(l)$ is just the total von Neumann information in $A$.
By the same token, one can also define the second moment of this distribution
\begin{equation}
    \Delta\zeta^2_A = \frac{1}{E_A}\sum_{l = 0}^{L_A} (l - \zeta_A)^2\;\mathcal I_A(l)  =\langle l^2 \rangle_A - \zeta^2_A,
    \label{eqn:var-length}
\end{equation}
which effectively quantifies the fluctuation in the length scale over which the information resides.
How $\zeta_A$ and $\Delta \zeta_A^2$ changes with time or the depth of the circuit will constitute a central quantity in the rest of the work.

\subsection{Non-local constraints on local information \label{sec:info-latt-constraints}}
The distribution of local information over the entire information lattice is equivalent to the set of von Neumann entropies of all contiguous intervals on the system.
At the same time, the total information contained in the state $\sum_{n,l}=I_n^l=L$ is conserved given the purity of the state. 
As we will show shortly, the dynamics of the local information $I_n^l$ is subject to 
constraints which, in principle, involve all the other information lattice points, and therefore are non-local on the information lattice.
These constraints are imposed fundamentally by the standard entanglement inequalities. 

The positivity of the QCMI, and hence the local information follows from the strong sub-additivity of the von Neumann entanglement entropy.
For a tripartition of the system into subsystems $A$, $B$, and $C$, we have \cite{lieb1973proof} 
\eq{
E_{A\cup B\cup C} + E_B \geq E_{A\cup B} + E_{A\cup C}\,.
}
Identifying $A\cup B\cup C$ as ${\cal C}_n^l$ with $A = \mathcal C_{n-l/2}^0$, $C = \mathcal C_{n+l/2}^0$, and $B = \mathcal C_{n}^{l-2}$, we immediately have $I_n^l \geq 0$. 
The local information is also manifestly constrained by the local Hilbert-space dimension which just gives $I_n^l \leq 2$; this is just the statement that a subsystem with finite Hilbert-space dimension can contain only a finite amount of information.

\begin{figure}
    \centering
    \begin{tikzpicture}[x=0.7cm, y=0.35cm]
\fill[fill=Apricot, rounded corners=6pt] (-0.5,-2) rectangle (2.5,-1);
\fill[fill=Apricot, rounded corners=6pt] (5.5,-2) rectangle (9.5,-1);
\fill[fill=Salmon, rounded corners=6pt] (2.5,-2) rectangle (5.5,-1);
\fill[fill=Salmon, rounded corners=7pt] (4,2.5+0.25) -- ++(-4*0.707/2-0.25,-4*0.707-0.5) -- (4+2*0.707+0.25,2.5 - 4*0.707-0.25) --  cycle;
\fill[fill=Gray!40,rounded corners=5pt] (4,3) -- (2.5,6) -- (1,3) -- (2.5,0) -- cycle;
\fill[fill=Gray!40,rounded corners=5pt] (4,3) -- (5.5,0) --(7.5,4) --(6,7) --cycle;
\foreach \l in {0,...,9}{
    \edef\maxN{\numexpr9-\l\relax}
    \foreach \n in {0,...,\maxN}{
        \filldraw (\n+0.5*\l, \l) circle (1.4pt);
    }   
}
\foreach \l in {0,...,2}{
\node at (-1.2+0.5*\l,\l) {\small $l=\l$};
}
\foreach \l in {0,...,9}{
\filldraw (\l, -1.5) circle (3pt);
}

\draw[thick, dotted] (0,2.8) -- (1-0.2,5-0.6);
\node at (1,-2.5) {{\color{Orange}{$C$}}};
\node at (4,-2.5) {{\color{red!70}{$A$}}};
\node at (7,-2.5) {{{$j$}}};
\node at (8,-2.5) {{\color{Orange}{$B$}}};
\filldraw[rounded corners=5pt, draw= OliveGreen, thick,dashed, fill=
OliveGreen, fill opacity=0.2] (7,-0.75) -- ++(0.25+0.125,0.75) -- ++(-3.75-0.125,7.75) -- ++(-0.25-0.125,-0.75) -- cycle;
\filldraw[rounded corners=5pt, draw= OliveGreen, thick,dashed, fill=
OliveGreen, fill opacity=0.2] (7,-0.75) -- ++(-0.25-0.125,0.75) -- ++(1.25+0.125,2.75) -- ++(0.25+0.125,-0.5-0.25) -- cycle;
\node at (2,6.5) {\small {\color{black!70}{$\mathfrak{I}(A:C)$}}};
\node at (6.8,7.5) {\small {\color{black!70}{$\mathfrak{I}(A:B)$}}};
\draw[thick](0,-1.5) -- (9,-1.5);
\end{tikzpicture}
    \caption{Non-local constraints on the information lattice. The sum of the $I_n^l$ over all points in the grey boxes yield the QMIs, $\mathfrak{I}(A:B)$ and $\mathfrak{I}(A:C)$, and sum over all points in the red triangle equals $E_A$. The sum of these three quantities is constrained by the length of the subsystem $A$ (see Eq.~\ref{eqn:constraint-interval}). The same notion but for a single site, $j$, constrains the total sum of all $I_n^l$ inside the two green boxes to 2 (see Eq.~\ref{eqn:constraint}). }
    \label{fig:constraints}
\end{figure}
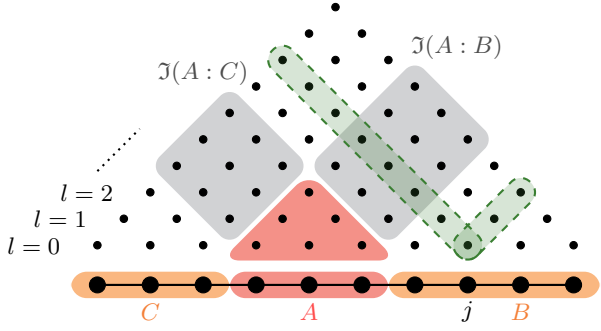

The purity of the state $\ket{\psi}$ leads to the most important constraint on the structure of local information.
For any contiguous interval $A$, let us denote the interval formed by all sites to its right by $B$ and to its left by $C$ (see Fig.~\ref{fig:constraints}). Since the state on $A\cup B\cup C$ is pure, we have 
\begin{equation}
    \mathfrak{I}(A:B) + \mathfrak{I}(A:C) = 2 S_A = 2L_A - 2E_A\,,
    \label{eqn:constraint-interval}
\end{equation}
quite generally.
However, for $A$ a single site, say $j$, such that $A\equiv {\cal C}_j^0$, Eq.~\ref{eqn:constraint-interval} takes the form 
\begin{equation}
    2I_j^0 + \left(\sum_{l=1}^{L-j}I_{j+l/2}^l + \sum_{l=1}^{j}I_{j-l/2}^l\right) = 2 \;\; \forall \;j.
    \label{eqn:constraint}
\end{equation}
Imposing the constraint above for all site $j\in A$ automatically satisfies Eq.~\ref{eqn:constraint-interval} for any $A$.
Diagrammatically, this is straightforward to see from Fig.~\ref{fig:constraints}.
The first term on the left-hand side just corresponds to the local information at the site $j$. 
The second and third terms corresponds to the mutual information between site $j$ and all sites to its right, and between $j$ and all sites to its left respectively -- they correspond to the sum of local informations on sites enclosed by the green boxes in Fig.~\ref{fig:constraints} barring the site $(j,l=0)$.
The key physical import of the constraint in Eq.\ref{eqn:constraint} is that it involves local information across several length scales, and is therefore non-local on the information lattice.
This in turn means that even if the dynamics is effected by a local quantum circuit, the dynamics of the local information is inherently non-local as it involves reorganisation of the information across several length scales in the system.

\subsection{Local information for slowly varying entanglement \label{sec:info-latt-homogeneous}}

The description of the information lattice, so far, has centred around its microscopic discrete structure. 
However, it will be insightful to have a continuum description as well which is meaningful in the limit of large $L$ and when the local information varies slowly in space.
In particular, if the latter is satisfied, it is natural to consider a coarse-grained, local information field which lives on the continuum limit of the information lattice. 
However, developing such a continuum description has two related subtleties. 
First, the effective total information content in the system goes to infinity and second, the effective `local' Hilbert-space dimension per unit length also diverges.
To take care of these two issues, we need to define appropriately rescaled coordinates for the information lattice as well as densities for the von Neumann information or entanglement entropy. 

To this end, we define rescaled coordinates
\eq{
\mathtt n = \epsilon n~~{\rm and}~~\mathtt l = \epsilon l\,,
}
and the length of the full system in these coordinates is given by $\mathtt L = \epsilon L$. 
The densities of von Neumann information and entanglement entropy are defined as
\eq{
\mathtt E(\mathtt n ,\mathtt l) = \epsilon E_{n = \mathtt n/\epsilon}^{l = \mathtt l /\epsilon}~~{\rm and}~~\mathtt S(\mathtt n,\mathtt l) = \epsilon S_{n = \mathtt n/\epsilon}^{l = \mathtt l /\epsilon}
}
respectively. 
In terms of these rescaled quantities, the total information in the entire system must satisfy the relation 
\begin{equation}
    \int_0^{\mathtt L} d\mathtt n\int_0^{\mathtt  L - |\mathtt L - 2\mathtt n|}d\mathtt l\;\mathtt I(\mathtt n,\mathtt l) = \mathtt L\,,
    \label{eqn:continuum-norm}
\end{equation}
which naturally mandates the local information density, defined as 
\eq{
\mathtt I(\mathtt n ,\mathtt l) = \epsilon^{-1} I_{n = \mathtt n/\epsilon}^{l = \mathtt l /\epsilon}\,,
}
as the continuum version of local information.
Even though $\mathtt{E(n,l)}$ and $\mathtt{S(n,l)}$ are, strictly speaking, densities, we will use the same terminology for the continuum and discrete versions of the quantities with the distinction understood from context and notation.

Within the continuum description, the relation between the local information the total information in contiguous segments, analogous to Eq.~\ref{eqn:local-info} is given by
\begin{equation}
    \mathtt I(\mathtt n ,\mathtt l) = \left(\partial^2_\mathtt{l} - \frac{1}{4}\partial_\mathtt{n}^2\right)\mathtt E(\mathtt n,\mathtt l)\,.
    \label{eqn:loc-info-cont}
\end{equation}
In addition, the non-local constraint in Eq.~\ref{eqn:constraint} takes the form 
\begin{equation}
    \int d\mathtt l d\mathtt n\;\frac{\mathtt I(\mathtt {n,l})}{2}\left[\delta\left(\mathtt n - \mathtt j + \frac{\mathtt l}{2}\right) + \delta\left(\mathtt n - \mathtt j -  \frac{\mathtt l}{2}\right)\right] = 1.
    \label{eqn:constr-cont}
\end{equation}
This sets up the formulation of the information lattice in the continuum limit which is expected to be meaningful when the entanglement varies slowly in space and across scales. In other words, $\mathtt{E(n,l)}$ is a smooth function of $\mathtt{(n,l)}$.

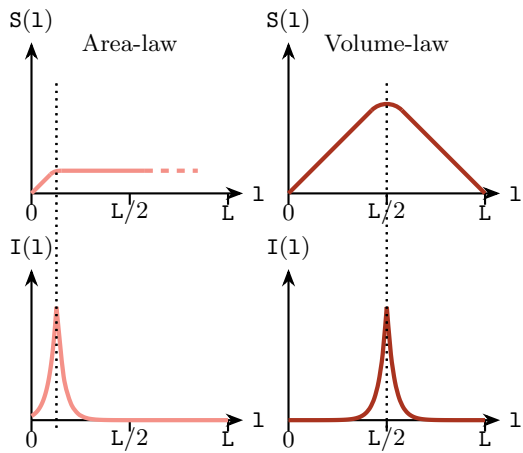
\begin{figure}
\centering
\begin{tikzpicture}
\draw[-Stealth,thick](0,0) -- (2.8,0) node[right]{$\mathtt{l}$};
\draw[-Stealth,thick](0,0) -- (0,2) node[above]{$\mathtt{S(l)}$};
\draw[-Stealth,thick](3.4,0) -- (6.2,0) node[right]{$\mathtt{l}$};
\draw[-Stealth,thick](3.4,0) -- (3.4,2) node[above]{$\mathtt{S(l)}$};
\draw[-Stealth,thick](0,-3) -- (2.8,-3) node[right]{$\mathtt{l}$};
\draw[-Stealth,thick](0,-3) -- (0,-1) node[above]{$\mathtt{I(l)}$};
\draw[-Stealth,thick](3.4,-3) -- (6.2,-3) node[right]{$\mathtt{l}$};
\draw[-Stealth,thick](3.4,-3) -- (3.4,-1) node[above]{$\mathtt{I(l)}$};
\foreach \x in {0,1.3,2.6}{
\draw[thick] (\x,0) -- ++(0,-0.15);
\draw[thick] (\x+3.4,0) -- ++(0,-0.15);
\draw[thick] (\x,-3) -- ++(0,-0.15);
\draw[thick] (\x+3.4,-3) -- ++(0,-0.15);
}
\draw[ultra thick,Salmon](0,0) -- (0.25,0.25);
\draw[ultra thick,Salmon](0.25,0.25) to[out=45,in=180] (0.4,0.3);
\draw[ultra thick,Salmon](0.4,0.3) -- (1.5,0.3);
\draw[dashed, ultra thick,Salmon](1.5,0.3) -- (2.2,0.3);
\draw[ultra thick,Mahogany](3.4,0) -- (4.5,1.1);
\draw[ultra thick,Mahogany](4.5,1.1) to[out=45,in=135] (4.9,1.1);
\draw[ultra thick,Mahogany](4.9,1.1) -- (6,0);
\draw[domain=0.33:2.6, smooth, variable=\x, Salmon, ultra thick] plot (\x, -3+1.5*exp{-10*(\x-0.33)});
\draw[domain=0:0.33, smooth, variable=\x, Salmon, ultra thick] plot (\x, -3+1.5*exp{-10*(0.33-\x)});
\draw[domain=3.4+1.3:3.4+2.6, smooth, variable=\x, Mahogany, ultra thick] plot (\x, -3+1.5*exp{-10*(\x-4.7)});
\draw[domain=3.4:4.7, smooth, variable=\x, Mahogany, ultra thick] plot (\x, -3+1.5*exp{-10*(4.7-\x)});
\draw[thick, dotted] (.33,-3.1) -- (.33,1.5);
\draw[thick, dotted] (.33,-3.1) -- (.33,1.5);
\draw[thick, dotted] (4.7,-3.1) -- (4.7,1.5);
\draw[thick, dotted] (4.7,-3.1) -- (4.7,1.5);
\foreach \x in {0,3.4}{
\foreach \y in {0,-3}{
\node at (\x,\y-0.25) {\small{$0$}};
\node at (\x+1.3,\y-0.25) {\small{$\mathtt{L}/2$}};
\node at (\x+2.6,\y-0.25) {\small{$\mathtt{L}$}};
}
}
\node at (1.3,2) {Area-law};
\node at (1.3+3.4,2) {Volume-law};

\end{tikzpicture}
\caption{The profiles for entanglement, $\mathtt{S(n,l)}$, in area-law and volume-law entangled states as a function of the subsystem size $\mathtt l$. Assuming these profiles are homogeneous, (negligible dependence on $\mathtt{n}$), Eq.~\ref{eqn:second-deriv} yields profiles for $\mathtt{I(l)}$ which are shown in the lower panels. }
\label{fig:sec-deriv}
\end{figure}

In fact, this formulation already leads to some useful insights if the entanglement is homogeneous in space, such that $\mathtt{E(n,l)}$ is a function only of $\mathtt{l}$.
In such a scenario, Eq.~\ref{eqn:loc-info-cont} reduces to 
\begin{equation}
    \mathtt I(\mathtt n,\mathtt l) \approx \partial_\mathtt{l}^2 \mathtt E(\mathtt l) = -\partial_\mathtt{l}^2 \mathtt S(\mathtt l)\,.
    \label{eqn:second-deriv}
\end{equation}
Note that the subadditivity of entanglement implies that $\mathtt E(\mathtt l)$ is a concave (downwards) function of $\mathtt l$ which in turn again guarantees the positivity of the local information field.
For a subsystem well inside the bulk of the system such that boundary effects can be neglected, the shape of $\mathtt S(\mathtt l)$ is qualitatively described by a rather generic forms which in turn lends itself to generic forms of $\mathtt{I(l)}$ as shown in Fig.~\ref{fig:sec-deriv}.
For area- and volume-law entangled states, $\mathtt S(\mathtt l)$ grows with $\mathtt l$, no faster than linearly, until it reaches an $O(1)$ or extensive ($\propto \mathtt L$) value,  respectively, at a corresponding $O(1)$ or extensive value of $\mathtt l$. 
For area-law state $\mathtt S(\mathtt l)$ saturates at this $O(1)$ value. 
For volume-law states in fact, $\mathtt S(\mathtt l)$ decays once $\mathtt l > \mathtt L /2$ as the profile of $\mathtt S(\mathtt l)$ is symmetric around $\mathtt l = \mathtt L/2$.
For such profiles, it is straightforward to see that Eq.~\ref{eqn:second-deriv} mandates a sharp peak in $\mathtt{I(l)}$.
The peak is located at an $O(1)$ value of $\mathtt l$ for area-law states, which corresponds precisely to the length scale at which the entanglement saturates.
On the other hand, for volume-law states, the peak is naturally at $\mathtt{l} = \mathtt{L}/2$.

To make this concrete, let us consider typical states from the Hilbert-space which on an average have maximal volume-law entanglement modulo the Page correction~\cite{page1993average}.
To leading order, the entanglement of a subsystem in such states scales linearly with the length of the subsystem,
\eq{
S_n^l = \mathrm{min}\left\{l+1,L-l -1\right\} + {\rm corrections}\,,
}
which immediately gives $I_n^l = 2\delta_{l, L/2} + {\rm corrections}$.
In the continuum formulation, the subsystem intervals $\mathtt l \ll \mathtt L/2$ or $\mathtt L-\mathtt l \ll \mathtt L/2$, the entanglement density is linear in $\mathtt l$ with corrections which are exponential in $\mathtt{l}$.
More specifically,
\begin{equation}
    \langle \mathtt S(\mathtt n , \mathtt l)\rangle_{\rm Haar}  \sim \mathtt l - \frac{\mathtt q^{-\vert 2\mathtt l - \mathtt L\vert}}{2\ln \mathtt q}\,,
\end{equation}
where $\mathtt q = q^{1/\epsilon}$ is the dimension of the Hilbert space of an interval of unit unit length. 
Using Eq.~\ref{eqn:second-deriv}, we therefore obtain
\begin{equation}
    \langle \mathtt I(\mathtt n,\mathtt l)\rangle_{\rm Haar} \sim 2 \ln \mathtt q\; e^{-2\ln \mathtt q \vert \mathtt L/2-\mathtt l \vert}\,,
    \label{eqn:loc-info-haar-state}
\end{equation}
which shows that the local information density is peaked at $\mathtt{l=L}/2$ and decays exponentially away from it. The latter stems from the exponential correction to the volume law in the entanglement entropy. 
An straightforward but important fallout of this is that in area-law states, the average length scale at which information resides is a finite, $\zeta\sim O(1)$ whereas in the volume-law states it is extensive, $\zeta\sim L$.
Looking ahead, this already suggests that in scrambling dynamics, the time-dependence of $\zeta$ can be used to define a {\it information velocity} as one one measure for quantifying how information spreads from smaller to larger scales. 

\section{Dynamics of local information for generic local circuits \label{sec:res-general}}
In this section, we discuss the dynamical anatomy of the information lattice and the define information currents thereon.
The discussion is completely general and holds for arbitrary local quantum circuits. 
We define a local random circuit to be one where all the gates specifying the circuit act on nearest-neighbour sites.
While in subsequent sections, we discuss specific ensembles from which the gates are drawn as well as specific architectures of the circuit, the results presented here are independent of those.

The motivation for considering circuits is that they have recently emerged as minimal models of information scrambling in locally interacting systems offering significantly more scope of tractability than Hamiltonian systems~\cite{nahum2017quantum,nahum2018operator,keyserlingk2018operator,rakovszky2018diffusive,khemani2018operator,zhou2019emergent,fisher2023random}. 
The locality of the circuit, as defined above, suggests that a theory for the dynamics of local information can be developed by understanding the effect of a single gate on the entire information lattice as the basic building block. 
As we will see shortly, even though such a gate is entirely local, its application affects the information lattice at all length scales due to the non-local constraints in Eq.~\ref{eqn:constraint}.

As long as the set of local unitaries is not tuned to do so, random circuits do not host any conservation quantities (not even energy as continuous time-translation symmetry is also broken). In the rest of what follows, we will consider circuits which have no conservation laws.
This leads to added simplifications as the dynamics of information and its steady state is free from any constraints that originate from the presence of conservation laws. There is no late time diffusive hydrodynamics and the subsystem entanglement at late times is that of an infinite temperature state over the entire Hilbert space.
Our setup consists of a system in an initially disentangled pure state which evolves under the action of random local unitaries where the nearest-neighbour two-site gates are placed randomly in spacetime. Since the evolution is unitary, the system continues to be in a pure state at all times, implying that the von Neumann information of the full system (or the total local information) remains conserved.

\subsection{Local unitary operations and information currents} \label{sec:loc-u-info-curr}

\begin{figure}
\begin{tikzpicture}[x=0.7cm, y=0.35cm]
\filldraw[draw= Mahogany, thick,dashed, fill=
Mahogany, fill opacity=0.2,rounded corners=5pt]  (3.25,6.5) to[out=135,in=135] (2.25,4.5) -- (4.75,-0.5) -- (3.25+3.5,-0.5) -- cycle;
\filldraw[draw= YellowOrange, thick,dashed, fill=
YellowOrange, fill opacity=0.2,rounded corners]  (7.75,2.5) to[out=45,in=45] (6.75,4.5) -- ++(-2.5,-5) -- (7.75-1.5,2.5-3) -- cycle;
\fill[rounded corners=7pt,fill=
gray, fill opacity=0.4] (8,-0.5-0.5) -- ++(0.5,1) -- ++(-4.5,9) -- ++(-0.5,-1) -- cycle;
\foreach \l in {0,...,9}{
    \edef\maxN{\numexpr9-\l\relax}
    \foreach \n in {0,...,\maxN}{
        \filldraw (\n+0.5*\l, \l) circle (1.4pt);
    }   
}
\foreach \l in {0,...,9}{
\filldraw (\l, -3) circle (3pt);
}
\foreach \i in {0,...,5}{
\draw[>=stealth, <->, Mahogany, thick] (5-\i*0.5+0.05,\i+0.1)-- ++ (0.4,0.8);
}
\foreach \i in {0,...,3}{
\draw[>=stealth, <->, Orange, thick] (6+\i*0.5-0.05,\i+0.1)-- ++ (-0.4,0.8);
}
\draw[>=stealth, <->, gray, thick] (5.1,0)-- ++ (0.8,0);
\draw[ thick] (5,-3) -- ++(0,2.25);
\draw[ thick] (6,-3) -- ++(0,2.25);
\filldraw[draw=black, thick, fill=Mahogany, rounded corners] (4.75,-2.25) rectangle (6.25,-1.25);
\draw[thick](0,-3) -- (9,-3);

\node at (5,-3.75) {$i$};
\node at (6,-3.75) {$i+1$};
\node at (8,-3.75) {$j$};

\node at (0,9) {(a)};
\end{tikzpicture}
\begin{tikzpicture}[x=0.7cm, y=0.35cm]
\def\sc{1.5};
\foreach \x in {1,2,3,4}{
\fill[black!20] (\sc*\x,\sc*2) circle (2.5pt);
\fill[black!20] (\sc*\x,\sc*4) circle (2.5pt);
}
\foreach \x in {1,2,3}{
\draw[>=stealth, ->, Mahogany!20, ultra thick] (\sc*\x+\sc*0.05,\sc*2+\sc*0.1)-- ++(\sc*0.4,\sc*0.8);
\draw[>=stealth, ->, Mahogany!20, ultra thick] (\sc*\x+\sc*0.05,\sc*4+\sc*0.1)-- ++(\sc*0.4,\sc*0.8);
\draw[>=stealth, ->, YellowOrange!20, ultra thick] (\sc*\x+\sc-\sc*0.05,\sc*2+\sc*0.1)-- ++(-\sc*0.4,\sc*0.8);
\draw[>=stealth, ->, YellowOrange!20, ultra thick] (\sc*\x+\sc-\sc*0.05,\sc*4+\sc*0.1)-- ++(-\sc*0.4,\sc*0.8);
}
\foreach \x in {1,2,3}{
\fill[black!20] (\sc*\x+\sc*0.5,\sc*1) circle (2.5pt);
\fill[black!20] (\sc*\x+\sc*0.5,\sc*3) circle (2.5pt);
\fill[black!20] (\sc*\x+\sc*0.5,\sc*5) circle (2.5pt);
\draw[>=stealth, ->, Mahogany!20, ultra thick] 
(\sc*\x+\sc*0.5+\sc*0.05,\sc*1+\sc*0.1)-- ++(\sc*0.4,\sc*0.8);
\draw[>=stealth, ->, Mahogany!20, ultra thick] (\sc*\x+\sc*0.5+\sc*0.05,\sc*3+\sc*0.1)-- ++(\sc*0.4,\sc*0.8);
\draw[>=stealth, ->, YellowOrange!20, ultra thick] 
(\sc*\x+\sc*0.5-\sc*0.05,\sc*1+\sc*0.1)-- ++(-\sc*0.4,\sc*0.8);
\draw[>=stealth, ->, YellowOrange!20, ultra thick] (\sc*\x+\sc*0.5-\sc*0.05,\sc*3+\sc*0.1)-- ++(-\sc*0.4,\sc*0.8);
}
\fill[black] (\sc*2.5,\sc*3) circle (2.5pt);
\draw[>=stealth, ->, ultra thick,YellowOrange] (\sc*2.5-\sc*0.05,\sc*3+\sc*0.1)-- ++(-\sc*0.4,\sc*0.8);
\draw[>=stealth, <-, ultra thick,YellowOrange] (\sc*2.5+\sc*0.05,\sc*3-\sc*0.1)-- ++(\sc*0.4,-\sc*0.8);
\draw[>=stealth, ->, ultra thick,Mahogany] (\sc*2.5+\sc*0.05,\sc*3+\sc*0.1)-- ++(\sc*0.4,\sc*0.8);
\draw[>=stealth, <-, ultra thick,Mahogany] (\sc*2.5-\sc*0.05,\sc*3-\sc*0.1)-- ++(-\sc*0.4,-\sc*0.8);

\node at (\sc*2.5+1.0,\sc*3-0.25) {\small $(n,l)$};
\node at (\sc*2.5+\sc,\sc*3+1) {{\color{Mahogany}\small $J_+(n,l)$}};
\node at (\sc*2.5-\sc,\sc*3+1) {{\color{Orange}\small $J_-(n,l)$}};

\foreach \x in {-10,...,-1}{
\draw[thick](\x*0.5+0.25,\sc*1) -- (\x*0.5+0.25,\sc*5);
}
\def\x{-10};
\def\t{1};
\filldraw[draw=black, thick, fill = Mahogany, rounded corners=2](\x*0.5-0.1+0.25,\sc*\t+0.25) rectangle (\x*0.5+0.5+0.1+0.25,\sc*\t+0.75);
\def\x{-5};
\def\t{1};
\filldraw[draw=black, thick, fill = Mahogany, rounded corners=2](\x*0.5-0.1+0.25,\sc*\t+0.25) rectangle (\x*0.5+0.5+0.1+0.25,\sc*\t+0.75);
\def\x{-3};
\def\t{1};
\filldraw[draw=black, thick, fill = Mahogany, rounded corners=2](\x*0.5-0.1+0.25,\sc*\t+0.25) rectangle (\x*0.5+0.5+0.1+0.25,\sc*\t+0.75);
\def\x{-8};
\def\t{1.5};
\filldraw[draw=black, thick, fill = Mahogany, rounded corners=2](\x*0.5-0.1+0.25,\sc*\t+0.25) rectangle (\x*0.5+0.5+0.1+0.25,\sc*\t+0.75);
\def\x{-2};
\def\t{1.5};
\filldraw[draw=black, thick, fill = Mahogany, rounded corners=2](\x*0.5-0.1+0.25,\sc*\t+0.25) rectangle (\x*0.5+0.5+0.1+0.25,\sc*\t+0.75);
\def\x{-2};
\def\t{1.5};
\filldraw[draw=black, thick, fill = Mahogany, rounded corners=2](\x*0.5-0.1+0.25,\sc*\t+0.25) rectangle (\x*0.5+0.5+0.1+0.25,\sc*\t+0.75);
\def\x{-6};
\def\t{1.75};
\filldraw[draw=black, thick, fill = Mahogany, rounded corners=2](\x*0.5-0.1+0.25,\sc*\t+0.25) rectangle (\x*0.5+0.5+0.1+0.25,\sc*\t+0.75);
\def\x{-10};
\def\t{2.25};
\filldraw[draw=black, thick, fill = Mahogany, rounded corners=2](\x*0.5-0.1+0.25,\sc*\t+0.25) rectangle (\x*0.5+0.5+0.1+0.25,\sc*\t+0.75);
\def\x{-4};
\def\t{2.5};
\filldraw[draw=black, thick, fill = Mahogany, rounded corners=2](\x*0.5-0.1+0.25,\sc*\t+0.25) rectangle (\x*0.5+0.5+0.1+0.25,\sc*\t+0.75);
\def\x{-2};
\def\t{2.6};
\filldraw[draw=black, thick, fill = Mahogany, rounded corners=2](\x*0.5-0.1+0.25,\sc*\t+0.25) rectangle (\x*0.5+0.5+0.1+0.25,\sc*\t+0.75);
\def\x{-8};
\def\t{3};
\filldraw[draw=black, thick, fill = Mahogany, rounded corners=2](\x*0.5-0.1+0.25,\sc*\t+0.25) rectangle (\x*0.5+0.5+0.1+0.25,\sc*\t+0.75);
\def\x{-4};
\def\t{3.2};
\filldraw[draw=black, thick, fill = Mahogany, rounded corners=2](\x*0.5-0.1+0.25,\sc*\t+0.25) rectangle (\x*0.5+0.5+0.1+0.25,\sc*\t+0.75);
\def\x{-6};
\def\t{3.5};
\filldraw[draw=black, thick, fill = Mahogany, rounded corners=2](\x*0.5-0.1+0.25,\sc*\t+0.25) rectangle (\x*0.5+0.5+0.1+0.25,\sc*\t+0.75);
\def\x{-10};
\def\t{3.8};
\filldraw[draw=black, thick, fill = Mahogany, rounded corners=2](\x*0.5-0.1+0.25,\sc*\t+0.25) rectangle (\x*0.5+0.5+0.1+0.25,\sc*\t+0.75);
\def\x{-3};
\def\t{4};
\filldraw[draw=black, thick, fill = Mahogany, rounded corners=2](\x*0.5-0.1+0.25,\sc*\t+0.25) rectangle (\x*0.5+0.5+0.1+0.25,\sc*\t+0.75);
\node at (-5.5,8) {(b)};
\node at (1,8) {(c)};

\def\x{-9};
\def\t{4.25};
\filldraw[draw=black, thick, fill = Mahogany, rounded corners=2](\x*0.5-0.1+0.25,\sc*\t+0.25) rectangle (\x*0.5+0.5+0.1+0.25,\sc*\t+0.75);

\def\x{-7};
\def\t{2.25};
\filldraw[draw=black, thick, fill = Mahogany, rounded corners=2](\x*0.5-0.1+0.25,\sc*\t+0.25) rectangle (\x*0.5+0.5+0.1+0.25,\sc*\t+0.75);

\end{tikzpicture}
\caption{(a) Flow of local information on the information lattice under the action of a 2-qudit local unitary gate sites $i$ and $i+1$. The flows are confined within the $\mathsf{V}$-shaped region enclosed by the red and yellow boxes. Conservation of the total information within the grey boxes (due to Eq.~\ref{eqn:constraint} being satisfied for all $j$) mandates the information flows only diagonally as indicated by the arrows except at $l=0$. (b) Illustration of a random circuit composed solely of nearest-neighbour gates. (c) For a generic circuit of the form in (b), currents on the information lattice flow diagonally which lets us define $J_\pm(n,l)$ with the arrow direction denoting the sign convention.}
\label{fig:gen-dynamics}
\end{figure}
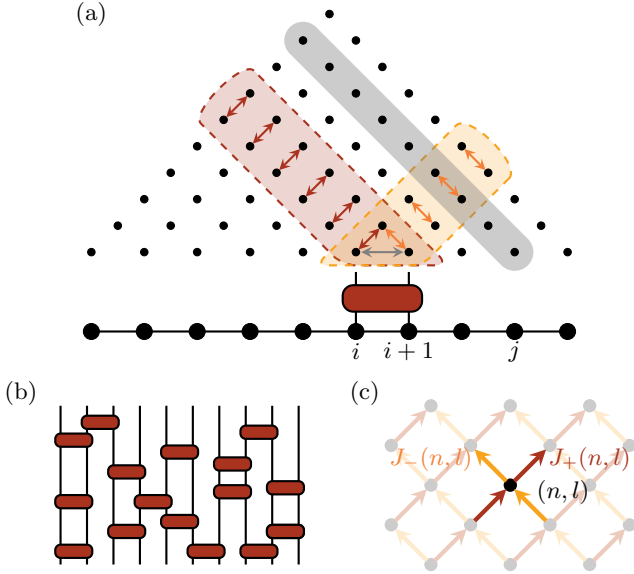

To understand the dynamics on the information lattice effected by a single gate, consider the action of a local unitary operator on sites $i$ and $i+1$ as shown in Fig.~\ref{fig:gen-dynamics}. 
This operation can bring about a change in the von Neumann information of only those subsystems which have their right boundary at site $i$ and those which have their left boundary at site $i+1$.
Formally, this implies that the change in the information, $\Delta E_{i-l/2}^l\neq 0$ and $\Delta E_{i+1+l/2}^l\neq 0$ for all admissible values $l$ constrained by the triangular shape of the information lattice. 
Using this in Eq.~\ref{eqn:local-info}, one concludes that the gate on sites $i$ and $i+1$ leads to a finite change in the local information
\eq{
\Delta I_n^l \neq 0~{\rm for}~n = i\pm \frac{l}{2}~{\rm and}~n=i+1 \pm \frac{l}{2}\,.
}
This defines a $\mathsf{V}$-shaped region on the information lattice, depicted in Fig.~\ref{fig:gen-dynamics} via the red and orange dashed boxes, such that it is only the points enclosed by them which see a finite change in their corresponding $I_n^l$ values.
This, in turn, means that all the information currents are confined within this $\mathsf{V}$-shaped region -- this constraint on the information flows arises purely from the locality of the dynamics due to a nearest-neighbour gate.

We next impose the constraint in Eq.~\ref{eqn:constraint} on the information currents. 
Note that irrespective of which nearest-neighbour sites the gate acts on, Eq.~\ref{eqn:constraint} must be satisfied for sites.
To this end, consider a site $j>i+1$ as shown in Fig.~\ref{fig:gen-dynamics}.
Imposing Eq.~\ref{eqn:constraint} for site $j$ implies that the total sum of $I_n^l$ on all the sites inside the grey boxes must be conserved. 
However, the local information changes only on the two points that lie in the intersection of the grey boxes and the yellow box and hence, the local information change on the points must be equal and opposite. 
Since this must hold for all $j>i+1$ we have,
\begin{equation}
    \Delta I_{i+1+l/2}^l = -\Delta I^{l+1}_{i+(l+1)/2} \;\;\;\;\forall\; l\geq 1\,.
\end{equation}
The upshot of this is that all the information currents in the right arm of the $\mathsf{V}$-shaped region (yellow box in Fig.~\ref{fig:gen-dynamics}) are oriented diagonally as $\NWarrow$.
Similarly, imposing Eq.~\ref{eqn:constraint} for all $j<i$, we obtain
\begin{equation}
    \Delta I_{i-l/2}^l = -\Delta I^{l+1}_{i+1-(l+1)/2} \;\;\;\;\forall\; l\geq 1\,.
\end{equation}
This amounts to all the currents in the left arm of the $\mathsf{V}$-shaped region being oriented diagonally as $\NEarrow$.

In addition, since the gate cannot change the entanglement of the subsystem $i\cup (i+1)$ with the rest of the system, the total information contained within the three points $(i,0)$, $(i+1,0)$ and $(i+1/2,1)$ is also conserved. 
However, it can reorganise the information within these sites and hence, there exists finite information current between any pair of these three sites, including a horizontal current between $(i,0)$ and $(i+1,0)$.
Using this conservation law along with invoking Eq.~\ref{eqn:constraint} leads to the following result. Considering a bipartition of the system by a cut placed between sites $i$ and $i+1$, the net change in the total information contained in the two subsystems is the same. 
This is consistent with the fundamental constrain placed by the unitarity of the dynamics, namely, the change in entanglement of the two subsystems across a bipartition must be the same.

The picture that emerges therefore is that under the action of a nearest-neighbour gate, the flows on the information lattice are confined to the arms of a $\mathsf{V}$-shaped region with its base supported on the two sites on which the gate acts. 
More importantly, the currents are constrained to point along the diagonal directions, $\NEarrow$ and $\NWarrow$. 
Therefore, for circuits constructed entirely out of nearest-neighbour gates (such as in Fig.~\ref{fig:gen-dynamics}(b)), there exist currents on the information lattice only between sites $(n,l)$ and $(n\pm 1/2,l\pm 1)$. This is represented schematically in Fig.~\ref{fig:gen-dynamics}(c), which shows that the change in local information with time can be described locally on the information lattice by defining the components of the information current $J_+(n,l)$ and $J_-(n,l)$ which point diagonally to the right and left respectively\footnote{We focus on the currents in the bulk of the information lattice and ignore the currents in the horizontal direction at $l=0$}. 
This leads to a difference equation for the local information
\eq{
\begin{split}
\Delta{I}_{n}^l = &-\left[J_+(n,l)+J_-(n,l)\right]
\\
&+J_+\left(n-\frac{1}{2},l-1\right)+J_-\left(n+\frac{1}{2},l-1\right)\,,
\end{split}
}
where use the convention that $J_{\pm}(n,l)$ denotes the current flowing from $(n,l)$ to $(n\pm 1/2,l+1)$.
The continuum limit of the above equation takes the form of a continuity equation
\begin{multline}
    \partial_t\mathtt I(\mathtt n , \mathtt l,t) = -\left(\partial_{\mathtt l} + \frac{1}{2}\partial_{\mathtt n}\right) J_+(\mathtt n,\mathtt l,t) \\- \left(\partial_{\mathtt l} - \frac{1}{2}\partial_{\mathtt n}\right) J_-(\mathtt n,\mathtt l,t)\,.
    \label{eq:eqn-continuity}
\end{multline}
Note that even though the flow of information itself is local, the currents $J_\pm$ are in general expected depend on the configuration of local information $\{I_n^l\}$
in a non-local manner due to the constraints imposed by Eq.~\ref{eqn:constraint}. So the hydrodynamic equation for the flow of local information  for a given class of unitary gates will inherently contain non-local terms. 
We will therefore focus primarily on understanding the dynamics of local information from a microscopic point of view and briefly return to the prospects of a hydrodynamic description in Sec.~\ref{sec:hydro}.

\subsection{Leading order behaviour} \label{sec:leading-order}

Before moving to a detailed microscopic analysis for certain classes of circuits, we develop a bare-bones picture at leading order for the flow of local information independent of the specific class of the circuit under consideration.
This is based on the generic features of entanglement dynamics in random circuit models~\cite{nahum2017quantum,jonay2018coarse,zhou2019emergent,fisher2023random}. 
At a coarse gained level, the bipartite entanglement of a finite interval of length $\mathtt l \ll \mathtt L$ located sufficiently away from the boundaries grows linearly with time as $2s_{\rm eq}v_Et$ \cite{nahum2017quantum}, where $s_{\rm eq}$ is the  equilibrium entropy density which is simply given by its infinite temperature value of $\log_q q = 1$, and $v_E$ is the entanglement velocity. 
The factor of 2 comes from the fact that the subsystem has two boundaries, each producing entanglement at rate $s_{\rm eq}v_E$. 
This linear in $t$ growth in entanglement continues until it saturates to an equilibrium value of $s_{\rm eq}\mathtt l$ at time  $\mathtt l/v_E$.

As a leading order approximation, let us assume that the crossover from ballistic growth to saturation is sharp, which implies $\mathtt S(\mathtt n,\mathtt l,t) = \text{min}\left\{2s_{eq}v_Et,\; s_{eq}\mathtt l\right\}$. 
Using Eq.~\ref{eqn:second-deriv}, we obtain $\mathtt I(\mathtt{n,l},t) \propto \delta (\mathtt l - 2v_Et)$ which is an infinitely sharp peak propagating with a velocity $2v_E$. This highlights a key feature generic to all setting where entanglement grows ballistically: at any intermediate time $t$, the local information peaks at a length scale which is the exactly the size of the subsystems whose entanglement saturates at $t$.

\begin{figure}
\begin{tikzpicture}[scale=2]
\def\x{0};
\def\y{0};
\draw[dashed] (\x+0.25,\y+0.5) -- ++(0.5,0);
\draw[line width=2, Mahogany!25] (\x,\y+0.025) -- ++(1,0);
\draw[line width=1.5, Mahogany!50] (\x,\y+0.015) -- ++(1,0);
\draw[line width=1, Mahogany!75] (\x,\y+0.0055) -- ++(1,0);
\draw[line width=0.5, Mahogany!100] (\x,\y+0.0055) -- ++(1,0);
\draw[thick] (\x,\y) -- ++(1,0) -- ++(-0.5,1) -- cycle;
\node at (\x+0.5,\y+1.2) {(a) $t=0$};
\node at (\x-0.2,\y) {$\mathtt{l}=0$};
\node at (\x-0.25+0.25,\y+0.5) {$\mathtt{l}=\frac{\mathtt{L}}{2}$};

\def\x{1.25};
\def\y{0};

\draw[line width=2, Mahogany!16] (\x+0.0625,\y+0.125) -- (\x+0.25,\y+0.25);
\draw[line width=1.5, Mahogany!33] (\x+0.0625,\y+0.125) -- (\x+0.25,\y+0.25);
\draw[line width=1, Mahogany!50] (\x+0.0625,\y+0.125) -- (\x+0.25,\y+0.25);
\draw[line width=0.5, Mahogany!66] (\x+0.0625,\y+0.125) -- (\x+0.25,\y+0.25);
\draw[line width=2, Mahogany!25] (\x+0.25,\y+0.25) -- (\x+0.75,\y+0.25); 
\draw[line width=1.5, Mahogany!50] (\x+0.25,\y+0.25) -- (\x+0.75,\y+0.25); 
\draw[line width=1, Mahogany!75] (\x+0.25,\y+0.25) -- (\x+0.75,\y+0.25); 
\draw[line width=0.5, Mahogany!100] (\x+0.25,\y+0.25) -- (\x+0.75,\y+0.25); 
\draw[line width=2, Mahogany!16] (\x+0.75,\y+0.25) -- (\x+1-0.0625,\y+0.125);
\draw[line width=1.5, Mahogany!33] (\x+0.75,\y+0.25) -- (\x+1-0.0625,\y+0.125);
\draw[line width=1, Mahogany!50] (\x+0.75,\y+0.25) -- (\x+1-0.0625,\y+0.125);
\draw[line width=0.5, Mahogany!66] (\x+0.75,\y+0.25) -- (\x+1-0.0625,\y+0.125);
\draw[thick] (\x,\y) -- ++(1,0) -- ++(-0.5,1) -- cycle;
\draw[dashed] (\x+0.25,\y+0.5) -- ++(0.5,0);
\node at (\x+0.5,\y+1.2) {(b) $t<\frac{\mathtt{L}}{4v_E}$};

\draw[gray, thick] (\x+.4,\y+0.15) rectangle (\x+0.6,\y+0.35);
\draw[>= stealth, ->, thick, gray] (\x+0.6,\y+0.35) to[out=45,in=180](\x+0.6+0.65,\y+0.35+0.2);

\def\x{2.5};
\def\y{0};
\draw[thick] (\x,\y+0.75) node[above]{\small $\mathtt{I(l)}$} -- ++(0,-0.75) -- ++(1,0) node[right]{\small $\mathtt{l}$};
\draw[domain=\x:\x+1, smooth, variable=\z, Mahogany, very thick] plot (\z, \y+0.65*exp{-500*(\z-\x-0.25)*(\z-\x-0.25)});
\draw[thick,dotted] (\x+0.25,-0.025) -- ++(0,0.5); 
\node at (\x+0.25,\y-0.1) {\small $2v_E t$};
\draw[>=stealth, ->, thick] (\x+0.25,\y+0.5) -- ++(0.25,0);
\node at (\x+0.65,\y+0.5) {\small $2v_E$};

\def\x{0};
\def\y{-1.5};

\draw[line width=2, Mahogany!16] (\x+0.175,\y+0.35) -- (\x+0.4,\y+0.5);
\draw[line width=1.5, Mahogany!33] (\x+0.175,\y+0.35) -- (\x+0.4,\y+0.5);
\draw[line width=1, Mahogany!50] (\x+0.175,\y+0.35) -- (\x+0.4,\y+0.5);
\draw[line width=0.5, Mahogany!66] (\x+0.175,\y+0.35) -- (\x+0.4,\y+0.5);
\draw[line width=2, Mahogany!25] (\x+0.4,\y+0.5) -- (\x+0.6,\y+0.5); 
\draw[line width=1.5, Mahogany!50] (\x+0.4,\y+0.5) -- (\x+0.6,\y+0.5); 
\draw[line width=1, Mahogany!75] (\x+0.4,\y+0.5) -- (\x+0.6,\y+0.5); 
\draw[line width=0.5, Mahogany!100] (\x+0.4,\y+0.5) -- (\x+0.6,\y+0.5); 
\draw[line width=2, Mahogany!16] (\x+1-0.175,\y+0.35) -- (\x+0.6,\y+0.5);
\draw[line width=1.5, Mahogany!33] (\x+1-0.175,\y+0.35) -- (\x+0.6,\y+0.5);
\draw[line width=1, Mahogany!50] (\x+1-0.175,\y+0.35) -- (\x+0.6,\y+0.5);
\draw[line width=0.5, Mahogany!66] (\x+1-0.175,\y+0.35) -- (\x+0.6,\y+0.5);
\draw[thick] (\x,\y) -- ++(1,0) -- ++(-0.5,1) -- cycle;
\draw[dashed] (\x+0.25,\y+0.5) -- ++(0.5,0);
\node at (\x+0.5,\y+1.2) {\small (c) $\frac{\mathtt{L}}{4v_E}<t<\frac{\mathtt{L}}{2v_E}$};
\draw[gray, thick] (\x+.4,\y+0.4) rectangle (\x+0.6,\y+0.6);
\draw[>= stealth, ->, thick, gray] (\x+0.6,\y+0.6) to[out=45,in=180](\x+0.6+1.9,\y+0.6+0.1);

\node at (\x-0.2,\y) {$\mathtt{l}=0$};
\node at (\x-0.25+0.25,\y+0.5) {$\mathtt{l}=\frac{\mathtt{L}}{2}$};

\def\x{1.25};
\def\y{-1.5};
\draw[line width=2, Mahogany!25] (\x+0.25,\y+0.5) -- (\x+0.75,\y+0.5); 
\draw[line width=1.5, Mahogany!50] (\x+0.25,\y+0.5) -- (\x+0.75,\y+0.5); 
\draw[line width=1, Mahogany!75] (\x+0.25,\y+0.5) -- (\x+0.75,\y+0.5); 
\draw[line width=0.5, Mahogany!100] (\x+0.25,\y+0.5) -- (\x+0.75,\y+0.5); 
\draw[thick] (\x,\y) -- ++(1,0) -- ++(-0.5,1) -- cycle;
\draw[dashed] (\x+0.25,\y+0.5) -- ++(0.5,0);
\node at (\x+0.5,\y+1.2) {\small (d) $\frac{\mathtt{L}}{2v_E}<t$};

\draw[gray, thick] (\x+.4,\y+0.4) rectangle (\x+0.6,\y+0.6);
\draw[>= stealth, ->, thick, gray] (\x+0.6,\y+0.6) to[out=45,in=180](\x+0.6+0.65,\y+0.6+0.1);

\def\x{2.5}
\def\y{-1.5}
\draw[thick] (\x,\y+0.75) node[above]{\small $\mathtt{I(l)}$} -- ++(0,-0.75) -- ++(1,0) node[right]{\small $\mathtt{l}$};
\draw[domain=\x:\x+0.5, smooth, variable=\z, Mahogany, very thick] plot (\z, {\y+0.65*exp(50.0*(\z-\x-0.5))});
\draw[domain=\x+0.5:\x+1, smooth, variable=\z, Mahogany, very thick] plot (\z, {\y+0.65*exp(-50*(\z-\x-0.5))});

\draw[thick,dotted] (\x+0.5,\y-0.025) -- ++(0,0.5); 
\node at (\x+0.5,\y-0.15) {\small $\frac{\mathtt{L}}{2}$};

\end{tikzpicture}
\caption{Leading order dynamics on the information lattice. At $t=0$ (a), all the information resides at $\mathtt l=0$ as we start with a product state. At $t< (4v_E)^{-1}\mathtt L$ (b), the profile is given by Eq.~\ref{eqn:loc-info-finite-form-a} where in general the profile of $\mathtt{I(n,l)}$ has a finite width and travels with velocity $2v_E$. (c) corresponds to the time regime  $\mathtt (4v_E)^{-1}\mathtt L<t<\mathtt (2v_E)^{-1}\mathtt L$ the results for which are given in Eq.~\ref{eqn:loc-info-finite-form-b}, and (d) corresponds to the late time saturation given in Eq.~\ref{eqn:loc-info-finite-sat}. For (c) and (d), the profile of $\mathtt{I(n,l)}$ has an exponential decay away from $\mathtt{l=L}/2$.}
\label{fig:leading-order}
\end{figure}
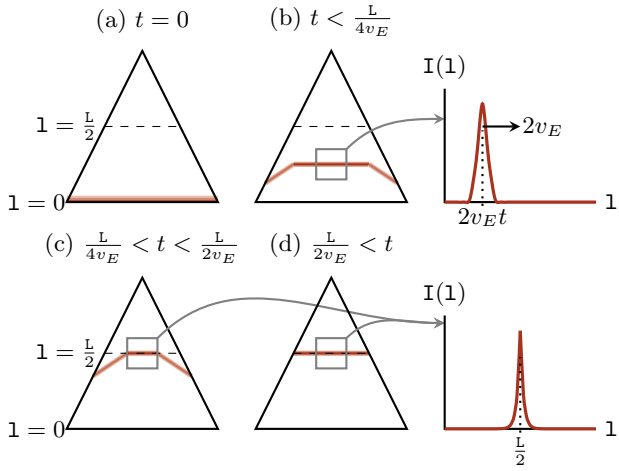

For a finite system, boundary effects can also be taken into account to obtain the leading order dynamics on the full information lattice. For the interval $\mathcal C_n^l$, the left boundary located at $\mathtt n-\mathtt l/2$ will stop producing entanglement when the interval runs out of degrees of freedom to get entangled to on the left. This happens when the entanglement produced by the left boundary reaches $(\mathtt n- \mathtt l/2)$. The same is true for the right endpoint, which can at most contribute $(\mathtt L -\mathtt n- \mathtt l/2)$ to the total entanglement entropy.  This behaviour is captured by the following form,
\begin{widetext}
\begin{equation}
    S(\mathtt{n,l},t) = {\rm min}\left\{2v_Et, \;v_Et + \left(\mathtt n-\frac{\mathtt l}{2}\right),\right. \left. v_Et + \left(\mathtt{L-n}-\frac{\mathtt l}{2}\right),\;\mathtt l,\; (\mathtt{L-l})\right\},
    \label{eqn:ent-finite-form}
\end{equation}
where the last entry deals with cases where $\mathtt l> \mathtt L/2$. 
The above expression can be used back in Eq.~\ref{eqn:loc-info-cont} to obtain the local information.
Since all the branches of the min function are linear functions of $\mathtt n$ and $\mathtt l$, their seconds derivates and hence, $\mathtt I(\mathtt{n,l},t)$ will vanish everywhere but on loci of intersections of different branches. 
On these points $I_n^l$ picks up a Dirac-delta contribution weighted by the discontinuous jump in slope of $S(\mathtt{n,l},t)$. An explicit calculation yields the following results. For $t<\mathtt (4v_E)^{-1}\mathtt L$, 
\begin{equation}
    \mathtt I(\mathtt {n,l},t) = \left\{ \begin{array}{ll}
         \frac{2}{3}\delta\left(\mathtt l - \frac{2}{3}(\mathtt n + v_Et)\right), & \mathtt n < 2v_Et \\
         \delta \left(\mathtt l - 2v_Et\right),& 2v_Et \leq n \leq \mathtt L - 2v_Et\\
         \frac{2}{3}\delta\left(\mathtt l - \frac{2}{3}(\mathtt L + v_Et - \mathtt n)\right), & \mathtt n > \mathtt L-2v_E t.
         \label{eqn:loc-info-finite-form-a}
    \end{array} \right.
\end{equation}
This characteristic timescale, $(4v_E)^{-1}\mathtt L$, is the one at which the entanglement entropy of the subsystem of length $\mathtt L/2$ centred at $\mathtt n = \mathtt L/2$ saturates.
On the information lattice, this corresponds to the timescale at which a packet of local information reaches $\mathtt l = \mathtt L/2$ for the first time. 

For $\mathtt (4v_E)^{-1}\mathtt L<t<\mathtt (2v_E)^{-1}\mathtt L$ we obtain,
\begin{equation}
    \mathtt I(\mathtt {n,l},t) = \left\{ \begin{array}{ll}
         \frac{2}{3}\delta\left(\mathtt l - \frac{2}{3}(\mathtt n + v_Et)\right), & \mathtt n < 3\mathtt L/4 - v_Et \\
         2\delta \left(\mathtt l - \mathtt L/2 \right),& 3\mathtt L/4-v_Et \leq n \leq \mathtt L/4 + v_Et\\
         \frac{2}{3}\delta\left(\mathtt l - \frac{2}{3}(\mathtt L + v_Et - \mathtt n)\right), & \mathtt n >\mathtt L/4 + v_Et.

         \label{eqn:loc-info-finite-form-b}
    \end{array} \right.
\end{equation}
\end{widetext}
Finally, for $t>(2v_E)^{-1}\mathtt L$, the entire profile of the local information saturates to its steady state form
\eq{
\mathtt I(\mathtt {n,l},t) = 2\delta(\mathtt l - \mathtt L/2)\,.
\label{eqn:loc-info-finite-sat}
}
This timescale, $(2v_E)^{-1}\mathtt{L}$ corresponds to the saturation of entanglement entropies of all subsystems.
This is naturally set by the time taken for the entanglement of the subsystems $\mathcal C_{\mathtt n = \mathtt L/4}^{\mathtt l = \mathtt L/2}$ and $\mathcal C_{\mathtt n =  3\mathtt L/4}^{\mathtt l = \mathtt L/2}$, 
to saturate.
This is simply because these two subsystems take the longest for their entanglement to saturate.
Since these subsystems only have one boundary where the entanglement grows it takes until $t=(2v_E)^{-1}\mathtt L$
for the entanglement to reach its saturation value of $\mathtt L/2$. 
On the information lattice, this corresponds to the timescale at which a packet of local information reaches $(\mathtt{L}/4,\mathtt{L}/2)$.
The entire dynamical behaviour of the $I_n^l$ over all these time regimes, encompassing Eq.~\ref{eqn:loc-info-finite-form-a}, Eq.~\ref{eqn:loc-info-finite-form-b} and Eq.~\ref{eqn:loc-info-finite-sat} is graphically summarised in Fig.~\ref{fig:leading-order}.

It is important to note that in this leading order analysis all local information at any instant of time lies on a curve on the information lattice. Microscopically, the transient from ballistic growth to saturation can be smooth, which means that the results provided above are only true in a coarse grained sense. For instance, if we consider a region on the information lattice near a fixed $\mathtt n$ far from boundaries, we still expect local information to form a packet propagating ballistically with $2v_Et$ albeit with non-zero width as depicted in the inset of Figure \ref{fig:leading-order}. Similarly, in the steady state, local information will exhibit a peak $\mathtt l = \mathtt L/2$ which may have a finite width. This can be understood for circuits composed of local gates sampled uniformly from the full unitary group, for which the time evolved state at long times is indistinguishable from a Haar-random state. This means that local information at saturation has the form given in Eq.~\ref{eqn:loc-info-haar-state} which has a $\mathtt L$-independent width which depends only on the local Hilbert-space dimension $\mathtt q$.
While these leading order behaviours are completely generic for any scrambling circuit, in the following two sections we will present (exact) results beyond this leading order analysis for two specific classes of circuits.

\section{Haar-random circuits \label{sec:res-haar}}

This section is dedicated to a detailed analysis of the information lattice dynamics for 
circuits composed of Haar-random unitary gates.
This class of circuits has recently been employed extensively as the minimal model for locally interacting quantum chaotic dynamics.
This is in large part due to the fact that they offer the practical advantage that the entanglement dynamics therein can be bounded in terms of a classical `minimal cut' through the circuit geometry. 
This minimal cut estimates exactly the Hartley entropy (zeroth R\'enyi entropy), $S_0$, which in turn provides an upper bound for the other R\'enyi entropies, $S_n$, with $n\geq 1$.
In fact, in the limit of infinite local Hilbert-space dimension ($q\to\infty$), this bound is saturated by all the R\'enyi entropies, thereby mapping the problem of entanglement dynamics in such a circuit to that a classical statistical mechanics one.
For finite but large $q$, more sophisticated methods using the toolbox of Haar-averaging~\cite{weingarten1978asymptotic} allow for the computation of R\'enyi entropies for $n\geq2$ \cite{zhou2019emergent,chan2018solution}.
Based on these approaches previous works have investigated both, the leading order ballistic growth and the subleading fluctuations in time of entanglement in Haar-random circuits.

We first study the limiting of case of $q \rightarrow \infty$ in Sec.~\ref{sec:haar-q-inf} where we leverage the exactness of the minimal cut bound for the von Neumann entropy to obtain an exact classical description of local information flow.
This is supplemented by numerical results for the case of qubits with $q=2$ in Sec.~\ref{sec:haar-q-2}.

\subsection{Large local Hilbert-space dimension: qudits with $q\to\infty$ \label{sec:haar-q-inf}}

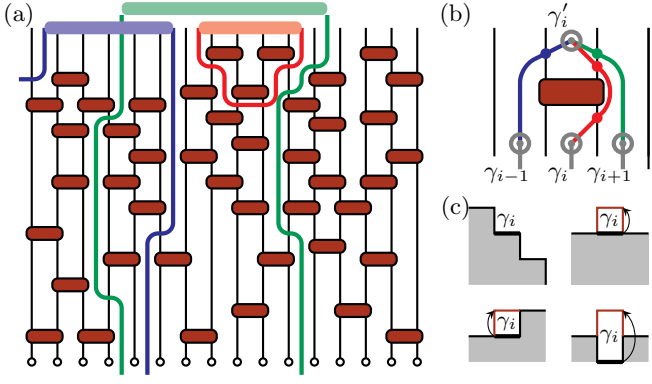
\begin{figure}
    \begin{tikzpicture}[scale=0.34]
\foreach \x in {0,...,15}{
\draw[thick](\x,-1) -- (\x,12);
\filldraw[draw=black,thick, fill=white] (\x,-1) circle (4pt);
}
\def\x{0};
\foreach \y in {0,4,9}{
\filldraw[draw=black,thick, fill=Mahogany, rounded corners=2] (\x-0.2,\y-0.25) rectangle (\x+1+0.2,\y+0.25);
}
\def\x{1};
\foreach \y in {2,6,8,10}{
\filldraw[draw=black,thick, fill=Mahogany, rounded corners=2] (\x-0.2,\y-0.25) rectangle (\x+1+0.2,\y+0.25);
}
\def\x{2};
\foreach \y in {0,9}{
\filldraw[draw=black,thick, fill=Mahogany, rounded corners=2] (\x-0.2,\y-0.25) rectangle (\x+1+0.2,\y+0.25);
}
\def\x{3};
\foreach \y in {3,6,8}{
\filldraw[draw=black,thick, fill=Mahogany, rounded corners=2] (\x-0.2,\y-0.25) rectangle (\x+1+0.2,\y+0.25);
}
\def\x{4};
\foreach \y in {5,7,9}{
\filldraw[draw=black,thick, fill=Mahogany, rounded corners=2] (\x-0.2,\y-0.25) rectangle (\x+1+0.2,\y+0.25);
}
\def\x{5};
\foreach \y in {3}{
\filldraw[draw=black,thick, fill=Mahogany, rounded corners=2] (\x-0.2,\y-0.25) rectangle (\x+1+0.2,\y+0.25);
}
\def\x{6};
\foreach \y in {0,7,9.5}{
\filldraw[draw=black,thick, fill=Mahogany, rounded corners=2] (\x-0.2,\y-0.25) rectangle (\x+1+0.2,\y+0.25);
}
\def\x{7};
\foreach \y in {6,8.5,11}{
\filldraw[draw=black,thick, fill=Mahogany, rounded corners=2] (\x-0.2,\y-0.25) rectangle (\x+1+0.2,\y+0.25);
}
\def\x{8};
\foreach \y in {1,5,7.5,9.5}{
\filldraw[draw=black,thick, fill=Mahogany, rounded corners=2] (\x-0.2,\y-0.25) rectangle (\x+1+0.2,\y+0.25);
}
\def\x{9};
\foreach \y in {3,11}{
\filldraw[draw=black,thick, fill=Mahogany, rounded corners=2] (\x-0.2,\y-0.25) rectangle (\x+1+0.2,\y+0.25);
}
\def\x{10};
\foreach \y in {5,7,9}{
\filldraw[draw=black,thick, fill=Mahogany, rounded corners=2] (\x-0.2,\y-0.25) rectangle (\x+1+0.2,\y+0.25);
}
\def\x{11};
\foreach \y in {3.5,6,8.25,10}{
\filldraw[draw=black,thick, fill=Mahogany, rounded corners=2] (\x-0.2,\y-0.25) rectangle (\x+1+0.2,\y+0.25);
}
\def\x{12};
\foreach \y in {1,5}{
\filldraw[draw=black,thick, fill=Mahogany, rounded corners=2] (\x-0.2,\y-0.25) rectangle (\x+1+0.2,\y+0.25);
}
\def\x{13};
\foreach \y in {3,6,8.5,11}{
\filldraw[draw=black,thick, fill=Mahogany, rounded corners=2] (\x-0.2,\y-0.25) rectangle (\x+1+0.2,\y+0.25);
}
\def\x{14};
\foreach \y in {0,7,10}{
\filldraw[draw=black,thick, fill=Mahogany, rounded corners=2] (\x-0.2,\y-0.25) rectangle (\x+1+0.2,\y+0.25);
}

\draw[ultra thick, Red, rounded corners] (6.5,12) -- ++(0,-1.5) --++(1,0) --++(0,-1.5) --++(2,0) --++(0,1.5) --++(1,0) --++(0,1.5) ;

\draw[ultra thick, Blue, rounded corners] (5.5,12) -- ++(0,-8) --++(-1,0) --++(0,-5.5);
\draw[ultra thick, Blue, rounded corners] (0.5,12) -- ++(0,-2) --++(-1,0);

\draw[ultra thick, ForestGreen, rounded corners] (3.5,12.5) -- ++(0,-4) --++(-1,0) --++(0,-7) --++(1,0) --++(0,-3);
\draw[ultra thick, ForestGreen, rounded corners] (11.5,12.5) -- ++(0,-2) --++(-1,0) --++(0,-1) -- ++(-0.9,0) --++(0,-5.5) -- ++(0.9,0) -- ++(0,-5.5);

\fill[Red!50, rounded corners=2] (6.5,11.75) rectangle (10.5,12.25);
\fill[Blue!50, rounded corners=2] (0.5,11.75) rectangle (5.5,12.25);
\fill[ForestGreen!50, rounded corners=2] (3.5,12.5) rectangle (11.5,13);

\fill[gray!50] (17,5) -- ++(1,0) --++(0,-1) -- ++(1,0) --++(0,-1) --++(1,0) --++(0,-1) -- ++(-3,0) --cycle;
\draw[thick] (17,5) -- ++(1,0) --++(0,-1) -- ++(1,0) --++(0,-1) --++(1,0) --++(0,-1);
\draw[ultra thick] (18,4) -- (19,4);
\node at (18.5,4.5) {\footnotesize $\gamma_i$};

\fill[gray!50] (22-1,5-1) -- (25-1,5-1) -- (25-1,3-1) -- (22-1,3-1) -- cycle;
\draw[Mahogany,thick] (23-1,5-1) rectangle (24-1,6-1);
\draw[thick] (22-1,5-1) -- (25-1,5-1);
\draw[ultra thick] (23-1,5-1) -- (24-1,5-1);
\draw[>=stealth, ->] (24-1,5-1) to[out=30, in =-45] (24-1,6-1);
\node at (23.5-1,5.5-1) {\footnotesize $\gamma_i$};

\fill[gray!50] (21,0) -- ++(1,0) -- ++(0,-1) -- ++(1,0) -- ++(0,1) --++(1,0) -- ++(0,-1) -- ++(-3,0) --cycle;
\draw[thick] (21,0) -- ++(1,0) -- ++(0,-1) -- ++(1,0) -- ++(0,1) --++(1,0);
\draw[ultra thick] (22,-1) -- (23,-1);
\draw[Mahogany, thick] (22,0) -- ++(0,1) -- ++(1,0) --++(0,-1);
\draw[>=stealth, ->] (23,-1) to[out=30, in =-45] (23,1);
\node at (22.5,0) {\footnotesize $\gamma_i$};

\fill[gray!50] (17,0) -- ++(2,0) -- ++(0,1) -- ++(1,0) -- ++(0,-2) --++(-3,0) --cycle;
\draw[thick] (17,0) -- ++(2,0) -- ++(0,1) -- ++(1,0);
\draw[ultra thick] (18,0) -- (19,0);
\draw[Mahogany, thick] (18,0) -- ++(0,1) -- ++(1,0) ;
\draw[>=stealth, ->] (18,0) to[out=150, in =-135] (18,1);
\node at (18.5,0.5) {\footnotesize $\gamma_i$};

\foreach \x in {18,20,22,24}{
\draw[thick] (\x,7) -- ++(0,5);
}

\draw[Blue, ultra thick, rounded corners=6] (19,7.5) -- (19,10.5) -- (21,11.5);
\draw[ForestGreen, ultra thick, rounded corners=6] (23,7.5) -- (23,10.5) -- (21,11.5);
\draw[ultra thick, Red, rounded corners] (21,7.5) -- (22.5,9) -- (22.5,10) -- (21,11.5);
\draw[thick] (24,6.5) -- ++(0,5.5);

\foreach \x in {18,20,22}{
\draw[gray, ultra thick] (\x+1,6.5) -- ++(0,1);
\draw[draw=gray, ultra thick] (\x+1,7.5) circle (10 pt);
\fill[gray] (\x+1,7.5) circle (4 pt);
}

\filldraw[draw=black,thick, fill=Mahogany, rounded corners=2](19.75,9) rectangle (22.25,10);

\draw[draw=gray, ultra thick] (21,11.5) circle (10 pt);
\fill[gray] (21,11.5) circle (4 pt);

\fill[Blue] (20,11) circle (6 pt);
\fill[ForestGreen] (22,11) circle (6 pt);
\fill[Red] (22,10.5) circle (6 pt);
\fill[Red] (22,8.5) circle (6 pt);

\node at (18.5,6.25) {$\gamma_{i-1}$};
\node at (20.5,6.25) {$\gamma_{i}$};
\node at (20.5,12.5) {$\gamma^\prime_{i}$};
\node at (22.5,6.25) {$\gamma_{i+1}$};

\node at (16.5,5.0) {(c)};
\node at (16.5,12.5) {(b)};
\node at (-0.5,12.5) {(a)};
\end{tikzpicture}
    \caption{(a) A realisation of the random circuit model. The red, green and blue bars are the top three different subsystems and curves in their respective colours denote a possible minimal cut for each of them. In this example the entanglement of the green subsystem continues to grow with contributions from both ends whereas for the blue subsystem, only its right subsystem contributes to entanglement growth. By contrast, the entanglement of the red subsystem has saturated. (b) Change in $\gamma_i$ resulting from the action of a unitary on sites $(i-1,i)$. (c) Examples of the resulting evolution of $\gamma_i$.}
    \label{fig:q-inf-model}
\end{figure}

The entanglement of a contiguous subsystem for any circuit realisation\footnote{A circuit realisation is specified by a set of spacetime points where the unitary gates act as well as the specific unitary operators defining the gates.} is bounded from above by the number of bonds or qudit worldlines which need to be broken to separate the qudits which belong to the subsystem at the final time. 
Viewing the state of the full system at the final time generated by the circuit as a planar graph in two dimensions [as shown in Fig.~\ref{fig:q-inf-model}(a)], the above process can be achieved by defining curves through this graph which emanate from the two boundaries of the contiguous subsystem. This curves split the entire graph into a part which contains the terminal points of the qudit worldlines corresponding to the subsystem, and its complement.
Any such curve carries a cost precisely equal to the number of worldlines it cuts.
Minimising this cost over the set of all such possible curves gives the cost associated to the ``minimal cut'' which is the best possible bound for the entanglement of the subsystem with the rest of the system. 
Depending on the depth of the circuit and the size of the subsystem, the minimal cuts can have different morphologies. 
At short times, or for very large subsystems, these cuts typically extend all the way to the bottom of the circuit and exit the circuit at $t=0$ which results in the entanglement growing linearly with $t$ at the leading order.
This corresponds to the regime where entanglement is produced across both the boundaries of the subsystem. An illustrative example of this scenario is shown in Fig.~\ref{fig:q-inf-model}(a) in green.
At a later time, or for subsystems close to one of the spatial boundaries of the system, the cut closer to the boundary may exit through the boundary of the circuit, which means that the associated subsystem boundary cannot produce more entanglement. This situation is exemplified by the blue subsystem in Fig.~\ref{fig:q-inf-model}(a) where the cut emanating from the left boundary exits the circuit through the side.
Finally, at late times, 
or for small subsystems,
either both cuts exit through the boundaries of the circuit or join to form a single minimal cut -- this corresponds to the saturation of entanglement. The cost of doing this is equal to $l+1$ for an interval of length $l < L/2$ and $L-l-1$ for and interval of length $l\geq L/2$ (recall that in our notation an interval of length $l$ contains $l+1$ qudits). This is illustrated in Fig.~\ref{fig:q-inf-model}(a) in red.
Note that for circuits with a regular spatiotemporal geometry, such as a brickwork circuit, since there are no fluctuations in the geometry of the minimal cuts, the entanglement growth is completely deterministic and is described exactly by Eq.~\ref{eqn:ent-finite-form} with $v_E = 1$. Fluctuations can be induced by either removing some gates randomly from the regular pattern, or by altogether applying gates randomly in a Poissonian fashion. We shall work with the latter.

\subsubsection{Local information from minimal cuts}

We denote by $\gamma_i$, the cost of a minimal cut which starts between sites $i-1$ and $i$ from the final time boundary, and exits the circuit from either the spatial or the initial time boundary. Since this cut can always fuse with a minimal cut which starts at a position adjacent to it, between sites $i$ and $i+1$, after cutting through a single bond, the variables $\left\{\gamma_i\right\}$ are constrained by
\begin{equation}
    \vert\gamma_i - \gamma_{i+1}\vert \leq 1\,.
    \label{eqn:cut-constr}
\end{equation}
If we also include $\gamma_0$ and $\gamma_L$ in this set, which being already in the exterior of the circuit geometry are always zero, the constraint in Eq.~\ref{eqn:cut-constr} translates to 
\eq{
\gamma_i \leq \min\{i,L-i\}\,.
}
Note that $\gamma_i$ is exactly the entanglement entropy of the subsystem ${\cal C}_{(i-1)/2}^{i+1}$ containing sites $0$ through $i-1$ with the rest of the system.
Therefore,
a unitary gate acting on sites $i-1$ and $i$ can only the change the entanglement of a subsystem which has a boundary site on $i-1$ or $i$.
For a given configuration of $\left\{\gamma_i\right\}$ satisfying Eq.~\ref{eqn:cut-constr}, 
this naturally means that such a gate can change only $\gamma_i$ and it leaves $\gamma_j~\forall j \neq i$ untouched.
The application of the gate can update the value of $\gamma_i$ in three possible ways. 
The cut originating between sites $i-1$ and $i$ can go around the unitary at the cost of cutting through two extra bonds and join with its prior configuration which gives $\gamma_i' = \gamma_i+2$.
This process is shown in red in Fig.~\ref{fig:q-inf-model}(b). 
Alternatively, the minimal cuts corresponding to $\gamma_{i\pm 1}$ can cut through one extra worldline of site $i\pm 1$ and join the updated minimal cut $\gamma_i^\prime$
resulting in either $\gamma_i' = \gamma_{i+1}+1$ or $\gamma_i' = \gamma_{i-1}+1$ respetively.
These processes are shown in blue and green in Fig.~\ref{fig:q-inf-model}(b). 
Any other configuration costs more and hence is not optimal.
Optimising over the three possibilities subject to the constraint in Eq.~\ref{eqn:cut-constr}, we obtain the dynamical rule for updating the configuration $\{\gamma_i\}$ upon the action of gate $U_{i-1,i}$, which can be summarised as
\begin{equation}
        \gamma_j \overset{U_{i-1,i}}{\longrightarrow} \gamma^\prime_j = \begin{cases}{\rm min}\left\{\gamma_{i-1},\gamma_{i+1}\right\} + 1;~~&j=i\\
        \gamma_j;~~&j\neq i
        \end{cases}\,.
        \label{eqn:cut-evolution}
\end{equation}
The dynamics thus generated for the configuration $\{\gamma_i\}$ therefore 
describes a growing interface which follows Kardar-Parisi-Zhang (KPZ) universality \cite{kardar1986dynamic} at large scales. 
The rules, as encoded in Eq.~\ref{eqn:cut-evolution} are diagrammatically summarised in Fig.~\ref{fig:q-inf-model}(c).
In the context of random unitary circuits, this model was part of the ``KPZ triumvirate"~\cite{nahum2017quantum} 
to pin down the universality class of entanglement growth.

Since $\gamma_i$ is the entanglement entropy of  the subsystem ${\cal C}_{(i-1)/2}^{i+1}$, the same for a contiguous region $\mathcal C_n^l$, is either equal to the combined cost $\gamma_{n-l/2} + \gamma_{n+l/2+1}$ or the cost of a single cut which goes from one endpoint to the other (such as the red curve in Fig.~\ref{fig:q-inf-model}(a)), whichever is more optimal. This gives 
\begin{equation}
    S_n^l = \text{min}\left\{l+1,\; \gamma_{n-l/2} + \gamma_{n+l/2+1}\right\}\,,
    \label{eqn:ent-cut}
\end{equation}
and subsequently,
\begin{equation}
    E_n^l = \left[l+1 - \gamma_{n-l/2} - \gamma_{n+l/2+1}\right]_+,
    \label{eqn:von-neumann-cut}
\end{equation}
where $\left[x\right]_+ = x \Theta(x)$ is the ramp function. 
The above equation explicitly relates the total information in a contiguous subsystem to the minimal cuts originating from its boundaries.
To relate the minimal cuts to the local information, we use the above result in Eq.~\ref{eqn:local-info},
which yields (see Appendix~\ref{appendix:loc-info-q-inf} for details) 
\begin{equation}
    I_n^l = \sum_{\alpha = 0}^{L-1} \;\delta(e^-_\alpha, n-l/2) \;\delta(e^+_\alpha, n+l/2)
    \label{eqn:loc-info-q-inf}
\end{equation}
where,
\begin{equation}
    e^\pm_\alpha = \sum_{i=0}^L\Theta(\alpha - i \pm \gamma_i) \;- 1\,.
    \label{eq:ealpha-pm}
\end{equation}
The quantities $e_\alpha^\pm$ have a following physical interpretation.
For a given $\alpha$, $e^-_\alpha$ tracks the position of the rightmost bipartition, say the cut between sites, $i_r-1$ and $i_r$, such that the bipartite entanglement, $\gamma_{i_r}$ is less than or equal to $i_r-\alpha$. 
This already hints towards the presence of local information in the subsystems to the right of the cut.
Analogously,  $e^+_\alpha$ tracks the leftmost bipartition, say between $i_l-1$ and $i_l$ such that the bipartite entanglement $\gamma_{i_l}$ is greater that $i_l-\alpha$; and this suggests the presence of local information in subsystems to the left of this cut. 
Together, the pair $(e_\alpha^-,e_\alpha^+)$ therefore implies the presence of local information in the subsystem spanned by sites $i_l$ through $i_r-1$ at a length scale equal to the length of the system.

\begin{figure}
\begin{tikzpicture}[scale=0.4]

\foreach \x in {0,...,9}{
    \draw[gray, very thin](\x,4.5) -- (\x,14);
}
\foreach \y in {5,...,14}{
    \draw[gray, very thin](-1,\y) -- (9,\y);
}

\foreach \x in {0,...,9}{
    \draw[gray, very thin](\x,16) -- (\x,20);
}
\foreach \y in {18,...,20}{
    \draw[gray, very thin](0,\y) -- (9,\y);
}

\filldraw[thick, fill=gray!30](-1,17)--++(1,0)--++(0,1)--++(1,0)--++(1,0)--++(0,1)--++(1,0)--++(0,1)--++(1,0)--++(0,-1)--++(1,0)--++(0,-1)--++(1,0)--++(0,1)--++(1,0)--++(0,-1)--++(1,0)--++(1,0)--++(0,-1)--++(1,0)--++(-10,0); 
\node at (-0.5,16.5) {\small{$\gamma_0$}};
\node at (0.5,16.5) {\small{$\gamma_1$}};
\node at (2,16.5) {\small{$\cdots\cdots$}};
\node at (8.5,16.5) {\small{$\gamma_9$}};
\node at (9.5,16.5) {\small{$\gamma_{10}$}};

\draw[ultra thick] (0,0+5) --(1,0+5);
\draw[ultra thick] (0,1+5) --(4,1+5);
\draw[ultra thick] (1,2+5) --(4,2+5);
\draw[ultra thick] (2,3+5) --(5,3+5);
\draw[ultra thick] (2,4+5) --(5,4+5);
\draw[ultra thick] (3,5+5) --(7,5+5);
\draw[ultra thick] (3,6+5) --(7,6+5);
\draw[ultra thick] (6,7+5) --(8,7+5);
\draw[ultra thick] (6,8+5) --(9,8+5);
\draw[ultra thick] (8,9+5) --(9,9+5);

\draw[dashed, thick, Red!90] (0,0+5)--(0,1+5)--(1,2+5)--(2,3+5)--(2,4+5)--(3,5+5)--(3,6+5)--(6,7+5)--(6,8+5)--(8,9+5);

\filldraw[draw=Red!90, thick, fill=white] (0,0+5) circle (6pt);
\filldraw[draw=Red!90, thick, fill=white] (0,1+5) circle (6pt);
\filldraw[draw=Red!90, thick, fill=white] (1,2+5) circle (6pt);
\filldraw[draw=Red!90, thick, fill=white] (2,3+5) circle (6pt);
\filldraw[draw=Red!90, thick, fill=white] (2,4+5) circle (6pt);
\filldraw[draw=Red!90, thick, fill=white] (3,5+5) circle (6pt);
\filldraw[draw=Red!90, thick, fill=white] (3,6+5) circle (6pt);
\filldraw[draw=Red!90, thick, fill=white] (6,7+5) circle (6pt);
\filldraw[draw=Red!90, thick, fill=white] (6,8+5) circle (6pt);
\filldraw[draw=Red!90, thick, fill=white] (8,9+5) circle (6pt);

\fill[Red!90] (0,0+5) circle (3pt);
\fill[Red!90] (0,1+5) circle (3pt);
\fill[Red!90] (1,2+5) circle (3pt);
\fill[Red!90] (2,3+5) circle (3pt);
\fill[Red!90] (2,4+5) circle (3pt);
\fill[Red!90] (3,5+5) circle (3pt);
\fill[Red!90] (3,6+5) circle (3pt);
\fill[Red!90] (6,7+5) circle (3pt);
\fill[Red!90] (6,8+5) circle (3pt);
\fill[Red!90] (8,9+5) circle (3pt);

\draw[dashed, thick, Blue!60](1,0+5)--(4,1+5)--(4,2+5)--(5,3+5)--(5,4+5)--(7,5+5)--(7,6+5)--(8,7+5)--(9,8+5)--(9,9+5);

\filldraw[draw=Blue!60, thick, fill=white] (1,0+5) circle (6pt);
\filldraw[draw=Blue!60, thick, fill=white] (4,1+5) circle (6pt);
\filldraw[draw=Blue!60, thick, fill=white] (4,2+5) circle (6pt);
\filldraw[draw=Blue!60, thick, fill=white] (5,3+5) circle (6pt);
\filldraw[draw=Blue!60, thick, fill=white] (5,4+5) circle (6pt);
\filldraw[draw=Blue!60, thick, fill=white] (7,5+5) circle (6pt);
\filldraw[draw=Blue!60, thick, fill=white] (7,6+5) circle (6pt);
\filldraw[draw=Blue!60, thick, fill=white] (8,7+5) circle (6pt);
\filldraw[draw=Blue!60, thick, fill=white] (9,8+5) circle (6pt);
\filldraw[draw=Blue!60, thick, fill=white] (9,9+5) circle (6pt);

\fill[Blue!60](1,0+5) circle (3pt);
\fill[Blue!60](4,1+5) circle (3pt);
\fill[Blue!60](4,2+5) circle (3pt);
\fill[Blue!60](5,3+5) circle (3pt);
\fill[Blue!60](5,4+5) circle (3pt);
\fill[Blue!60](7,5+5) circle (3pt);
\fill[Blue!60](7,6+5) circle (3pt);
\fill[Blue!60](8,7+5) circle (3pt);
\fill[Blue!60](9,8+5) circle (3pt);
\fill[Blue!60](9,9+5) circle (3pt);
\draw[>=Stealth, ->, thick] (-1,4.5) -- (10,4.5) node[below]{\small{${\color{Red!75}{e_\alpha^-}},{\color{Blue!75}{e_\alpha^+}}$}};
\draw[>=Stealth, ->, thick] (-1,4.5) -- (-1,15) node[right]
{\small{$\alpha$}};

\foreach \l in {0,...,9}{
    \edef\maxN{\numexpr9-\l\relax}
    \foreach \n in {0,...,\maxN}{
        \filldraw[draw=black, thick,fill=white] (11+\n+0.5*\l, \l+5) circle (7pt);
    }   
}
\fill[black] (11.5,6)-- ++(180:7pt) arc (180:360:7pt) -- cycle;
\fill[black] (19.5,6)-- ++(180:7pt) arc (180:360:7pt) -- cycle;
\fill[black] (18,7)-- ++(180:7pt) arc (180:360:7pt) -- cycle;
\fill[black] (13.5,8)-- ++(180:7pt) arc (180:360:7pt) -- cycle;
\fill[black] (14.5,8) circle (7pt);
\fill[black] (18.5,8)-- ++(180:7pt) arc (180:360:7pt) -- cycle;
\fill[black] (13,9)-- ++(180:7pt) arc (180:360:7pt) -- cycle;
\fill[black] (16,9) circle (7pt);

\fill[OliveGreen, fill opacity=0.3, rounded corners=4] (0.5,5+1.5) rectangle (4.5,5+2.5);
\draw[OliveGreen, thick, dashed, rounded corners=4] (0.5,5+1.5) rectangle (4.5,5+2.5);

\fill[YellowOrange, fill opacity=0.3, rounded corners=4] (2.5,5+4.5) rectangle (7.5,5+6.5);
\draw[YellowOrange, thick, dashed, rounded corners=4]  (2.5,5+4.5) rectangle (7.5,5+6.5);

\fill[OliveGreen, fill opacity=0.3, rounded corners=3] (13.5,7.25) -- ++(0.75,0.75) --++(-0.75,0.75) --++(-0.75,-0.75) -- cycle;

\fill[YellowOrange, fill opacity=0.3, rounded corners=3] (16,8.25) -- ++(0.75,0.75) --++(-0.75,0.75) --++(-0.75,-0.75) -- cycle;

\fill[white, fill opacity=0.8](4.5,5.5) rectangle (10.5,7.5);

\draw[OliveGreen, >=Stealth, ->, thick,dashed, rounded corners=6] (4,6.5) -- (10,6.5) -- (13.25,7.75);

\draw[YellowOrange, >=Stealth, ->, thick,dashed, rounded corners=6] (6.5,11.5) -- (10,11.5) -- (15.75,9.25);

\node at (7.5,7.1) {\footnotesize{\color{OliveGreen} $(e_2^-,e_2^+)=(1,4)$}};
\node at (7.5,5.9) {\footnotesize{\color{OliveGreen} $(n,l)=(5/2,3)$}};

\filldraw[draw=black,fill=gray] (4-0.1,16) --++(0.2,0) --++(0,-1) -- ++(0.25,0) -- ++(-0.35,-0.35) -- ++(-0.35,0.35) --++(0.25,0) -- ++(0,1) --cycle;

\filldraw[draw=black,fill=gray] (10,10) --++(0,0.2) --++(1,0) -- ++(0,0.25) -- ++(0.35,-0.35) -- ++(-0.35,-0.35) --++(0,0.25) -- ++(-1,0) --cycle;

\node at (-0.5,20) {(a)};

\node at (11.5,15) {(b)};

\end{tikzpicture}
\caption{(a) An example configuration of $\left\{\gamma_i\right\}$ and the corresponding values of $e^\pm_\alpha$ with the red and blue point denoting $e_\alpha^{-}$ and $e_\alpha^+$ respectively. Each pair of red and blue points connected by a horizontal black line denotes the bigram at the respective $\alpha$ value. The grid lines denote integer steps of 1. 
(b) The corresponding configuration of $\{I_l^n\}$ on the information lattice where empty, half-filled and fully-filled circles denote $I_n^l=0,1,2$ respectively.
Every bigram $(e^-_\alpha,e^+_\alpha)$ corresponds to one unit of local information on the information lattice at $(n,l) = ((e^+_\alpha + e^-_\alpha)/2,\; e^+_\alpha - e^-_\alpha))$ as shown by a couple of examples. }
\label{fig:bigram-mapping}
\end{figure}
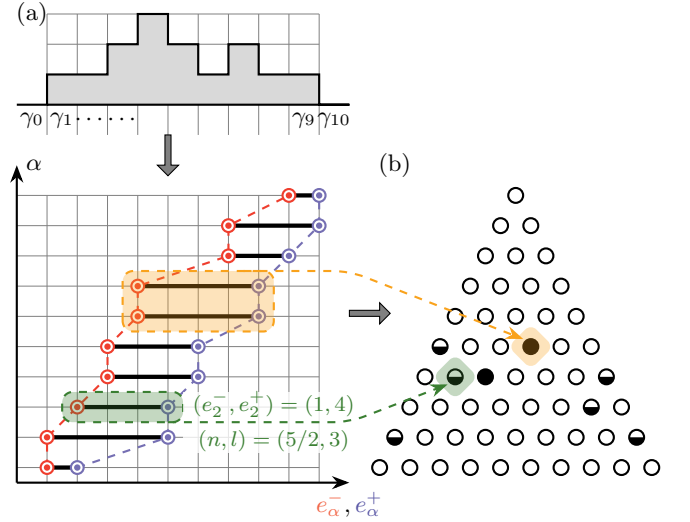

The above picture can be put on entirely concrete formal footing by noting that Eq.~\ref{eq:ealpha-pm} constitutes an explicit map\footnote{From a practical point of view, this map has a convenient diagrammatic interpretation.
To construct $e_\alpha^-$, we assign values in a chronological order in $\alpha$: starting from $i = 0$ and moving towards the right, whenever $\Delta\gamma_i = \gamma_{i+1} - \gamma_i  = 1$ we assign the next two subsequent values $e^-_\alpha = e_{\alpha+1}^-$ = i. If $\Delta\gamma_i = 0$ we set only one value $e^-_\alpha = i$. For $\Delta\gamma_i = -1$ we do nothing. This assigns all $e^-_\alpha$ for $\alpha=0,1,... L-1$. Similarly, to construct $e^+_\alpha$ we follow the same protocol with the rules for $\Delta\gamma_i = 1$ and $\Delta\gamma_i = -1$ interchanged.} between the minimal cuts and $\{e_\alpha^\pm\}$.
which leaves us with the set
\begin{equation}
    \mathcal B = \left\{(e^-_0, e^+_0),(e^-_1, e^+_1), ... ,(e^-_{L-1}, e^+_{L-1})\right\}\,.
\end{equation}
An illustrative example of the configuration $\{e_\alpha^\pm\}$ obtained from a $\{\gamma_i\}$ is shown in Fig.~\ref{fig:bigram-mapping}(a).
Each element of the set ${\cal B}$ is a bigram which can be interpreted as an extended one-dimensional object with left and right endpoints $e^\pm_\alpha$. This leads to an interpretation of Eq.~\ref{eqn:loc-info-q-inf} that each bigram $(e^-_\alpha,e^+_\alpha)$ contributes a single unit of local information with the length scale $l$ determined by the length of the bigram $e^+_\alpha-e^-_\alpha$ and the location $n$ determined by the of the centre of the bigram $(e^+_\alpha+e^-_\alpha)/2$. 
Alternatively, $I^l_n$ is contributed to one unit of information by those bigrams which satisfy $e_\alpha^{\pm} = n\pm l/2$; this is illustrated in Fig.~\ref{fig:bigram-mapping}(b).
The collection of all the bigrams therefore uniquely specifies the configuration $\{I_n^l\}$ on the information lattice and therefore constitutes an explicit map between the minimal cuts and the local information.

Note that in this model, entanglement of all contiguous regions is determined exactly by the variables $\gamma_i$. As a result, the local information for all $(n,l)$ can be obtained entirely from the knowledge of entanglement across all bipartitions made across single bonds. This simplification is specific to large-$q$ Haar-random circuits and is only valid approximately at large scales in general random circuit models.

\subsubsection{Exact stochastic process for local information}
\label{sec:haar-q-inf-classical}

Having mapped the local information configuration on the information lattice to a configuration of bigrams, we next describe the dynamical rules governing the evolution of the bigrams. This leads to an exact classical stochastic process for the dynamics of local information.

However before describing the aforementioned rules, it is useful to recast the constraints on the local information flow in terms of the bigrams. 
First, the non-local constraint in Eq.~\ref{eqn:constraint} translates to the constraint that every site hosts exactly two bigram endpoints irrespective of their type (left or right). 
To see this, consider the constraint on site $j$ in terms of the sum of $I_n^l$'s enclosed by the green boxes in Fig.~\ref{fig:constraints}. 
Any point $(j\mp l/2,l)$ inside the left (right) green box having a unit of information corresponds to there being a bigram with its right (left) endpoint on site $j$ and its left (right) endpoint at $j\mp l$.
From Eq.~\ref{eqn:constraint}, the sum of the local information on all such points must equal 2 which automatically implies that there must be two bigram endpoints on site $j$. 
Second, the bigrams are already constrained by $e^\pm_\alpha \geq e^\pm_\beta$ for $\alpha>\beta$, by construction in Eq.~\ref{eq:ealpha-pm}.

To update the bigrams, since the unitary gates act locally on nearest-neighbour sites, it is natural to derive their action on the bigram endpoints. 
To this end, we define the densities of left and right endpoints as
\begin{equation}
    \rho^\pm_i = \sum_{\alpha = 0}^L \delta(e^\pm_\alpha,i)\,,\label{eq:rho-def}
\end{equation}
and obtain the action of the gates on these variables. 

The central guiding principle which determines these rules is that the Haar-random gate always tries to the increase the entanglement across the bipartition it acts on, subject to the constraints placed by subadditivity. 
In terms of the minimal cuts, this translates to $\gamma_i$ always increasing under the action of the gate modulo the constraint in Eq.~\ref{eqn:cut-constr}; this was shown schematically in Fig.~\ref{fig:q-inf-model}(c).
Finally, using the explicit relation between the minimal cuts and the bigram endpoints, this implies that the unitary gates always try to increase the length of the bigrams wherever possible, again subject to the constraints $\rho_i^++\rho_i^- = 2$ for all $i$ and $e^\pm_\alpha \geq e^\pm_\beta$ for $\alpha>\beta$.
This results in dynamical rules for the bigram endpoints where if sites $i$ and $i+1$ have right and left endpoints respectively, obviously of different bigrams, a gate acting on sites $(i,i+1)$ tries to increase the length of the bigrams by swapping the two endpoints.
Concomitant with these principles, for infinite-$q$ Haar-random circuits the rules can be derived exactly.
The detailed derivation of the rules for updating the $e_\alpha^{\pm}$ is presented in Appendix~\ref{appendix:loc-info-q-inf}, and here we only describe the rules and discuss their physical implication.

\begin{figure}
\begin{tikzpicture}[scale=0.33]
\foreach \x in {0,5,10,15,20}{
\draw (\x+0.1,0) -- ++(4.8,0) -- ++(-2.4,5) -- cycle;
\draw(\x+0.1+1.9,-0.25) -- ++(0,-0.8);
\draw(\x+0.1+2.5,-0.25) -- ++(0,-0.8);
\filldraw[draw=black, fill=Mahogany, rounded corners=1] (\x+1.9,-1.+0.15) rectangle (\x+0.2+2.5,-0.75+0.3);
\foreach \j in {0,1,2}{
\draw(\x+0.1+1.9-0.3*\j,0.25+0.6*\j) circle (6pt);
}
\foreach \j in {0,1,2,3,4}{
\draw(\x+0.1+1.9+0.3*\j,0.25+0.6*\j) circle (6pt);
}
\foreach \j in {0,1,2,3}{
\draw(\x+0.1+2.5-0.3*\j,0.25+0.6*\j) circle (6pt);
\draw(\x+0.1+2.5+0.3*\j,0.25+0.6*\j) circle (6pt);
}

\fill[OliveGreen, fill opacity=0.3,rounded corners=2] (\x+0.1+1.9+0.4-1.1,2.2)-- (\x+0.1+1.9+0.4,0) -- (\x+0.1-0.15+1.9,0) -- ++(-0.9,1.8)--cycle;

\fill[OliveGreen, fill opacity=0.3,rounded corners=2] (\x+0.1+2.5-0.4+1.4,2.8)-- (\x+0.1+2.5-0.4,0) -- (\x+0.1+0.15+2.5,0) -- ++(1.1,2.2)--cycle;

\fill[YellowOrange, fill opacity=0.3,rounded corners=2] (\x+0.1+2.2,0.3)-- ++(1.2,2.4) -- ++(-0.32,0.64 ) --++(-1.2+0.32,-2.4+0.64) -- ++(-0.6,1.2)--++(-0.32,-0.64)--cycle;

\draw[thick,dotted] (4+\x,7)--++(0.8,0);

}

\node at (2.5,8.5) {\footnotesize $\gamma_i$};
\draw[thick] (1,7) -- ++(1,0) --++(0,1) -- ++(1,0) --++(0,-1) --++(1,0);
\draw[thick,dotted] (4-3,7)--++(-0.8,0);

\draw[thick] (6,9) -- ++(1,0) --++(0,-1) -- ++(1,0) --++(0,-1) --++(1,0);
\node at (7.5,8.5) {\footnotesize $\gamma_i$};
\draw[thick,dotted] (4-3+5,9)--++(-0.8,0);

\node at (12.5,8.5) {\footnotesize $\gamma_i$};
\draw[thick] (11,8) -- ++(1,0) --++(0,-1) -- ++(1,0) --++(1,0);
\draw[thick, Mahogany](12,8)--++(1,0)--++(0,-1);
\draw[>=stealth,->, gray] (12.5,7.1) -- ++(0,0.8);
\draw[thick,dotted] (4-3+10,8)--++(-0.8,0);

\node at (17.5,8.5) {\footnotesize $\gamma_i$};
\draw[thick] (16,7) -- ++(3,0);
\draw[thick, Mahogany](17,7)--++(0,1)--++(1,0)--++(0,-1);
\draw[>=stealth,->, gray,] (17.5,7.1) -- ++(0,0.8);
\draw[thick,dotted] (4-3+15,7)--++(-0.8,0);

\node at (22.5,8.5) {\footnotesize $\gamma_i$};
\draw[thick] (21,7) -- ++(1,0) --++(0,-1) -- ++(1,0) --++(0,1) --++(1,0);
\draw[thick, Mahogany](22,7)--++(0,1)--++(1,0)--++(0,-1);
\draw[>=stealth,->, gray] (22.5,6.2) -- ++(0,1.6);
\draw[thick,dotted] (4-3+20,7)--++(-0.8,0);

\def\x{2};
\draw[Blue!60,thick](1,10) -- (3,10);
\draw[Blue!60,thick](1,11) -- (3,11);
\draw[Blue!60,thick,dotted](1,10) -- (0.2,10);
\draw[Blue!60,thick,dotted](1,11) -- (0.2,11);

\draw[Red!60,thick](2,12) -- (4,12);
\draw[Red!60,thick](2,13) -- (4,13);
\draw[Red!60,thick,dotted](4,12) -- (4.8,12);
\draw[Red!60,thick,dotted](4,13) -- (4.9,13);

\filldraw[draw=Blue!60, thick, fill = white] (\x+1,10) circle (7pt);
\filldraw[draw=Blue!60, thick, fill = white] (\x+1,11) circle (7pt);
\fill[Blue!60] (\x+1,10) circle (4pt);
\fill[Blue!60] (\x+1,11) circle (4pt);
\filldraw[draw=Red!60, thick, fill = white] (\x,12) circle (7pt);
\filldraw[draw=Red!60, thick, fill = white] (\x,13) circle (7pt);
\fill[Red!60] (\x,12) circle (4pt);
\fill[Red!60] (\x,13) circle (4pt);

\def\x{7};
\draw[Blue!60,thick](\x,10) -- ++(-1,0);
\draw[Blue!60,thick](\x,11) -- ++(-1,0);
\draw[Blue!60,thick,dotted](\x-1,10) -- ++(-0.8,0);
\draw[Blue!60,thick,dotted](\x-1,11) -- ++(-0.8,0);

\draw[Blue!60,thick](\x+1,12) -- ++(-2,0);
\draw[Blue!60,thick](\x+1,13) -- ++(-2,0);
\draw[Blue!60,thick,dotted](\x-1,12) -- ++(-0.8,0);
\draw[Blue!60,thick,dotted](\x-1,13) -- ++(-0.8,0);

\filldraw[draw=Blue!60, thick, fill = white] (\x,10) circle (7pt);
\filldraw[draw=Blue!60, thick, fill = white] (\x,11) circle (7pt);
\fill[Blue!60] (\x,10) circle (4pt);
\fill[Blue!60] (\x,11) circle (4pt);
\def\x{8};
\filldraw[draw=Blue!60, thick, fill = white] (\x,12) circle (7pt);
\filldraw[draw=Blue!60, thick, fill = white] (\x,13) circle (7pt);
\fill[Blue!60] (\x,12) circle (4pt);
\fill[Blue!60] (\x,13) circle (4pt);

\def\x{12};
\draw[Blue!60,thick](\x,10) -- ++(-1,0);
\draw[Blue!60,thick](\x,11) -- ++(-1,0);
\draw[Blue!60,thick,dotted](\x-1,10) -- ++(-0.8,0);
\draw[Blue!60,thick,dotted](\x-1,11) -- ++(-0.8,0);
\draw[Blue!60,thick](\x+1,12) -- ++(-2,0);
\draw[Blue!60,thick,dotted](\x-1,12) -- ++(-0.8,0);
\draw[Red!60,thick](\x+1,13) -- ++(1,0);
\draw[Red!60,thick,dotted](\x+2,13) -- ++(0.8,0);

\filldraw[draw=Blue!60, thick, fill = white] (\x,10) circle (7pt);
\filldraw[draw=Blue!60, thick, fill = white] (\x,11) circle (7pt);
\filldraw[draw=Blue!100, fill = white] (\x+1,11) circle (6pt);
\fill[Blue!60] (\x,10) circle (4pt);
\fill[Blue!60] (\x,11) circle (4pt);
\fill[Blue!100] (\x+1,11) circle (3pt);
\def\x{13};
\filldraw[draw=Blue!60, thick, fill = white] (\x,12) circle (7pt);
\filldraw[draw=Red!60, thick, fill = white] (\x,13) circle (7pt);
\fill[Blue!60] (\x,12) circle (4pt);
\fill[Red!60] (\x,13) circle (4pt);
\filldraw[draw=Red!90, fill = white] (\x-1,13) circle (6pt);
\fill[Red!90] (\x-1,13) circle (3pt);

\draw[>=stealth,->,Red!80] (\x,13.25) to[out=135,in=45] (\x-1,13.25);
\draw[>=stealth,->,Blue!80] (\x-1,10.75) to[out=-45,in=180+45] (\x,10.75);
\def\x{17};
\draw[Blue!60,thick](\x,10) -- ++(-1,0);
\draw[Blue!60,thick](\x+1,11) -- ++(-2,0);
\draw[Blue!60,thick,dotted](\x-1,10) -- ++(-0.8,0);
\draw[Blue!60,thick,dotted](\x-1,11) -- ++(-0.8,0);
\draw[Red!60,thick](\x,12) -- ++(2,0);
\draw[Red!60,thick,dotted](\x+2,12) -- ++(0.8,0);
\draw[Red!60,thick](\x+1,13) -- ++(1,0);
\draw[Red!60,thick,dotted](\x+2,13) -- ++(0.8,0);

\filldraw[draw=Blue!60, thick, fill = white] (\x,10) circle (7pt);
\fill[Blue!60] (\x,10) circle (4pt);
\filldraw[draw=Red!60, thick, fill = white] (\x,12) circle (7pt);
\fill[Red!60] (\x,12) circle (4pt);
\filldraw[draw=Red!90, fill = white] (\x,13) circle (6pt);
\fill[Red!90] (\x,13) circle (3pt);
\def\x{18};
\filldraw[draw=Blue!90, fill = white] (\x,10) circle (6pt);
\fill[Blue!90] (\x,10) circle (3pt);
\filldraw[draw=Blue!60, thick, fill = white] (\x,11) circle (7pt);
\fill[Blue!60] (\x,11) circle (4pt);
\filldraw[draw=Red!60, thick, fill = white] (\x,13) circle (7pt);
\fill[Red!60] (\x,13) circle (4pt);
\draw[>=stealth,->,Red!80] (\x,13.25) to[out=135,in=45] (\x-1,13.25);
\draw[>=stealth,->,Blue!80] (\x-1,9.75) to[out=-45,in=180+45] (\x,9.75);

\def\x{22};
\draw[Blue!60,thick](\x,10) -- ++(-1,0);
\draw[Blue!60,thick](\x,11) -- ++(-1,0);
\draw[Blue!60,thick,dotted](\x-1,10) -- ++(-0.8,0);
\draw[Blue!60,thick,dotted](\x-1,11) -- ++(-0.8,0);
\draw[Red!60,thick](\x+1,12) -- ++(1,0);
\draw[Red!60,thick,dotted](\x+2,12) -- ++(0.8,0);
\draw[Red!60,thick](\x+1,13) -- ++(1,0);
\draw[Red!60,thick,dotted](\x+2,13) -- ++(0.8,0);
\filldraw[draw=Blue!60, thick, fill = white] (\x,10) circle (7pt);
\fill[Blue!60] (\x,10) circle (4pt);
\filldraw[draw=Blue!60, thick, fill = white] (\x,11) circle (7pt);
\fill[Blue!60] (\x,11) circle (4pt);
\filldraw[draw=Red!90, fill = white] (\x,12) circle (6pt);
\fill[Red!90] (\x,12) circle (3pt);
\filldraw[draw=Red!90, fill = white] (\x,13) circle (6pt);
\fill[Red!90] (\x,13) circle (3pt);
\def\x{23};
\filldraw[draw=Red!60, thick, fill = white] (\x,12) circle (7pt);
\fill[Red!60] (\x,12) circle (4pt);
\filldraw[draw=Red!60, thick, fill = white] (\x,13) circle (7pt);
\fill[Red!60] (\x,13) circle (4pt);
\filldraw[draw=Blue!90, fill = white] (\x,11) circle (6pt);
\fill[Blue!90] (\x,11) circle (3pt);
\filldraw[draw=Blue!90, fill = white] (\x,10) circle (6pt);
\fill[Blue!90] (\x,10) circle (3pt);
\draw[>=stealth,->,Red!80] (\x,13.25) to[out=135,in=45] (\x-1,13.25);
\draw[>=stealth,->,Red!80] (\x,12.25) to[out=135,in=45] (\x-1,12.25);
\draw[>=stealth,->,Blue!80] (\x-1,10.75) to[out=-45,in=180+45] (\x,10.75);
\draw[>=stealth,->,Blue!80] (\x-1,9.75) to[out=-45,in=180+45] (\x,9.75);

\def\x{0}
\fill(\x+0.1+2.5-0.3*2,0.25+0.6*2) circle (6pt);
\fill[black] (\x+0.1+1.9+0.3*2,0.25+0.6*2)-- ++(180:7pt) arc (180:360:7pt) -- cycle;
\fill[black] (\x+0.1+1.9+0.3*3,0.25+0.6*3)-- ++(180:7pt) arc (180:360:7pt) -- cycle;
\def\x{5}
\fill(\x+0.1+2.5-0.3*2,0.25+0.6*2) circle (6pt);
\fill(\x+0.1+1.9-0.3*2,0.25+0.6*2) circle (6pt);

\def\x{10}
\fill[black] (\x+0.1+1.9-0.3*1,0.25+0.6*1)-- ++(180:7pt) arc (180:360:7pt) -- cycle;
\fill[black] (\x+0.1+1.9-0.3*2,0.25+0.6*2)-- ++(180:7pt) arc (180:360:7pt) -- cycle;
\fill[black] (\x+0.1+2.5+0.3*2,0.25+0.6*2)-- ++(180:7pt) arc (180:360:7pt) -- cycle;
\fill[black] (\x+0.1+2.5-0.3*3,0.25+0.6*3)-- ++(180:7pt) arc (180:360:7pt) -- cycle;
\draw[>=stealth,->,Red!90, semithick](\x+0.1+2.5+0.3*2,0.25+0.6*2) -- ++(-0.3,0.6);
\draw[>=stealth,->,Blue!90, semithick](\x+0.1+1.9-0.3*1,0.25+0.6*1) -- ++(0.3,0.6);

\def\x{15}
\fill[black] (\x+0.1+1.9-0.3*2,0.25+0.6*2)-- ++(180:7pt) arc (180:360:7pt) -- cycle;
\fill[black] (\x+0.1+1.9+0.3*3,0.25+0.6*3)-- ++(180:7pt) arc (180:360:7pt) -- cycle;
\fill[black] (\x+0.1+2.5+0.3*2,0.25+0.6*2)-- ++(180:7pt) arc (180:360:7pt) -- cycle;
\fill[black] (\x+0.1+2.5-0.3*2,0.25+0.6*2)-- ++(180:7pt) arc (180:360:7pt) -- cycle;
\draw[>=stealth,->,Red!90, semithick](\x+0.1+2.5+0.3*2,0.25+0.6*2) -- ++(-0.3,0.6);
\draw[>=stealth,->,Blue!90, semithick](\x+0.1+1.9-0.3*2,0.25+0.6*2) -- ++(0.3,0.6);

\def\x{20}
\fill[black] (\x+0.1+1.9-0.3*2,0.25+0.6*2) circle (7pt);
\fill[black] (\x+0.1+2.5+0.3*2,0.25+0.6*2)-- ++(180:7pt) arc (180:360:7pt) -- cycle;
\fill[black] (\x+0.1+2.5+0.3*3,0.25+0.6*3)-- ++(180:7pt) arc (180:360:7pt) -- cycle;
\draw[>=stealth,->,Red!90, semithick](\x+0.1+2.5+0.3*2,0.25+0.6*2) -- ++(-0.3,0.6);
\draw[>=stealth,->,Red!90, semithick](\x+0.1+2.5+0.3*3,0.25+0.6*3) -- ++(-0.3,0.6);
\draw[>=stealth,->,Blue!90, semithick](\x+0.1+1.9-0.3*2-0.1,0.25+0.6*2) to[out=135,in=180] ++(0.3,0.6);
\draw[>=stealth,->,Blue!90, semithick](\x+0.1+1.9-0.3*2+0.1,0.25+0.6*2) to[out=0,in=-45] ++(0.3,0.6);

\node at (0+0.5,13.5) {\footnotesize{(i,a)}};
\node at (5+0.5,13.5) {\footnotesize{(i,b)}};
\node at (10+0.5,13.5) {\footnotesize{(ii)}};
\node at (15+0.5,13.5) {\footnotesize{(iii)}};
\node at (20+0.5,13.5) {\footnotesize{(iv)}};
\end{tikzpicture}
    \caption{Summary of the rules for updating the bigram endpoints under the action of an infinite-$q$ Haar-random gate. The red and blue circles denote the left and right endpoints, ($e_\alpha^-$ and $e_\beta^+$) respectively and the arrows show their hopping due to the gate. The associated configurations of $\{\gamma_i\}$ and updates thereof are also shown where the brown lines denote the updated configuration. The bottom row shows the corresponding configurations on the information lattice and the flows therein. The red and blue arrows denote the flows due to the motion of the right and left endpoints respectively. }
    \label{fig:q-inf-rules}
\end{figure}
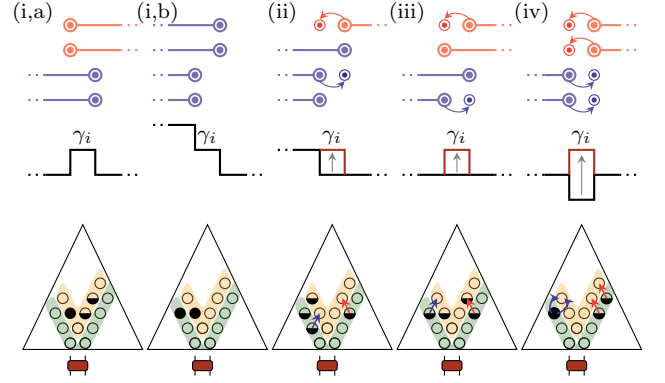

A gate acting on sites $(i,i+1)$ updates the $(\rho_i^\pm,\rho_{i+1}^\pm)$ as follows. 
\begin{enumerate}[label=(\roman*)]
    \item If $\rho_i^-=2$ or $\rho_{i+1}^+ = 2$, then any change decreases the length of at least one bigram and hence such a configuration is left untouched.
    \item If $\rho_{i}^+ = 2$ and $\rho_{i+1}^- = 1$, we chose the $+$ endpoint on the left with larger index $\alpha$  and swap its position with the one $-$ endpoint on the right.
    This increases the length of two of the bigrams each by one. 
    Similarly, if $\rho_{i}^+ = 1$ and $\rho_{i+1}^- = 2$, we chose the $-$ endpoint on the right with smaller index $\alpha$  and swap its position with the one $+$ endpoint on the left.
    
    \item If $\rho_{i}^+ = 1$ and $\rho_{i+1}^- = 1$, two of the bigrams can grow in length. we swap the one $+$ endpoint on the left with the $-$ endpoint on the right, again two of the bigrams growing in length by 1.
    \item If $\rho_{i}^+ = 2$ and $\rho_{i+1}^- = 2$, we swap the positions of both $+$ endpoints on the left with the positions of both $-$ endpoints on the right, which grows all the four bigrams by length 1.
\end{enumerate}
The rules (i)-(iv) are graphically shown correspondingly in panels (i)-(iv) in Fig.~\ref{fig:q-inf-rules}. 
In addition, the corresponding change to the $\gamma_i$ is also shown for clarity. 
Finally, Fig.~\ref{fig:q-inf-rules} also shows the information flow effected by the gate on the information lattice for each of the cases. 
The pattern of the flows is entirely consistent with the predictions made in Sec.~\ref{sec:loc-u-info-curr}, see Fig.~\ref{fig:constraints}.

Note that, on the information lattice, every bigram $(e_\alpha^-,e_\alpha^+)$ corresponds to `particle' at $((e^+_\alpha+e^-_\alpha)/2,e^+_\alpha-e^-_\alpha)$ carrying a unit of local information. However, since the gates only move the endpoints of the bigram, the dynamics of local information  
maps onto a classical stochastic model with nearest-neighbour hoppings subject to a non-local constraint. At each time-step, particles hop in the north-east and north-west directions, with the only restriction being the constraints in Eq.~\ref{eqn:constraint}. 
This reflects the underlying physics that the gates always try to move information from smaller to larger length scales.
In case of ambiguity, where only one of two eligible particles can hop, the particle at lower $l$ is selected as a consequence of maintaining the monotonic nature of $e^\pm_\alpha$ (see Fig.~\ref{fig:q-inf-rules}(iii) for an example). This means that on the information lattice, information currents are sensitive to the length scale at which information is encoded. We note that in the absence of this feature, 
the dynamics of the right endpoints will become independent of the location of the left endpoint, 
making the fluctuations in bigram length increase $\sim t^{1/4}$ which, as we show shortly, is incorrect.

Through all of the above, we have exactly and explicitly mapped the dynamics of local information onto a classical stochastic process which is local but governed by non-local constraints. This naturally makes it amenable to numerical simulations for arbitrarily large systems. 
We present results for such simulations in the next subsection but before that we can already make some general comments here.
At the leading order, if we ignore the subleading fluctuations in $\gamma_i$, Eq.~\ref{eqn:ent-cut} reduces to Eq.~\ref{eqn:ent-finite-form} with the asymptotic value of $v_E = 1$. 
Thus, at a coarse-grained level, the microscopic evolution rules outlined above furnish the local information flow described in Sec.~\ref{sec:leading-order}. 
In the long time limit, the configuration of $\left\{\gamma_i\right\}$ takes the form, $\gamma_i = {\rm min}\left\{i,L-i\right\}$ due to finite boundaries. This gives $I_n^l = 2\delta_{l,L/2}$. This is in agreement with Eq.~\ref{eqn:loc-info-haar-state} since in the limit $q\rightarrow \infty$, the constant associated with the exponential decay away from $l=L/2$ diverges.

It is also possible to make some remarks about the subleading effects in the local information, in particular, the broadening (in the direction of $l$) of the peak in the local information profile. As pointed out before, $\gamma_i$ as a function of $i$ acts as a KPZ height field (subject to appropriate boundary constraints) which fluctuates as $\Delta\gamma_i \sim t^{1/3}$ at the subleading order. 
This means that microscopically, the local information profile will have a finite width even after averaging 
which also must scale as $\sim t^{1/3}$. 
This result can be heuristically justified by the following argument. 
As a direct consequence of the evolution rules stated above, note that an increase in length of one of the minimal cuts by one (two) unit(s) implies that two (four) units of local information observe a unit increase in their $l$ coordinate. This means that if we ignore the spatial correlations in the height field, the width must scale with the same exponent that governs fluctuations in $\gamma_i$. As a cautionary remark, we mention that this exponent does not stem from the local information profile itself behaving like a KPZ height field (the dynamics of local information has non-local correlations which are unlikely to give the KPZ correlation structure), but just from the fact that the fluctuations carry over when the local information is computed from the height field. We shall return to this connection in Sec.~\ref{sec:hydro}.

\subsubsection{Numerical results from stochastic process}

\begin{figure}
    \centering
    \includegraphics[width=\linewidth]{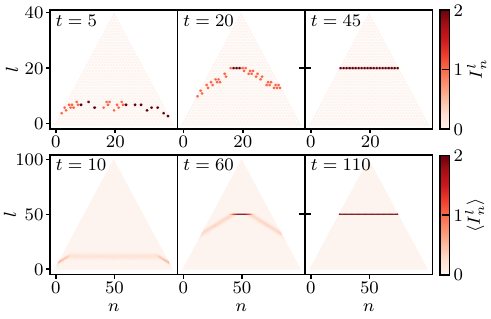} 
    \caption{Numerical results for the local information profile at different times in the $q\rightarrow \infty$ Haar-random circuit. (a) Results for a single circuit realisation and $L = 40$. For a single circuit realisation, $I_{n}^l$ can take values only 0, 1 and 2. (b) Results averaged over 1000 circuit realisations for $L=100$.}
    \label{fig:q-inf-res-1}
\end{figure}

We now turn to numerical results for the dynamics of the local information obtained from simulating the stochastic process discussed above. 
Specifically, we consider circuits with random geometry, where at every step, we randomly select a site $i$ with uniform probability and apply a gate between sites $i$ and $i+1$.
One unit of time is defined by the application of $L-1$ such gates.
We start with an initial product state such that at $t=0$, all the information resides on level $l=0$ such that $I_n^l = \delta_{l,0}$.
In Fig.~\ref{fig:q-inf-res-1}(a), we show the information lattice for a single circuit realisation at different times which shows the typically upward motion of particles containing a unit information on the lattice. 
Averaging the results over several circuit realisations leads to the profiles of local information shown in Fig.~\ref{fig:q-inf-res-1}(b),
which agree with the analysis of Sec.~\ref{sec:leading-order} at the leading order including the boundary effects. This is evinced by the similarity in the information lattices in Fig.~\ref{fig:q-inf-res-1}(b) compared to those in Fig.~\ref{fig:leading-order} where the latter was obtained from Eqs.~\ref{eqn:loc-info-finite-form-a} through Eq.~\ref{eqn:loc-info-finite-sat}.
In addition, at saturation, we have $I_n^l = 2\delta_{l,L/2}$ as expected.

\begin{figure}
    \centering
    \includegraphics[width=\linewidth]{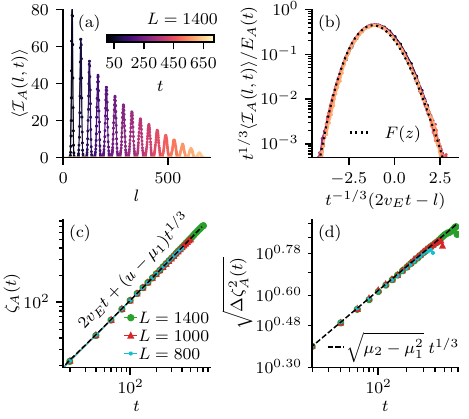}
    \caption{Numerical results for the subsystem information lattice corresponding to a subsystem $A$ of length $L/2$ located at the centre of the chain for the inifnite-$q$ Haar-random circuit. (a) The average information per scale in $A$ as a function of $l$ for different times as indicated by the colour scale. (b) Scale collapse of the data in (a) onto a common curve which shows the validity of the scaling form in Eq.~\ref{eq:IAlt-scaling-qinf}. The dashed line denotes the normalised Tracy-Widom distribution with a shifted argument, see Eq.~\ref{eq:tw-dist}. (c)-(d) The behaviour of the average information length scale and its fluctuations with time $t$. The dashed lines denote the predictions in Eq.~\ref{eq:zeta-mean} and Eq.~\ref{eq:zeta-fluc}.
    }
    \label{fig:q-inf-res-2}
\end{figure}

To put these results on a quantitative footing we compute the information per scale, $\langle \mathcal I_A(l,t) \rangle$ (defined in Eq.~\ref{eqn:info-per-scale-A}), as a function of time within the subsystem $A = \mathcal C_{(L-1)/2}^{\lfloor L/2-1\rfloor}$, where the $\braket{\cdot}$ denotes the averaging over the circuit realisations.
This subsystem --- of half the system length and located exactly at the centre --- is chosen such that it just avoids the boundary effects and as such, captures the middle piece of Eq.~\ref{eqn:loc-info-finite-form-a} 
at the leading order. 
In Fig.~\ref{fig:q-inf-res-2}(a), we show the results for  $\langle \mathcal I_A(l,t) \rangle$ as a function of $l$ at different values of $t$.
The data show that 
$\braket{\mathcal I_A(l,t)}$ takes the form of a wave packet with propagating ballistically in positive $l$ direction with a velocity of $2v_E$ with a width which also grows in time. 

Recall that the profile, $\braket{\mathcal I_A(l,t)}/E_A(t)$, where $E_A(t) = \sum_l \braket{\mathcal I_A(l,t)}$ is the total information in $A$ at time $t$, can be interpreted as a probability distribution over $l$.
We find that this distribution for different $t$ can be collapsed onto a common universal curve (see Fig.~\ref{fig:q-inf-res-2}(b)),
\begin{equation}
    \frac{\langle \mathcal I_A(l,t) \rangle}{E_A(t)} =   t^{-1/3}F(z)\,;~~z = \frac{2v_Et-l}{t^{1/3}}\,,
    \label{eq:IAlt-scaling-qinf}
\end{equation}
for $t < (4v_E)^{-1}L_A$, given that $A$ is chosen to be free of boundary effects at all times\footnote{This can be relaxed to include subsystems that do see an onset of boundary effects from either boundary at some time; in this case the scaling form will be valid before said onset.}.
In Eq.~\ref{eq:IAlt-scaling-qinf}, $E_A(t) = L_A-2v_Et$ which reflects simply the ballistic growth of entanglement. 
More interestingly, we find that the scaling function $F$ is in an excellent agreement with the Tracy-Widom distribution
\eq{
F(z) = P_{{\rm TW}, \beta=2}(z+u)\,,
\label{eq:tw-dist}
}
where $u\approx 0.7$ is a constant. 
It is worth reiterating the disclaimer that the Tracy-Widom distribution of $\braket{{\cal I} (l,t)}$ does not arise from the local information itself behaving like a KPZ height field and instead, it is the KPZ nature of the $\gamma_i$ height field that possibly lies at the root of it.

Given the scaling forms, it is now straightforward to estimate the time dependence of the average length scale at which information lives and its variance.
Using the definition in Eq.~\ref{eqn:expected-length} for $\zeta_A(t)$ in Eq.~\ref{eq:IAlt-scaling-qinf} we obtain
\eq{
\zeta_A(t) = 2v_Et + (u-\mu_1)t^{1/3}\,,
\label{eq:zeta-mean}
}
where $\mu_1\approx -1.77$ is the mean of $P_{{\rm TW}, \beta=2}$. Note that the computation not only captures the ballistic growth of the information length scale but also the subleading correction to it. 
Similar calculations lead to the width of the profile as 
\eq{
\sqrt{\Delta \zeta_A^2(t)} = \sqrt{\mu_2-\mu_1^2}\times t^{1/3}\,,\label{eq:zeta-fluc}
}
where $\mu_2\approx 3.95$ is the second moment of the Tracy-Widom distribution. 
These temporal scalings of both $\zeta_A(t)$ and $\sqrt{\Delta \zeta_A^2(t)}$ are confirmed by the numerical results in Fig.~\ref{fig:q-inf-res-2}(c)-(d).

For completeness, we note that the above scalings are valid only for $t<(4v_E)^{-1}L_A$. For $t>(4v_E)^{-1}L_A$, the dynamics of the local information saturates such that $\zeta_A = L/2$ and $\sqrt{\Delta \zeta_A^2(t)}\sim O(1)$ due to the exponential decay of the profile away from $L/2$.

\subsection{Finite local Hilbert-space dimension: qubits with $q=2$ \label{sec:haar-q-2}}

A natural question that arises is the fate of the information lattice dynamics for, arguably, the physically (more) relevant cases of qubits with $q=2$.
However, unlike their $q\to\infty$ counterparts, Haar-random circuits with finite $q$ do not lend themselves to exact solvability.
As such, we are restricted to numerical results limited by finite system sizes. 
Nonetheless, in this section, we show that much of the qualitative features of the local information dynamics does carry over from ${q\to\infty}$ to ${q=2}$.

Specifically, we consider Haar-random circuits for a chain of $L$ qubits, in the same setting as before, where dynamics is generated by random 2-site gates drawn uniformly from $U(4)$, acting on a randomly chosen nearest neighbour pair. 
At $t = 0$, we start with a random product state $\vert\psi_0\rangle = \bigotimes_{i=0}^{L-1} \vert \phi_i\rangle$ where $\vert\phi_i\rangle$ are independent single qubit Haar random states. This is then evolved with the circuit geometry described above and local information is computed directly by computing von Neumann information for all contiguous subsystems. 

\begin{figure}
    \centering
    \includegraphics[width=\linewidth]{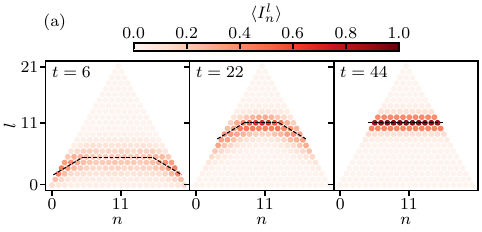}
    \includegraphics[width=0.9\linewidth]{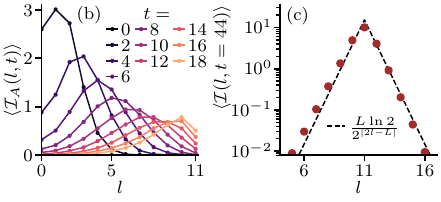}
    \caption{ Numerical results for the local information dynamics for the $q=2$ Haar-random circuits, averaged over 50 realisations and $L=22$. (a) The averaged local information on the information lattice at three different times which shows the general transfer of information from smaller to larger scales. The leading order results from Eq.~\ref{eqn:loc-info-finite-form-a} and Eq.~\ref{eqn:loc-info-finite-form-b} are superposed on the numerical data. (b) The average information per scale, $\braket{{\mathcal I_A}(l)}$ as a function of $l$ for different times (indicated by the colour scale) shows a travelling wave form. (c) $\braket{{\cal I}(l)}$ at times greater than the saturation timescale shows a peak at $l=L/2$ and exponential decay away from it.}
    \label{fig:q-2-res}
\end{figure}

Numerical results for the profiles of $I_n^l$ at different times, averaged over circuit realisations, are shown in Fig.~\ref{fig:q-2-res}(a). For clarity and comparison, the leading order profiles in Eq.~\ref{eqn:loc-info-finite-form-a} and Eq.~\ref{eqn:loc-info-finite-form-b} are superposed on the numerical data with $v_E = 0.4$. The profiles look qualitatively identical to those of the ${q\to\infty}$ limit shown in Fig.~\ref{fig:q-inf-res-1}(b).
We also show the total information per length scale, $\braket{{\cal I}(l)}$ as a function of $l$ for different times in Fig.~\ref{fig:q-2-res}(b).
Analogous to the data in Fig.~\ref{fig:q-inf-res-2}(a), the profile again has a travelling wave form with a width which seems to grow with time.
However, note that limited system sizes and dynamical temporal range lead to a contamination of the data from the boundaries at all times. This renders an accurate and faithful analysis of the data for $\zeta(t)$ and $\Delta\zeta^2(t)$ unfeasible.

That being said, the late time saturated dynamics is independent of the boundaries. In Fig.~\ref{fig:q-2-res}(c), we show the profile $\braket{{\cal I}(l)}$ at very late times, $t\gg L$. At these times, the state of the system is expected to be very well described as a Haar-random state. Consistent with the prediction in Eq.~\ref{eqn:loc-info-haar-state}, in this case the profile is again peaked at $l=L/2$ and decays exponentially away it.

\section{Random Clifford circuits \label{sec:res-cliff}}

Given the lack of tractability, analytical or numerical, in generic circuits with finite $q$, it is natural to look for specific classes of models which may offer significantly more solvability or simulability\footnote{In generic random-circuit models with finite $q$, while some analytical progress can be made for higher R\'enyi entropies~\cite{zhou2019emergent}, the von Neumann entropy remains intractable. Stabiliser states offer the simplification of a flat entanglement spectrum such that all R\'enyi entropies and the von Neumann entropy of subsystems are the same.}. 
With this motivation, in this section, we study the dynamics of local information in random Clifford circuits within the stabiliser formalism, as the dynamics of stabiliser states under unitary operators from the Clifford lend themselves to efficient classical simulation~\cite{gottesman1997stabilizer,gottesman1998heisenberg,aaronson2004improved}. 
This allows for numerical computations on system sizes significantly larger than those allowed by generic quantum systems and facilitates the development of a phenomenological understanding of local information flow, thereby extending out understanding to finite $q$, albeit within a restrictive model. For the remainder of this section, we consider qubits ($q=2$).

\subsection{Stabiliser formalism and Clifford circuits}

A stabiliser state, $\ket{\psi_{\mathcal S}}$, on $L$ qubits is specified uniquely by a set of $L$ mutually commuting and independent Pauli strings (with an overall phase of $\pm 1$), each of length $L$. 
This set, $\left\{\mathcal Q_1,\mathcal Q_2,...,\mathcal Q_L\right\}$, serves as minimal generating set
for a group $\mathcal S$ which is a subgroup of the Pauli group. 
The stabiliser state satisfies
\begin{equation}
    \mathcal Q_\alpha\vert\psi_{\mathcal S}\rangle = \vert \psi_{\mathcal S}\rangle\;\;\forall\; \mathcal Q_\alpha \in \mathcal S\,.
    \label{eqn:stabilizer-state}
\end{equation}
The Pauli strings $\mathcal Q_\alpha$ are called stabiliser generators and $\vert \psi_S\rangle$ is the associated stabiliser state. 
The full description of a quantum state in terms of only $L$ Pauli strings, each of length $L$, naturally allows for their efficient numerical storage.
Starting from a stabiliser state and evolving it with a time-evolution unitary,
\eq{
\vert \psi_{\mathcal S}(t)\rangle = \mathcal U_t\vert \psi_{\mathcal S}\rangle\,,
}
the time-evolved state remains a stabiliser state if the unitary ${\cal U}_t$ is such that it evolves a single Pauli string into another single Pauli string,
\eq{
{\cal Q}_\alpha^\prime={\cal U}_t{\cal Q}_\alpha {\cal U}_t^\dagger\,.
}
In such a case, $\vert \psi_{\mathcal S}(t)\rangle$ is stabilised the group $\mathcal S^\prime$ which is generated by the set $\{{\cal Q}_\alpha^\prime\}$.

An ubiquitous example of a family of such unitaries are ${{\cal U}_t\in{\rm CL}(L)}$
where ${\rm CL}(L)$ denotes the Clifford group, which is a subgroup of $U(L)$ and maps Pauli strings to Pauli strings. 
For any $L$,
${\rm CL}(L)$ can be generated by a set of one- and two-qubit unitary gates  given by
\begin{align}
\begin{split}
    H &= \frac{1}{\sqrt{2}}\left(X + Y\right)\,,\;\;\;P = \sqrt Z\,, \\
\text{CNOT} &= \frac{1}{2}\left[(\mathbb I+Z)\otimes \mathbb I + (\mathbb I-Z)\otimes X\right]\,,
\end{split}
    \label{eqn:cliff-gen}
\end{align}
where $X,Y,Z$ denote the three Pauli operators. 
The gates in Eq.~\ref{eqn:cliff-gen} are the familiar Hadamard ($H$), Phase ($P$) and Controlled-NOT (with first qubit as control and the second as target) operations. Note that the CNOT is the only two qubit gate in this set. 

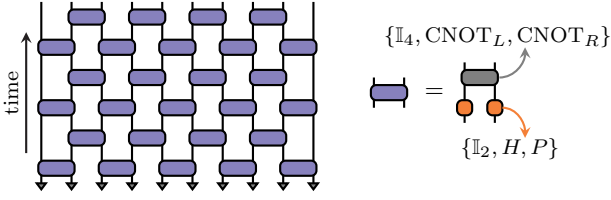
\begin{figure}
\begin{tikzpicture}[scale=0.4]

\foreach \x in {0,2,4,6,8}{
    \draw[thick](\x,-0.5) -- (\x,5.5);
    \draw[thick](\x+1,-0.5) -- (\x+1,5.5);
    \filldraw[draw=black,thick,fill=gray](\x-0.15,-0.5)--(\x+0.15,-0.5)--(\x,-0.7)--cycle;

    \filldraw[draw=black,thick,fill=gray](\x-0.15+1,-0.5)--(\x+0.15+1,-0.5)--(\x+1,-0.7)--cycle;
    \foreach \y in {0,2,4}{
        \filldraw[draw=black,thick,fill=Blue!50, rounded corners=2](\x-0.1,\y-0.25) rectangle (\x+1+.1,\y+0.25);
    }
}
\foreach \x in {1,3,5,7}{
    \foreach \y in {1,3,5}{
\filldraw[draw=black,thick,fill=Blue!50, rounded corners=2](\x-0.1,\y-0.25) rectangle (\x+1+.1,\y+0.25);
    }
}

\draw[thick](11,2)--(11,3);
\draw[thick](12,2)--(12,3);
\filldraw[draw=black,thick,fill=Blue!50, rounded corners=2](11-0.1,2.5-0.25) rectangle (11+1+.1,2.5+0.25);

\node at (13,2.5) {$=$};

\draw[thick](14,1.5)--(14,3.5);
\draw[thick](15,1.5)--(15,3.5);
\filldraw[draw=black,thick,fill=gray, rounded corners=2](14-0.1,3-0.25) rectangle (14+1+.1,3+0.25);

\draw[>=stealth, ->, thick, gray] (15.15,3) to[out=0,in=270] (16,4);

\node at (15.1,4.35) {\footnotesize{$\{\mathbb{I}_4,{\rm CNOT}_L,{\rm CNOT}_R\}$}};

\filldraw[draw=black,thick,fill=Orange, rounded corners=2](14-0.25,2-0.25) rectangle (14+.25,2+0.25);
\filldraw[draw=black,thick,fill=Orange, rounded corners=2](15-0.25,2-0.25) rectangle (15+.25,2+0.25);

\draw[>=stealth, ->, thick, Orange] (15.25,2) to[out=0,in=90] (16,1);
\node at (15.5,0.65) {\footnotesize{$\{\mathbb{I}_2,H,P\}$}};

\draw[>=stealth, ->, thick, Black] (-0.5,0.5) -- ++(0,4.0);
\node[rotate=90] at (-1,2.5) {\footnotesize{time}};

\end{tikzpicture}
\caption{A brickwork circuit where each gate is a Clifford gate constructed as shown on the right. Each gate is constructed out of two one-qubit gates chosen independently from $\{\mathbb{I}_2,H,P\}$ and one two-qubit gate chosen randomly from $\{\mathbb{I}_4,{\rm CNOT}_L,{\rm CNOT}_R\}$.}
    \label{fig:cliff-circuit}
\end{figure}

A circuit composed entirely of the gates in Eq.~\ref{eqn:cliff-gen} naturally constitutes a Clifford circuit and retains the stabiliser-ness of an initial stabiliser state.
Since our interest is in local circuits, we consider a random brickwork Clifford circuit with open boundary conditions (see Fig.~\ref{fig:cliff-circuit}),
\eq{
{\cal U}_t = \prod_{\tau=1}^t \left[\left(\bigotimes_{i=1}^{L/2-1} u^{(\tau)}_{2i-1,2i}\right)\left(\bigotimes_{i=0}^{L/2-1} u^{(\tau)}_{2i,2i+1}\right)\right]\,.
\label{eq:cliff-circ}
}
Each nearest-neighbour gate is given by 
\eq{
u^{(\tau)}_{i,i+1} = B^{(\tau)}_{i,i+1}\cdot \left(A^{(\tau)}_{i}\otimes A^{(\tau)}_{i+1}\right)\,,
}
where each one-qubit generator $A$ is chosen randomly and independently from $\{\mathbb{I}_2, H,P\}$ and each two-qubit generator $B$ from 
$\left\{\mathbb I_2, \text{CNOT}_L,\text{CNOT}_R\right\}$. Here $\text{CNOT}_{L(R)}$ acts on two nearest-neighbour qubits where the first (second) qubit is the control and the second (first) qubit is the target. 
Efficient classical simulability is achieved by noting that a stabiliser state can be evolved simply by updating its set of stabiliser generators successively under the action of each of the gates in Eq.~\ref{eq:cliff-circ}.
We give the details of the numerical algorithm and the computation of entanglement entropy in Appendix~\ref{appendix:cliff-num}.

In earlier studies of this model~\cite{nahum2017quantum}, it was observed that the von Neumann entropy across a cut in a finite system behaves similarly to less restrictive models such as the Haar-random circuit --- growing ballistically with subleading KPZ fluctuations and saturating to a near maximal value. This implies that the analysis of Section \ref{sec:leading-order} continues to be valid for this model. 
However, it is important to note that entanglement in stabiliser states is highly structured, manifested in a flat entanglement spectrum and integer valued entanglement entropies~\footnote{This is of course contingent on the logarithms being taken with base $q=2$}. 
This implies that the local information $I_n^l$ takes discrete values from the set $\left\{0,1,2\right\}$. This hints towards the possibility of describing the local information flow on the information lattice in terms of classical particles, each carrying one unit of information, hopping thereon where each site can host 0, 1 or 2 particles, reminiscent of the infinite-$q$ Haar-random circuit studied in Sec.~\ref{sec:haar-q-inf}.

In the following we first present numerical results for the dynamics of $I_n^l$ in Sec.~\ref{sec:numerics-cliff}. We then provide a phenomenological theory for the same, in Sec.~\ref{sec:stab-pheno}, in terms of a classical Markov process for the end points of bigrams where the bigrams now encode the geometry of the Pauli strings in an appropriate gauge.

\subsection{Numerical results \label{sec:numerics-cliff}}

\begin{figure}
    \centering
    \includegraphics[width=\linewidth]{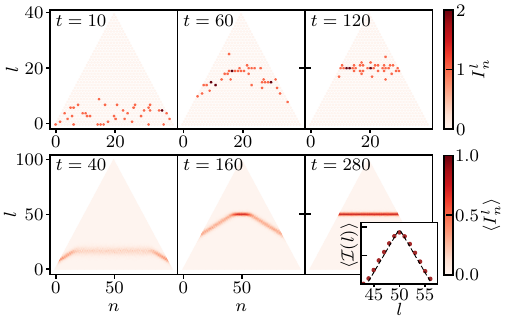}
    \caption{Numerical results for the local information profile for Clifford dynamics. The top rows shows results for a single realisation with $L=40$ and the bottom row shows results averaged over $500$ realisations with $L= 100$. The inset shows the profile of $\braket{{\cal I}(l)}$ as a function of $l$ at $t=280$, which is representative of the profiles at and after saturation. The expected form $\sim L\ln2e^{-2\ln 2 \vert L/2-l\vert}$ is also superposed as a dashed line.}
    \label{fig:cliff-res}
\end{figure}

By exploiting the efficient classical simulability of stabiliser states under Clifford dynamics and computation of entanglement therein, we numerically compute the dynamics of the local information, $\{I_n^l\}$, on the information lattice for system sizes much larger than those afforded by direct simulations of quantum systems.

The results for the profile of $I_n^l$ at different times are shown in Fig.~\ref{fig:cliff-res} where the top row shows the data for a single circuit realisation whereas the bottom row shows the same but averaged over a few thousand circuit realisations.
The 
results appear qualitatively identical 
to the results for Haar-random circuits discussed in Sec.~\ref{sec:haar-q-inf}.
The results for single circuit realisation shows that 
there is a general drift of the local information from smaller to larger scales.
The circuit realisation averaged results make it evident that this flow happens ballistically with a velocity $v_E$ which is smaller than that in the Haar-random circuits.
At timescales beyond saturation, the information per scale, $\braket{{\cal I}(l)}$, peaks at $l=L/2$ and decays exponentially away from it
with a decay constant which is evidently the same as in Eq.~\ref{eqn:loc-info-haar-state}. 
Evidence for this is provided in the inset in Fig.~\ref{fig:cliff-res}.
Since Clifford circuits cannot produce Haar-random states, this suggests that the said feature on the information lattice is a property of near maximal, volume-law entangled (across any bipartition) states and not necessarily restricted to Haar-random states.

\begin{figure}
    \centering
    \includegraphics[width=\linewidth]{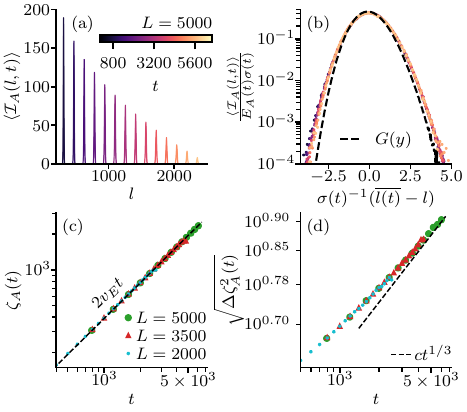}
    \caption{Numerical results for the subsystem information lattice corresponding to a subsystem $A$ of length $L/2$ located at the centre of the chain for the Clifford circuit model. (a) The average information per scale in $A$ as a function of $l$ for different times as indicated by the colour scale. (b) Scale collapse of the data in (a) onto a stable distribution which shows the validity of the scaling form in Eq.~\ref{eq:cliff-scaling}. The dashed line denotes the normalised Tracy-Widom distribution with a shifted argument. (c)-(d) The behaviour of the average information length scale and its fluctuations with time $t$ compared to the predicted behavior (dashed lines) in Eq. \ref{eq:zeta-fluc-cliff}.}
    \label{fig:cliff-res-l}
\end{figure}

To the put the local information dynamics on a quantitative footing, analogous to the Sec.~\ref{sec:haar-q-inf}, we consider the information per scale in the subsystem $A = \mathcal C_{(L-1)/2}^{\lfloor L/2-1\rfloor}$. 
The results for $\braket{{\cal I}_A(l,t)}$ as a function of $l$ for different $t$ are shown in Fig.~\ref{fig:cliff-res-l}(a).
The data can be reasonably well collapsed to a scaling form
\eq{
\frac{\braket{{\cal I}_A(l,t)}}{E_A(t)} = \sigma(t) G(y)\,; ~~~y = \frac{\overline{l(t)}-l}{\sigma(t)}\,,\label{eq:cliff-scaling}
}
where $G(y)$ is again approximately described by a shifted Tracy-Widom distribution, $P_{{\rm TW}, \beta=2}(y+\mu_1)$ where the shift ensures that the distribution has mean zero concomitant with the variable $y$ in Eq.~\ref{eq:cliff-scaling}. 
This is shown in Fig.~\ref{fig:cliff-res-l}(b).
From the scaling form above it follows straightforwardly that
\eq{
\zeta_A(t) = \overline{l(t)}\;;~\sqrt{\Delta \zeta_A^2(t)} = \sqrt{\mu_2-\mu_1^2}\sigma(t)\,,\label{eq:zeta-fluc-cliff}
}
where again $\mu_1\approx-1.77$ and $\mu_2\approx 3.95$ are the first and second moments of $P_{{\rm TW}, \beta=2}$.
The results in Fig.~\ref{fig:cliff-res-l}(c) and (d) show that 
\eq{
\begin{split}
\overline{l(t)} &= 2v_E t\\
\sigma(t) &= c\,t^{1/3}
\end{split}\,,
}
with $v_E\approx 0.39$ which confirms that the spatiotemporal anatomy of the local information for Clifford circuits is similar to that of the Haar-random circuits~\footnote{Given all evidence, it is almost certain that $\zeta_A(t)$ has a subleading correction, $\propto t^{1/3}$. However, since it is negligible on the long timescales we can access relative to the ballistic leading contribution, it can be safely ignored for the numerical analysis.}.

However, a few words of caution are in order. 
The collapse of data as well as a the match with the Tracy-Widom distribution for the Clifford circuits is not as perfect as it is for the Haar-random circuits. However, the data for later $t$ do tend towards the Tracy-Widom distribution.
This is also consistent with the observation that the $t^{1/3}$ scaling of $\sqrt{\zeta_A^2(t)}$ sets in only at very late times. 
We will provide further evidence for the this via a phenomenologically obtained classical Markov process in the next subsection, where significantly larger sizes and timescales are accessible.

We also remark that, since it was earlier shown that the dynamical behaviour of entanglement entropy across a bond for random Clifford circuits also falls into the KPZ universality class~\cite{nahum2017quantum}, one might expect that the entanglement fluctuations which scale as $t^{1/3}$ and the exponent appearing in the broadening of the local information might share a common origin. However, unlike for Haar random circuits in the $q\rightarrow\infty$ limit, we are not aware of such a connection for random Clifford circuits. 
In what follows we build on the theory of entanglement in Clifford circuits~\cite{nahum2017quantum} and develop a phenomenological picture which unifies the description of  
entanglement fluctuations and the broadening of the sample averaged local information.

\subsection{Phenomenological theory \label{sec:stab-pheno}}

Having established the numerical results for the local information dynamics in random Clifford circuits, we now turn towards a phenomenological theory for the same in terms of classical Markov process.

\subsubsection{Stabiliser bigrams and information lattice}

\begin{figure}
    \centering
    \begin{tikzpicture}[scale=0.4]

\foreach \x in {0,...,7}{
    \draw[gray, very thin](\x,4.5) -- (\x,10);
}

\foreach \l in {0,...,7}{
    \edef\maxN{\numexpr7-\l\relax}
    \foreach \n in {0,...,\maxN}{
        \filldraw[draw=black, thick,fill=white] (11+\n+0.5*\l, \l+4) circle (7pt);
    }   
}
\fill[black] (13,4)-- ++(180:7pt) arc (180:360:7pt) -- cycle;
\fill[black] (11.5,5)-- ++(180:7pt) arc (180:360:7pt) -- cycle;
\fill[black] (15.5,5)-- ++(180:7pt) arc (180:360:7pt) -- cycle;
\fill[black] (17.5,5) circle (7pt);
\fill[black] (13,6)-- ++(180:7pt) arc (180:360:7pt) -- cycle;
\fill[black] (14,6)-- ++(180:7pt) arc (180:360:7pt) -- cycle;
\fill[black] (13,8)-- ++(180:7pt) arc (180:360:7pt) -- cycle;

\draw[ultra thick, Blue!50] (1.85,5) -- (2.15,5);
\node at (1.85,5) {$\mathbf{\mathsf{x}}$};
\node at (2.15,5) {$\mathbf{\mathsf{x}}$};
\draw[thick] (1.85,5) -- (2.15,5);

\def\x{0};
\def\y{1};
\draw[ultra thick, Blue!50] (\x,6) -- (\y,6);
\node at (\x,6) {$\mathbf{\mathsf{x}}$};
\node at (\y,6) {$\mathbf{\mathsf{x}}$};
\draw[thick] (\x,6) -- (\y,6);

\def\x{4};
\def\y{5};
\draw[ultra thick, Blue!50] (\x,6) -- (\y,6);
\node at (\x,6) {$\mathbf{\mathsf{x}}$};
\node at (\y,6) {$\mathbf{\mathsf{x}}$};
\draw[thick] (\x,6) -- (\y,6);

\def\x{6};
\def\y{7};

\filldraw[draw=YellowOrange, dashed, thick, fill=YellowOrange!20, rounded corners=4](\x-0.35,6-0.35) rectangle (\y+0.35,6.5+0.35);

\draw[ultra thick, Blue!50] (\x,6) -- (\y,6);
\node at (\x,6) {$\mathbf{\mathsf{x}}$};
\node at (\y,6) {$\mathbf{\mathsf{x}}$};
\draw[thick] (\x,6) -- (\y,6);

\def\x{6};
\def\y{7};
\draw[ultra thick, Blue!50] (\x,6.5) -- (\y,6.5);
\node at (\x,6.5) {$\mathbf{\mathsf{x}}$};
\node at (\y,6.5) {$\mathbf{\mathsf{x}}$};
\draw[thick] (\x,6.5) -- (\y,6.5);

\def\x{3};
\def\y{5};
\draw[ultra thick, Blue!50] (\x,7) -- (\y,7);
\node at (\x,7) {$\mathbf{\mathsf{x}}$};
\node at (\y,7) {$\mathbf{\mathsf{x}}$};
\draw[thick] (\x,7) -- (\y,7);

\def\x{1};
\def\y{3};
\draw[ultra thick, Blue!50] (\x,7.5) -- (\y,7.5);
\node at (\x,7.5) {$\mathbf{\mathsf{x}}$};
\node at (\y,7.5) {$\mathbf{\mathsf{x}}$};
\draw[thick] (\x,7.5) -- (\y,7.5);

\def\x{0};
\def\y{4};

\filldraw[draw=OliveGreen, dashed, thick, fill=OliveGreen!20, rounded corners=4](\x-0.35,9-0.35) rectangle (\y+0.35,9+0.35);

\draw[ultra thick, Blue!50] (\x,9) -- (\y,9);
\node at (\x,9) {$\mathbf{\mathsf{x}}$};
\node at (\y,9) {$\mathbf{\mathsf{x}}$};
\draw[thick] (\x,9) -- (\y,9);

\draw[thick] (0,4) -- (7,4);
\foreach \x in {0,...,7}{
\fill (\x,4) circle (3pt);
\node at (\x,3.5) {{\footnotesize{\x}}};
}
\node at (-1,3.5) {{\footnotesize{$i=$}}};



\fill[OliveGreen, fill opacity=0.3, rounded corners=3] (13.5-0.5,7.25) -- ++(0.75,0.75) --++(-0.75,0.75) --++(-0.75,-0.75) -- cycle;

\fill[YellowOrange, fill opacity=0.3, rounded corners=3] (16+1.5,8.25-4) -- ++(0.75,0.75) --++(-0.75,0.75) --++(-0.75,-0.75) -- cycle;


\draw[OliveGreen, >=Stealth, ->, thick,dashed, rounded corners=6] (4.5,9) to[out=0,in=180-30] (13.25-0.5,8);

\draw[YellowOrange, >=Stealth, ->, thick,dashed, rounded corners=6] (7.35,6.25) to[out=-30,in=180+15] (16+1.25,5);


\node at (7.5,9.5) {\footnotesize{\color{OliveGreen} $(e_l,e_r)=(0,4)$}};
\node at (7.5,8.5) {\footnotesize{\color{OliveGreen} $(n,l)=(3,4)$}};





\end{tikzpicture}
    \caption{Correspondence between stabiliser bigrams and information lattice. Each horizontal dashed line on the left denotes a stabiliser Pauli string in the clipped gauge with the crosses marking its endpoints. The corresponding information lattice on thr right where empty, half-filled and fully-filled circles denote $I_n^l=0,1,2$. A couple of bigrams and their contribution to the information lattice are highlighted as examples.}
    \label{fig:stab-info-lat}
\end{figure}
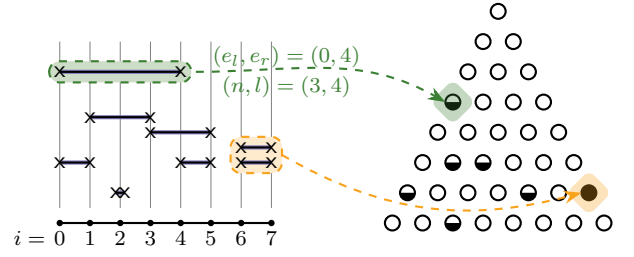

To set this up, we first show that the information lattice for a stabiliser state can be mapped exactly onto a configuration of bigrams (not dissimilar to the case of infinite-$q$ Haar-random states in Sec.~\ref{sec:haar-q-inf}) where the bigrams now correspond to the stabiliser generators corresponding to the state in a specific gauge.   
This mapping is rooted in the fact that the 
entanglement structure of a stabiliser state is contained entirely in the set of stabiliser generators $\left\{\mathcal Q_\alpha\right\}$.
However, there is a `gauge freedom' in the  
choice of the generating set of $\mathcal S$ which can be exploited to chose a set of generators which make the 
entanglement structure more transparent. The gauge transformation here  just $\mathcal Q_\alpha \rightarrow \mathcal Q_\alpha\mathcal Q_\beta$ for $\beta \neq \alpha$. Every generator $\mathcal Q_\alpha$ will have its leftmost and rightmost 
site with a non-trivial Pauli operator ($X$, $Y$, or $Z$),
which we call the left and right endpoints, denoted $e^-_\alpha$ and $e^+_\alpha$, respectively. The gauge transformation can be repeatedly applied so as to make sure that every site has exactly two of these endpoints (irrespective of their type). which is referred to as the  `clipped gauge'\footnote{For completeness, we review it briefly in Appendix \ref{appendix:clipped-gauge}.}~\cite{nahum2017quantum,li2019measurement}.  The locations of the endpoints represented as a set of bigrams,
\begin{equation}
    \mathcal B (S) = \left\{(e^-_1,e^+_1),\;(e^-_2,e^+_2),..., (e^-_{L-1},e^+_{L-1})\right\}
\end{equation}
are unique once the gauge is fixed to the clipped gauge.
Of course, this still leaves residual gauge freedom
within the clipped gauge which as far as the Pauli content of the generators is concerned but does not alter end points of the generators and hence the bigram configuration.
Within this gauge choice, the entanglement structure becomes transparent: we show in Appendix \ref{appendix:clipped-gauge}, that the von Neumann information of a region $\mathcal C_n^l$ is given by the number of clipped bigrams contained entirely inside $\mathcal C_n^l$. Formally,
\begin{equation}
    E_n^l = \sum_{\alpha = 0}^{L-1} \Theta(n+l/2 - e^+_\alpha)\Theta(e^-_\alpha - n+l/2).
    \label{eqn:von-Neumann-info-stab-state}
\end{equation}
Consequentially, the local information $I_n^l$ is equal to the number of clipped stabilizer bigrams of length $l$ contained in $\mathcal C_n^l$,
\begin{equation}
     I_n^l = \sum_{\alpha=1}^L \delta(e_\alpha^-,\;n-l/2)\; \delta(e_\alpha^+,\;n+l/2).
     \label{eqn:loc-info-stab-state}
\end{equation}
In other words, the clipped bigrams correspond to the local information content in the same way as for the bigrams defined for Haar-random circuit model in Sec.~\ref{sec:haar-q-inf}. 
Each bigram,
$(e_\alpha^-,e_\alpha^+)$,
corresponds to one unit of local information on the information lattice site located at $(n,l) = ((e^+_\alpha+e^-_\alpha)/2,\; e_\alpha^+-e_\alpha^-)$.
This is schematically shown in Fig.~\ref{fig:stab-info-lat}.
Also, the clipped gauge condition that each site hosts exactly two bigram endpoints is just the constraint in Eq.~\ref{eqn:constraint} in disguise. 

However, we also note some crucial differences. 
As discussed in Sec.~\ref{sec:haar-q-inf}, for infinite-$q$ Haar-random circuits, the bigram configuration was completely slaved to the minimal cuts (Eq.~\ref{eq:ealpha-pm}) which in turn constituted an exact representation of the entanglements across each single-cut bipartitions.
This is not the case for finite-$q$ states and hence 
$e^\pm_\alpha$ are no longer required to be monotonic functions of the bigram index $\alpha$. 
The bigrams simply encode the extent of the stabiliser generators in the clipped gauge.
In addition, 
for infinite-$q$ Haar-random states, the bigram configuration was in direct correspondence to an underlying KPZ height field of the since the information about entanglement entropy across all single-cut bipartitions (which was related to the endpoint densities) determined the local information exactly. This is no longer true for general stabiliser states. 
This naturally means hat the dynamical rules in terms of a classical Markov process cannot be derived unambiguously.
However, based on very similar guiding principles as in Haar-random circuits, a set of dynamical rules can nonetheless be obtained albeit phenomenologically, which we discuss next.

\subsubsection{Phenomenological dynamical rules for bigrams}

The evolution of the stabiliser bigrams under the action of a Clifford unitary takes place in two steps. 
First, the action of the gate locally modifies the bigram endpoints on the sites where the unitary gate acts and typically grows the stabilisers. However, this action can break the clipped gauge and hence, the stabilisers need to be clipped back in the second step.
This effectively amounts to a reshuffling of the bigram endpoints on the sites where the unitary acts.

\begin{figure}
\begin{tikzpicture}
\def\x{0};

\fill[YellowOrange!20] (\x-0.5,-0.5) rectangle (\x+0.75+.5,2.75);
\draw[YellowOrange, dashed,thick] (\x-0.5,-0.5) -- ++(1.75,0) -- ++(0,3.25) --++(0,0.75+0.125) --++(-0.5,-0.75-0.125) --++(-1.25,0)--cycle;

\draw[thick](\x,1.5) -- ++(-0.75,0);
\draw[thick](\x,2.25) -- ++(-0.75,0);
\draw[thick](\x+.75,0) -- ++(0.75,0);
\draw[thick](\x+.75,0.75) -- ++(0.75,0);
\filldraw[thick,black, fill=YellowOrange!20] (\x,1.5) circle (8pt);
\node at (\x,1.5) {\footnotesize{$Y$}};
\filldraw[thick,black, fill=YellowOrange!20] (\x,2.25) circle (8pt);
\node at (\x,2.25) {\footnotesize{$X$}};
\filldraw[thick,black, fill=YellowOrange!20] (\x+.75,0) circle (8pt);
\node at (\x+0.75,0) {\footnotesize{$Y$}};
\filldraw[thick,black, fill=YellowOrange!20] (\x+.75,0.75) circle (8pt);
\node at (\x+0.75,0.75) {\footnotesize{$X$}};

\def\x{3};
\draw[thick,gray](\x+0.75,1.5) -- ++(-1.5,0);
\draw[thick,gray](\x+.75,2.25) -- ++(-1.5,0);
\draw[thick,gray](\x,0) -- ++(1.5,0);
\draw[thick,gray](\x+.75,0.75) -- ++(0.75,0);

\filldraw[thick,,gray, fill=white] (\x,0) circle (8pt);
\node at (\x,0) {{\color{gray}{\footnotesize{$Z$}}}};
\filldraw[thick,,gray, fill=white!20] (\x,1.5) circle (8pt);
\node at (\x,1.5) {{\color{gray}{\footnotesize{$Y$}}}};
\filldraw[thick,,gray, fill=white!20] (\x+.75,2.25) circle (8pt);
\node at (\x+.75,2.25) {{\color{gray}{\footnotesize{$X$}}}};
\filldraw[thick,,gray, fill=white!20] (\x+0.75,1.5) circle (8pt);
\node at (\x+0.75,1.5) {{\color{gray}{\footnotesize{$X$}}}};
\filldraw[thick,,gray, fill=white!20] (\x,2.25) circle (8pt);
\node at (\x,2.25) {{\color{gray}{\footnotesize{$X$}}}};
\filldraw[thick,,gray, fill=white!20] (\x+.75,0) circle (8pt);
\node at (\x+0.75,0) {{\color{gray}{\footnotesize{$Y$}}}};
\filldraw[thick,,gray, fill=white!20] (\x+.75,0.75) circle (8pt);
\node at (\x+0.75,0.75) {{\color{gray}{\footnotesize{$X$}}}};

\def\x{6};
\fill[YellowOrange!20] (\x-0.5,-0.5) rectangle (\x+0.75+.5,2.75);
\draw[YellowOrange, dashed,thick] (\x-0.5,-0.5) -- ++(0,3.25) --++(0,0.75+0.125) --++(0.5,-0.75-0.125) --++(1.25,0)--++(0,-3.25)--cycle;

\draw[thick](\x+0.75,1.5) -- ++(-1.5,0);
\draw[thick](\x,2.25) -- ++(-0.75,0);
\draw[thick](\x,0) -- ++(1.5,0);
\draw[thick](\x+.75,0.75) -- ++(0.75,0);

\filldraw[thick,black, fill=YellowOrange!20] (\x,0) circle (8pt);
\node at (\x,0) {\footnotesize{$Z$}};
\filldraw[thick,black, fill=YellowOrange!20] (\x,1.5) circle (8pt);
\node at (\x,1.5) {\footnotesize{$Y$}};
\filldraw[thick,black, fill=YellowOrange!20] (\x+0.75,1.5) circle (8pt);
\node at (\x+0.75,1.5) {\footnotesize{$X$}};
\filldraw[thick,black, fill=YellowOrange!20] (\x,2.25) circle (8pt);
\node at (\x,2.25) {\footnotesize{$Z$}};
\filldraw[thick,black, fill=YellowOrange!20] (\x+.75,0) circle (8pt);
\node at (\x+0.75,0) {\footnotesize{$Y$}};
\filldraw[thick,black, fill=YellowOrange!20] (\x+.75,0.75) circle (8pt);
\node at (\x+0.75,0.75) {\footnotesize{$X$}};

\filldraw[draw=black,fill=Blue!40] (1.5,1) --++(0,0.2*0.4) --++(1*0.4,0) -- ++(0,0.25*0.4) -- ++(0.35*0.4,-0.35*0.4) -- ++(-0.35*0.4,-0.35*0.4) --++(0,0.25*0.4) -- ++(-1*0.4,0) --cycle;
\node at (1.75,1.35) {{\color{Blue}\footnotesize{grow}}}; 

\filldraw[draw=black,fill=Red!40] (4.5,1) --++(0,0.2*0.4) --++(1*0.4,0) -- ++(0,0.25*0.4) -- ++(0.35*0.4,-0.35*0.4) -- ++(-0.35*0.4,-0.35*0.4) --++(0,0.25*0.4) -- ++(-1*0.4,0) --cycle;
\node at (1.75+3,1.35) {{\color{Red}\footnotesize{clip}}}; 

\filldraw[draw=YellowOrange, dashed, fill=YellowOrange!20] (3.5*0.3,3.625) rectangle (5.5*0.3,4.625);

\def\a{0};
\def\b{6};
\def\y{3.25};
\draw[thick](\a*0.3,\y)--(\b*0.3,\y);
\foreach \i in {\a,...,\b}{
\fill[black](\i*0.3,\y) circle (1pt);
}
\node at (\a*0.3,\y) {\footnotesize{$\mathsf{x}$}};
\node at (\b*0.3,\y) {\footnotesize{$\mathsf{x}$}};

\def\a{3};
\def\b{6};
\def\y{3.5};
\draw[thick](\a*0.3,\y)--(\b*0.3,\y);
\foreach \i in {\a,...,\b}{
\fill[black](\i*0.3,\y) circle (1pt);
}
\node at (\a*0.3,\y) {\footnotesize{$\mathsf{x}$}};
\node at (\b*0.3,\y) {\footnotesize{$\mathsf{x}$}};

\def\a{5};
\def\b{8};
\def\y{3.75};
\draw[thick](\a*0.3,\y)--(\b*0.3,\y);
\foreach \i in {\a,...,\b}{
\fill[black](\i*0.3,\y) circle (1pt);
}
\node at (\a*0.3,\y) {\footnotesize{$\mathsf{x}$}};
\node at (\b*0.3,\y) {\footnotesize{$\mathsf{x}$}};

\def\a{5};
\def\b{7};
\def\y{4};
\draw[thick](\a*0.3,\y)--(\b*0.3,\y);
\foreach \i in {\a,...,\b}{
\fill[black](\i*0.3,\y) circle (1pt);
}
\node at (\a*0.3,\y) {\footnotesize{$\mathsf{x}$}};
\node at (\b*0.3,\y) {\footnotesize{$\mathsf{x}$}};

\def\a{1};
\def\b{4};
\def\y{4.25};
\draw[thick](\a*0.3,\y)--(\b*0.3,\y);
\foreach \i in {\a,...,\b}{
\fill[black](\i*0.3,\y) circle (1pt);
}
\node at (\a*0.3,\y) {\footnotesize{$\mathsf{x}$}};
\node at (\b*0.3,\y) {\footnotesize{$\mathsf{x}$}};

\def\a{0};
\def\b{4};
\def\y{4.5};
\draw[thick](\a*0.3,\y)--(\b*0.3,\y);
\foreach \i in {\a,...,\b}{
\fill[black](\i*0.3,\y) circle (1pt);
}
\node at (\a*0.3,\y) {\footnotesize{$\mathsf{x}$}};
\node at (\b*0.3,\y) {\footnotesize{$\mathsf{x}$}};

\def\a{7};
\def\b{8};
\def\y{4.25};
\draw[thick](\a*0.3,\y)--(\b*0.3,\y);
\foreach \i in {\a,...,\b}{
\fill[black](\i*0.3,\y) circle (1pt);
}
\node at (\a*0.3,\y) {\footnotesize{$\mathsf{x}$}};
\node at (\b*0.3,\y) {\footnotesize{$\mathsf{x}$}};

\def\a{1};
\def\b{3};
\def\y{3.75};
\draw[thick](\a*0.3,\y)--(\b*0.3,\y);
\foreach \i in {\a,...,\b}{
\fill[black](\i*0.3,\y) circle (1pt);
}
\node at (\a*0.3,\y) {\footnotesize{$\mathsf{x}$}};
\node at (\b*0.3,\y) {\footnotesize{$\mathsf{x}$}};

\filldraw[draw=YellowOrange, dashed, fill=YellowOrange!20] (3.5*0.3+4,3.625) rectangle (5.5*0.3+4,4.625);

\def\a{0};
\def\b{6};
\def\y{3.25};
\draw[thick](\a*0.3+4,\y)--(\b*0.3+4,\y);
\foreach \i in {\a,...,\b}{
\fill[black](\i*0.3+4,\y) circle (1pt);
}
\node at (\a*0.3+4,\y) {\footnotesize{$\mathsf{x}$}};
\node at (\b*0.3+4,\y) {\footnotesize{$\mathsf{x}$}};

\def\a{3};
\def\b{6};
\def\y{3.5};
\draw[thick](\a*0.3+4,\y)--(\b*0.3+4,\y);
\foreach \i in {\a,...,\b}{
\fill[black](\i*0.3+4,\y) circle (1pt);
}
\node at (\a*0.3+4,\y) {\footnotesize{$\mathsf{x}$}};
\node at (\b*0.3+4,\y) {\footnotesize{$\mathsf{x}$}};

\def\a{4};
\def\b{8};
\def\y{3.75};
\draw[thick](\a*0.3+4,\y)--(\b*0.3+4,\y);
\foreach \i in {\a,...,\b}{
\fill[black](\i*0.3+4,\y) circle (1pt);
}
\node at (\a*0.3+4,\y) {\footnotesize{$\mathsf{x}$}};
\node at (\b*0.3+4,\y) {\footnotesize{$\mathsf{x}$}};

\def\a{5};
\def\b{7};
\def\y{4};
\draw[thick](\a*0.3+4,\y)--(\b*0.3+4,\y);
\foreach \i in {\a,...,\b}{
\fill[black](\i*0.3+4,\y) circle (1pt);
}
\node at (\a*0.3+4,\y) {\footnotesize{$\mathsf{x}$}};
\node at (\b*0.3+4,\y) {\footnotesize{$\mathsf{x}$}};

\def\a{1};
\def\b{5};
\def\y{4.25};
\draw[thick](\a*0.3+4,\y)--(\b*0.3+4,\y);
\foreach \i in {\a,...,\b}{
\fill[black](\i*0.3+4,\y) circle (1pt);
}
\node at (\a*0.3+4,\y) {\footnotesize{$\mathsf{x}$}};
\node at (\b*0.3+4,\y) {\footnotesize{$\mathsf{x}$}};

\def\a{0};
\def\b{4};
\def\y{4.5};
\draw[thick](\a*0.3+4,\y)--(\b*0.3+4,\y);
\foreach \i in {\a,...,\b}{
\fill[black](\i*0.3+4,\y) circle (1pt);
}
\node at (\a*0.3+4,\y) {\footnotesize{$\mathsf{x}$}};
\node at (\b*0.3+4,\y) {\footnotesize{$\mathsf{x}$}};

\def\a{7};
\def\b{8};
\def\y{4.25};
\draw[thick](\a*0.3+4,\y)--(\b*0.3+4,\y);
\foreach \i in {\a,...,\b}{
\fill[black](\i*0.3+4,\y) circle (1pt);
}
\node at (\a*0.3+4,\y) {\footnotesize{$\mathsf{x}$}};
\node at (\b*0.3+4,\y) {\footnotesize{$\mathsf{x}$}};

\def\a{1};
\def\b{3};
\def\y{3.75};
\draw[thick](\a*0.3+4,\y)--(\b*0.3+4,\y);
\foreach \i in {\a,...,\b}{
\fill[black](\i*0.3+4,\y) circle (1pt);
}
\node at (\a*0.3+4,\y) {\footnotesize{$\mathsf{x}$}};
\node at (\b*0.3+4,\y) {\footnotesize{$\mathsf{x}$}};

\filldraw[draw=black,fill=gray] (3.,3.75) --++(0,0.2*0.4) --++(1*0.4,0) -- ++(0,0.25*0.4) -- ++(0.35*0.4,-0.35*0.4) -- ++(-0.35*0.4,-0.35*0.4) --++(0,0.25*0.4) -- ++(-1*0.4,0) --cycle;

\draw[thick] (4*0.3,4.75) -- ++(0,0.5);
\draw[thick] (5*0.3,4.75) -- ++(0,0.5);
\filldraw[draw=black,thick, fill=Blue!50,rounded corners=2] (1.1,4.88) rectangle (1.6,5.12);

\end{tikzpicture}
\caption{Illustration of growth and clipping of stabiliser generators. 
In the first step, the action of the Clifford operation ($\text{CNOT}_R$ in this case) grows the generators. However, this leads to a violation of the clipped gauge; in the example here the right site has three endpoints. In the second step, the new generators are restored to the clipped gauge; in the example here this is achieved by multiplying the top two generators.
Together this leads to a reshuffling of the clipped gauge bigram endpoints within the two sites the gate acts on as shown on the top.}
\label{fig:stab-clip-example}
\end{figure}
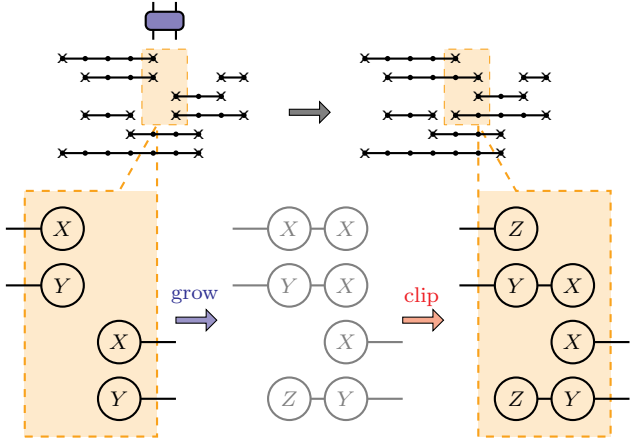

To makes this concrete, consider the application of a two-qubit Clifford unitary on sites $i$ and $i+1$.
Since such a gate can only change the entanglement across the cut between sites $i$ and $i+1$, it only changes the bigram configuration within these two sites. Specifically, the bigrams with their left ($-$) endpoints on site $i+1$ can either grow or stay of the same length, which is equivalent to saying the left endpoints can move to site $i$ or stay on $i+1$.
The same argument for the bigrams with their right ($+$) endpoints on site $i$ implies these endpoints can either move to $i+1$ or stay at $i$.
An example is shown in the step labelled `grow' in Fig.~\ref{fig:stab-clip-example}.
Note that at this step, the generators violate the clipped gauge condition
The next step is to clip the new generators back to the clipped gauge. 
This clipping procedure must satisfy the following conditions.
\begin{enumerate}[label=(\roman*)]
\item The clipping can always be done locally. This is a manifestation of the fact that the clipped gauge always exists for any given set of stabiliser generators and associated state is invariant under gauge transformations.
Since the generators in the clipped gauge unambiguously specify the entanglement of each subsystem and a gate acting on sites $i$ and $i+1$ can only change the entanglement of subsystems with their boundaries at $i$ or $i+1$, it naturally implies that such a gate cannot modify the clipped gauge bigram endpoints away from these two sites. 

\item The clipping can always be done between the same kinds of endpoints, {\it i.e.}, a right (left) endpoint is clipped by a right (left) endpoint on the same site.
To understand this, we note that if the endpoints are of the different kind, the gauge transformation $\mathcal Q_\alpha \rightarrow \mathcal Q_\alpha \mathcal Q_\beta$ will  increase the support of $\mathcal Q_\alpha$ and lead to $Q_\alpha$ straddling the sites $i$ and $i+1$. This would result in a non-local change in the bigram configuration which is forbidden by the action of a local unitary as well as break the clipped gauge condition as two bigram endpoints will disappear from the sites $i$ and $i+1$ in general.

\item The clipping must be carried out so that the longer of the two generators is replaced after the transformation, {\it i.e.}, if the length of $\mathcal Q_\alpha$ is less than $\mathcal Q_\beta$, the allowed transformation which restores the clipped gauge is $\mathcal Q_\beta \rightarrow \mathcal Q_\beta \mathcal Q_\alpha$. This prevents a change in the location of the other endpoint of the shorter generator, which would have lead to the violation of the gauge constraint elsewhere.
\end{enumerate}
An example of the clipping process is shown in the step labelled `clip' in Fig.~\ref{fig:stab-clip-example}. Together the action of the Clifford gate followed by the clipping leads to a reshuffling of the bigram endpoints within the two sites the gate acts on; this is shown at the top of Fig.~\ref{fig:stab-clip-example}.

While a completely faithful simulation of an instance of a Clifford circuit requires the knowledge of both, the unitary gate and the stabiliser generators, we are only interested in the information dynamics averaged over the Clifford circuits with gates chosen randomly as discussed earlier (see Fig.~\ref{fig:cliff-circuit}).
We can therefore neglect the precise Pauli operator content of the generators and build a classical stochastic process of the bigram configurations in the clipped gauge at the cost of obtaining only averaged (over Clifford circuits) results. 
Given the conditions that the action of the Clifford operator and the clipping must satisfy, as discussed above, the classical stochastic description essentially has the following two features.
\begin{enumerate}[label=(\roman*)]
    \item A gate acting on sites $i$ and $i+1$ can only reshuffle the four endpoints on the two sites with a bias towards increase bigram lengths.
    \item Shorter bigrams are favoured for growth over longer ones.
\end{enumerate}
Here, (i) is just the result of the 
locality of the gate which translates to the non-local constraint on the information lattice 
while (ii) mandates the presence of real-space non-local correlations and makes the role of length scales explicit. 
Note that these features were also present in the analysis of the infinite-$q$ Haar random circuit model. However, for $q=2$ Clifford dynamics, the number of allowed rules are greater (for instance, moves which decrease the generator lengths are also allowed) and there is inherent stochasticity in the choice of these moves, resulting from the randomness of the gates themselves.

Using the above guiding principles, we propose 
a classical stochastic process describing the time evolution of bigrams (or equivalently, particles on the information lattice) 
which captures 
the universal features of the averaged dynamics of local information under random Clifford evolution. 
Since each gate leads to a reshuffling of the bigram endpoints on the two site it acts on, to define the classical process we need to assign probabilities for the various possibilities of the reshuffling.
This can be specified via the following rules which we lay out for a gate acting on sites $i$ and $i+1$.
\begin{enumerate}[label=(\roman*)]
    \item If site $i$ hosts two $+$ endpoints, then the endpoint belonging to the shorter bigram is chosen over the the one belonging to the longer one with a probabilistic bias $p_1>1/2$. If the corresponding bigrams are of equal length, then either of the endpoints can be picked with equal probability. If there are two $-$ endpoints on site $i$, the bias is reversed. If there is one $+$ and one $-$ endpoint, $+$ is picked over the $-$ endpoint with a bias of $p_2>1/2$.
    
    \item Another endpoint is picked from the site $i+1$ with the biases of rule (i) reversed.
    
    \item If both the selected endpoints are of different type, they are exchanged. If they are of the same type, then the exchange is accepted with probability 1 if it causes the shorter bigram to increase in length or if both the corresponding bigrams are of equal length. Otherwise, the move is only accepted with a probability $p_3<1$.
\end{enumerate}
We simulate this classical model numerically starting from a product state which corresponds to each site hosting both the endpoints of a bigram.
While two of the parameters can be fixed by ensuring the classical model captures some of the universal features accurately, all the parameters cannot be uniquely fixed indicating the phenomenological nature of the model.
However, the asymptotic (in time) scaling of $\zeta_A(t)=2v_E t$ and $\sqrt{\Delta\zeta_A^2(t)}=ct^{1/3}$ with $t$ is robust to the choice of the parameters $(p_1,p_2,p_3)$ as long as $1/2<p_1,p_2\leq 1$  and $p_3<1$ as this is sufficient to incorporate all the microscopic effects that govern the dynamics. Of course the non-universal features such as the velocity $v_E$ and the constant $c$ depend on the parameter values. Fitting these to the exact numerical results from Clifford simulations lets us fix these to $(p_1,p_2,p_3) = (0.8,0.7,0.7)$. We provide additional details in  Appendix~\ref{appendix:pheno}.
The steps described above are applied to nearest neighbour pairs $(i,i+1)$ chosen in a brickwork pattern to mimic the geometry of the Clifford circuit (see Fig.~\ref{fig:cliff-circuit}). Moreover, for a chosen pair $(i,i+1)$ the evolution is applied with a probability of $2/3$, to account for the presence of the $\mathbb I_4$ gate in the two qubit operations in the quantum model, which occurs with a probability of $1/3$. 

\begin{figure}
    \centering
    \includegraphics[width=1\linewidth]{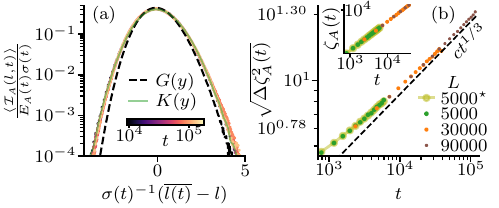}
    \caption{Numerical results from the classical stochastic process and comparison with exact Clifford evolution. (a) Information per scale in the subsytem $A = \mathcal C_{(L-1)/2}^{\lfloor L/2-1\rfloor}$ at different times but rescaled according to Eq.~\ref{eq:cliff-scaling}. The data is for a total system size of $L=90000$ sites. The data for exact Clifford evolution from Fig.~\ref{fig:cliff-res-l}(b) (for $L=5000$ at $t = 4000$, represented by the curve $K(y)$) and the Tracy-Widom distribution are superposed in green and black respectively.
    (b) The main panel shows the evolution of $\sqrt{\Delta\zeta_A^2(t)}$ which shows a perfect match with the data from exact Clifford evolution (indicated by a `$\star$' in the legend) until the timescale allowed by the latter. Extending the results from the classical model to much longer timescales clearly evinces the $t^{1/3}$ scaling. The inset shows the same thing for $\zeta_A(t)$. }
    \label{fig:cliff-pheno}
\end{figure}

The results from the simulation of the classical model are shown in Fig.~\ref{fig:cliff-pheno}.  The classical evolution rules allow access to far greater system sizes than what are accessible with exact Clifford evolution
In Fig.~\ref{fig:cliff-pheno}(a) we show the averaged local information at length scale $l$ in the subsystem $A$ at different $t$ but in it rescaled form according to Eq.~\ref{eq:cliff-scaling}. For comparison, we also superpose the data from the exact Clifford evolution, albeit for a smaller system size, shown in Fig.~\ref{fig:cliff-res-l}(b).
We note that the two sets of data are virtually indistinguishable suggesting that the classical model not only captures the qualitative features of the local information dynamics in Clifford circuits but is also quantitatively accurate with the parameters chosen.
In Fig.~\ref{fig:cliff-pheno}(b) the time dependence of the expected information length scale, $\zeta_A(t)$, and its fluctuation, $\sqrt{\zeta_A^2(t)}$ are shown. 
Reassuringly again, the data are in complete agreement with those from the exact Clifford simulations until the length scales and timescales that the latter allows.
This allows us to reinforce the conclusions drawn in Sec.~\ref{sec:numerics-cliff} about the scalings $\zeta_A(t) = 2v_Et$ and $\sqrt{\Delta\zeta_A^2(t)}=c\,t^{1/3}$.
In fact, the latter is much more compellingly visible in the results from the classical model due to the much larger timescales and system sizes accessible.
We provide additional details of the analysis of this model and comparison with the Clifford evolution in Appendix \ref{appendix:pheno}.

While the quantitative agreement between the results of the exact Clifford evolution and the classical stochastic model is useful for validating the latter, the more significant aspect is that it establishes concretely the key physical mechanisms which control the local information dynamics in Clifford circuits. 
The classical model shows that the inherent randomness in the Clifford gates can lead to various kinds of reorganisations of the stabiliser generators in the clipped gauge, each with its own probability.
The interplay of these probabilities along with the effect of these reorganisations on the local information leads to the ballistic growth of the average information length scale along with a $\sim t^{1/3}$ fluctuation in it. 

In addition, this classical model also explains why the entanglement velocity $v_E$ must be less than the butterfly velocity $v_B$ in Clifford circuits. 
To see this, note that under random Clifford evolution, the stabiliser generators independently exhibit operator growth, evolving into spatially extended Pauli strings. The dynamics can be described as biased random walks of endpoints, with the left (right) endpoint performing a biased random walk to the left (right) with the average velocity of $v_B$~\cite{nahum2018operator}. Due to the locality of the unitary operations the walks of both endpoints are effectively random at late times and hence the length of the generator grows as $\sim 2v_Bt + \eta(t)t^{1/2}$ where $\eta(t)$ denotes white noise. 
However, to track the local information we necessarily need to work in the clipped gauge
and track the time evolution of the clipped generators (see Eq.~\ref{eqn:loc-info-stab-state}).
The clipping of the generators effectively arrests the growth of the stabiliser generators, which is now controlled by a smaller velocity which controls the entanglement growth and also appears in the dynamics of local information. This is just the entanglement velocity $v_E < v_B$. Moreover, the clipping process can induce additional correlations among different stabiliser generators and the left and right endpoints of the same generator. The effective dynamics now is a constrained growth of operators. This means that the subleading fluctuations of the clipped generators are no longer diffusive. The numerical results above suggest that the effective dynamics of the clipped bigrams at long times is $l_\alpha \sim 2v_Et + \eta(t)t^{1/3}$.

Before concluding this section, we note that this construction roughly reduces to the model of Sec.~\ref{sec:haar-q-inf} when all moves which lead to a decrease in any of the bigram lengths are forbidden and bias for shorter lengths to grow is exact. This translates to $p = q = 1$. Additionally, if only the entanglement across a single cut is to be computed, we can ignore the role of bigram lengths entirely and treat endpoints of the same type as identical. The respective densities of left and right endpoints of the clipped generators evidently obey the noisy Burgers equation~\cite{nahum2017quantum}. The entanglement entropy then obeys the KPZ equation. Thus, this construction also correctly reproduces the fluctuations in bipartite entanglement.

\section{Towards a hydrodynamic theory \label{sec:hydro}}
Given that the analyses presented in the previous sections present a compelling picture for the universal properties of the dynamics of local information in terms of their flow on the information lattice, a natural question that arises is does it admit a hydrodynamic description in the continuum limit.
In Sec.~\ref{sec:info-latt-homogeneous}, we had already defined the appropriate rescaling of the coordinates and the local information to define the continuum limit and results at leading order in this limit were presented in Sec.~\ref{sec:leading-order}.
However, the results for Haar-random circuits in Sec.~\ref{sec:haar-q-inf} and Clifford circuits Sec.~\ref{sec:res-cliff} reveal set of universal features beyond leading order.
The same features also make explicit, the challenges concerned with understanding the dependence of the currents $J_\pm$ on the local information profile within a hydrodynamic treatment.
Nevertheless, in this section we make some speculative remarks about how such theory can be arrived at and present the first steps towards that.

\begin{figure}
    \centering
    \begin{tikzpicture}[scale=2.5]
\def\x{-0.1};
\def\y{0.3};

\draw[line width=5, cap=round, opacity=0.4, Orange] (\x+0.4,\y+0.02) -- ++(0.3,0.6);
\draw[line width=5, cap=round, opacity=0.4, Mahogany] (\x+1,\y+0.02) -- ++(-0.3,0.6);

\draw[thick] (\x,\y) -- ++(1.5,0) -- ++(-0.5,1);
\draw[thick] (\x,\y) -- ++(0.5,1);
\node at (\x+0.15,\y+1.1) {(a)};
\node at (\x+3,\y+1.1) {(b)};

\draw[very thick] (\x,\y-0.15) -- ++(1.5,0);
\draw[line width = 5, cap=round, opacity = 0.7, Gray] (\x+0.4,\y-0.15) -- ++(0.6,0);

\draw[thick] (\x+0.4,\y-0.15-0.03) -- ++(0,0.06);
\draw[thick] (\x+1,\y-0.15-0.03) -- ++(0,0.06);

\node at (\x+0.38,\y-0.26) {$\mathtt{x}_-$};
\node at (\x+1.125,\y-0.26) {$\mathtt{x}_+$};


\draw[>= stealth, ->, very thick, Mahogany] (\x+0.7,\y+0.6)to (\x+0.8,\y+0.8);
\draw[>= stealth, ->, very thick, Orange] (\x+0.7,\y+0.6)to (\x+0.6,\y+0.8);

\fill (\x+0.7,\y+0.61) circle (0.55pt);

\node at (\x+1.035,\y+0.35) {\textcolor{Mahogany}{$\mathcal E_+$}};
\node at (\x+0.42,\y+0.35) {\textcolor{Orange}{$\mathcal E_-$}};

\node at (\x+0.855,\y+0.66) {\textcolor{Mahogany}{$J_+$}};
\node at (\x+0.575,\y+0.66) {\textcolor{Orange}{$J_-$}};

\def\x{1.9};
\def\c{1.05/2}
\def\y{0.2};
\draw[thick] (\x,\y+2*\c) node[above]{\small $\chi (u)$} -- ++(0,-2*\c);
\draw[thick] (\x,\y+\c) -- ++(1.1,0) node[right]{\small $u$};
\draw[domain=\x:\x+1, smooth, variable=\z, Mahogany, very thick] plot (\z, {\c-(\z-\x-0.5) + 1.3*(\z-\x-0.5)^3 + \y});
\draw[domain=\x:\x+1, variable=\z, Black, thick, dashed] plot (\z, {\c-(\z-\x-0.5)*(1-2*0.175) + \y});

\draw[thick] (\x+0.5,{\y+\c-0.05})node[below]{$1$} -- ++ (0,0.1);
\draw[thick] (\x+1,{\y+\c-0.05})node[below]{$2$} -- ++ (0,0.1);
\draw[thick] (\x-0.05,{\y+2*\c-0.195})node[left]{$v_B$} -- ++ (0.1,0);
\draw[thick] (\x-0.05,{\y+0.195})node[left]{$-v_B$} -- ++ (0.1,0);


\end{tikzpicture}
    \caption{Information currents in the continuum description. (a) Non-local dependence of $J_\pm(\mathtt x_-,\mathtt x_+)$. In the continuum limit, the currents $J_\pm$ on the information lattice depend on the total local information ${\cal E}_\pm$ contained within the shaded regions, see Eqs.~\ref{eq:J-nonlocal-dep}-\ref{eq:J-nonlocal-dep1}. (b) Schematic description of forms of $\chi(u)$ for generic models (solid curve) and random circuits (dashed line).}
    \label{fig:hydro}
\end{figure}
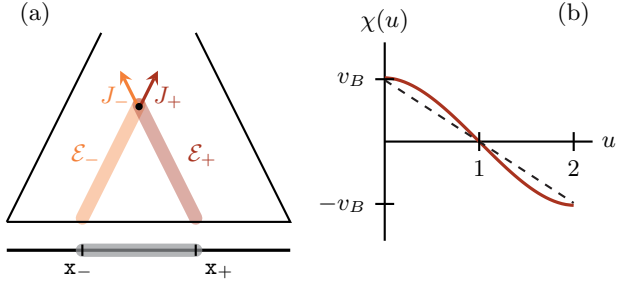

Consider a contiguous spatial interval $\left(\mathtt x_+,\mathtt x_-\right)$ and let us denote its entanglement as $\mathtt{S(x_-,x_+)}$. The information lattice coordinates corresponding to this subsystem are $\mathtt l = \mathtt x_+ - \mathtt x_-$ and $2\mathtt n = \mathtt x_+ +\mathtt x_-$. From the definition of local information we have
\begin{equation}
    \mathtt{I(x_-,x_+)}\equiv \mathtt I(\mathtt n,\mathtt l) = \partial_{\mathtt x_-\mathtt x_+}^2 \mathtt S(\mathtt x_-,\mathtt x_+)\,.
    \label{eq:cont-I-S}
\end{equation}
As a starting point for the dynamics, we consider the coarse-grained theory of entanglement in chaotic systems, developed in Ref.~\cite{jonay2018coarse}, wherein it was proposed that the rate of growth of entanglement of a contiguous spatial interval 
can be written in terms of local gradients of the entanglement at the boundaries. Assuming the entanglement production at different endpoints to be uncorrelated, this gives\footnote{The entanglement production rate at an endpoint is also proportional to equilibrium entropy density, which yields $\partial_t\mathtt S = s_{eq}\Gamma(\partial_x S)$. We restrict ourselves to models without any conservation laws which saturate to infinite temperature so that $s_{eq} =1$.}
\begin{multline}
      \partial_t\mathtt S(\mathtt x_-,\mathtt x_+,t) \approx \Gamma\left(\partial_{\mathtt x_-}\mathtt S(\mathtt x_-,\mathtt x_+,t)\right) \\+ \Gamma\left(\partial_{\mathtt x_+}\mathtt S(\mathtt x_-,\mathtt x_+,t)\right)\,.
      \label{eqn:ent-dynamics-gradient}
\end{multline}
Also from Eq.~\ref{eq:cont-I-S}, we have
\begin{equation}
    \partial_{\mathtt x_+ \mathtt x_-}^2\Gamma(\partial_{\mathtt x_\pm}\mathtt S(\mathtt x_-,\mathtt x_+))\! =\!\partial_{\mathtt x_\pm}\left[\Gamma^\prime(\partial_{\mathtt x_\pm}\mathtt S)\;\mathtt I(\mathtt x_-,\mathtt x_+)\right].
\end{equation}
In addition, integrating Eq.~\ref{eq:cont-I-S} over either $\mathtt{x_+}$ or $\mathtt{x_-}$ with appropriate boundary conditionss yields
\eq{
\begin{split}
    \partial_{\mathtt x_+} \mathtt S(\mathtt x_-,\mathtt x_+) &=1- \int_{0}^{\mathtt x_+-\mathtt x_-} d\mathtt y \; \mathtt I(\mathtt x_-+\mathtt y,\mathtt x_+)\,,\\
    -\partial_{\mathtt x_-} \mathtt S(\mathtt x_-,\mathtt x_+) &=1- \int_{\mathtt x_- - \mathtt x_+}^{0} d\mathtt y \; \mathtt I(\mathtt x_-,\mathtt x_+ + \mathtt y)\,.
\end{split}
}
We define $\chi(u) = -\Gamma^\prime(1-u)$ for $u \in [0,2]$, which is a decreasing function with $\chi(0) = -\chi(2) = v_B$ and $\chi(1) = 0$ (shown schematically in Fig.~\ref{fig:hydro}(b)). Using the above results in the continuity equation for $\mathtt{I}(\mathtt n,\mathtt l)$ in Eq.~\ref{eq:eqn-continuity} and transforming to $(\mathtt n,\mathtt l)$ coordinates, we see that the information currents take the forms
\begin{equation}
      J_\pm(\mathtt n,\mathtt l) \approx \chi\left(\mathcal E_\pm(\mathtt n,\mathtt l)\right) \mathtt I(\mathtt n,\mathtt l)\,,
      \label{eq:J-nonlocal-dep}
\end{equation}
with the positive valued functions
\begin{multline}
    \mathcal E_\pm = 1\mp\partial^2_{\mathtt x_- \mathtt x_+}\mathtt S(\mathtt x_-,\mathtt x_+) \\=  \int_{\mathtt n \pm \mathtt l/2}^{\mathtt n} \mathtt I(\mathtt n^\prime,\mathtt l \pm 2(\mathtt n -\mathtt n^\prime))\;d\mathtt n^\prime\,.
    \label{eq:J-nonlocal-dep1}
\end{multline}

At a formal level, this constitutes a set of hydrodynamic equations governing the flow of local information on the information lattice.
These equations seem to have expected properties. For instance, $J_+(\mathtt n,\mathtt l)$ depends not only on the local information content at $(\mathtt n,\mathtt l)$, but also on the total local information in the diagonal ranging from $(\mathtt n+\mathtt l/2,0)$ to $(\mathtt n,\mathtt l)$, as depicted in Fig.~\ref{fig:hydro}(a). The decreasing nature of $\chi$ then implies that larger local information content in this diagonal region will lead to a smaller $J_+$. This is reminiscent of the role of length-scales in the models discussed above. However, note that equations of this form might not be well behaved in general: starting with a local information profile uniform in $\mathtt n$ and peaked around $\mathtt l \sim \zeta$, information at $\mathtt l < \zeta$ moves faster than at larger $\mathtt l > \zeta$ and the solution evolves to become singular (Dirac-$\delta$-like in $\mathtt{l}$). This is expected since we started with a coarse grained description which on its own leads to Eq.~\ref{eqn:ent-finite-form} and subsequently gives propagating singular solutions on the information lattice, as described in Section \ref{sec:leading-order}. 
\par

The leading order form of $\Gamma(s)$ is known explicitly for random circuit models discussed above~\cite{jonay2018coarse}. For circuits which saturate to maximal entanglement (up to exponentially small corrections in subsystem size) and have an entanglement velocity of $v_E$,
\begin{equation}
    \Gamma(s) = \frac{v_E}{2}(1-s^2)\,.
\end{equation}
Strictly speaking, this form is exact for Haar random circuits in the $q\rightarrow \infty$ limit while being a leading order approximation for generic chaotic circuits. This gives
\begin{equation}
    \chi(u) = v_E(u-1),\;\;u \in\left[0,2\right].
\end{equation}
Using this functional form in the expressions of currents derived above and plugging the currents back in the continuity equation, we obtain an `Euler' equation
\begin{equation}
    \partial_t\mathtt I + 2v_E\bigg( \partial_{\mathtt l} \big[\left(\mathcal E_+ +  \mathcal E_-\right) \mathtt I\big] + \frac{1}{2}\partial_{\mathtt n} \big[\left(\mathcal E_+ -  \mathcal E_-\right) \mathtt I\big]\bigg) = 0\,.
\end{equation}
Since we expect singular solutions, we take $\mathtt I(\mathtt n,\mathtt l,t) = a\delta(\mathtt l-vt)$ in a region far from boundaries and for $vt\ll \mathtt L$ where the constraint mandates that $a=1$. For such a form f $\mathtt{I(n,l)}$, we obtain $\mathcal E_\pm = \Theta(\mathtt l-vt)$. This leads to 
\begin{equation}
    \partial_t\mathtt \delta(\mathtt l - vt) + 4v_E\;\partial_{\mathtt l}\left[\chi(\Theta(\mathtt l - vt)) \delta(\mathtt l - vt)\right] = 0,
\end{equation}
which, on using the symmetric convention $\Theta(x)\delta(x) =\delta(x)/2 $ leads to
\begin{equation}
    -v\delta^\prime(\mathtt l - vt) = -2v_E\delta^\prime(\mathtt l - vt)\,.
\end{equation}
Therefore, singular solutions of this form must propagate with $v = 2v_E$ as expected from the leading order analysis in Sec.~\ref{sec:leading-order}. This is expected since $\Gamma(s)$ of the form given above leads to Eq.~\ref{eqn:ent-finite-form}, and the Euler equation is nothing more than a reformulation of this on information lattice.

We speculate that to regularise the equation, one will need to include higher order corrections to the currents derived above. It was indeed argued in Ref.~\cite{jonay2018coarse} that at the sub-leading order, the function $\Gamma$ picks up corrections from higher derivatives of $\mathtt S$. A second derivative correction will lead to corrections in $J_\pm$ which include spatial gradients of $\mathtt I(\mathtt n,\mathtt l,t)$. Diffusive terms of this sort can prospectively lead to regular solutions. However, the supposed connection between the temporal broadening of the propagating packet of local information and the fluctuations in bipartite entanglement indicates that to correctly capture the sub-leading behaviour on the information lattice, one might need to further include stochastic noise in a manner which is consistent with the constraints. Alternatively, one may try to arrive at a noisy continuum equation starting from the classical processes discussed above. The analysis in this section was provided as a starting point to a more comprehensive continuum theory, which we leave as a direction for further research.

\section{Summary and Outlook \label{sec:discussion}}
In summary, we showed that the information lattice allows a local description of entanglement dynamics in one dimensional chaotic systems even though the equations governing the dynamics of the local information are non-locally constrained. This local description further simplifies into a particle description for random circuit models in certain limits. 
For Haar-random circuits in the limit of $q\rightarrow\infty$, we obtained a classical stochastic process composed of nearest neighbour particle hoppings which were correlated in a non-local fashion. A similar, albeit phenomenological, description was also motivated for random Clifford circuits. 
A key to arriving at these descriptions was the emergent interpretation of local information as extended one dimensional classical objects called `bigrams' which capture both, the length scale and the spatial location of a unit of local information. The bigram description serves an important role in both models discussed in this work by highlighting the microscopic nature of the rearrangement of local information as exchange of bigram endpoints with with length dependent biases. 

Moreover, in the Clifford circuit model, the bigram description reveals how information scrambling differs from operator growth. Boundaries of the region of support of an operator generally perform uncorrelated random walks, leading to linear growth with a well defined velocity and diffusive fluctuations. Information scrambling in Clifford circuits on the other hand, is described by the dynamics of a set of $L$ Pauli string operators which generate stabiliser subgroup corresponding to the stabiliser state.
The dynamics of these strings must satisfy constraints respected by local information. These constraints arrest the  growth of these strings and induce additional correlations leading to a smaller velocity and subdiffusive fluctuation of the string endpoints. 

Our results indicate that for the models discussed above, a culmination of the non-local constraints on the local information profile, and a bias for the local information at smaller scales to scramble faster lead to a ballistic growth of the average information length scale along with a KPZ-like $t^{1/3}$ scaling of the width of the distribution of local information length scale in presence of noise which comes from the randomness of the unitary evolution. Additionally, the classical descriptions we obtained also highlight the inherent non-locality of the microscopic mechanisms responsible for the effects in information scrambling beyond the leading order. This non-locality may cease to be important in certain exceptional cases like for integrable systems where entanglement is carried by local `quasiparticles' \cite{calabrese2005evolution}, leading to a hydrodynamic description. However, we argued that that for non-integrable systems a hydrodynamic description which goes beyond the leading order will necessarily need to include non-local terms and possibly noise.

Some aspects of the scaling behaviour obtained in this work can also be expected for general chaotic models in the absence of conserved charges without the need for explicit averaging. Local information can also be used to understand if and when the nature of information scrambling might differ at a fine grained level in models which yield the same universal behaviour in the dynamics of bipartite entanglement. A primary challenge in studying the local information in general systems stems from a lack of analytical tractability of the von Neumann entanglement entropy. The current understanding of dynamics of bipartite entanglement in random circuits at finite $q$ employs R\'enyi entropies $S_{n\neq 1}$ which can be computed by a replica trick \cite{zhou2019emergent,zhou2020entanglement}. Since R\'enyi entropies are not strongly sub-additive, positivity of local information defined in terms of them is not guaranteed. However, positivity may be expected on average in certain models and one can then hope to use `local R\'enyi information' as a proxy to true local information.

In this work, we only considered models which were statistically homogeneous in space.
A next natural question of immediate interest will be to understand the fate of the local information dynamics in the presence of spatial inhomogeneities~\cite{nahum2018dynamics}.
Specifically, quenched randomness or disorder can lead to a non-trivial profile of the local information and their flows across different regions of the system.
However, within our framework of bigrams, this physics can be incorporated directly by making the dynamical rules for the bigrams also spatially inhomogeneous in accordance with the disorder configuration.

In fact, the formulation of local information via the information lattice is completely general and not restricted to chaotic systems. 
It will therefore be interesting to study the anatomy of the information lattice for strongly disordered systems across the many-body localisation transition~\cite{nandkishore2015many,abanin2019colloquium,sierant2024manybody}. 
In particular, one can ask how do the multiscale entanglement clusters at the transition as well as the rare resonances in the many-body localised phase manifest themselves on the information lattice~\cite{herviou2019multiscale,garratt2021resonances,morningstar2021avalanches,garratt2022resonant,crowley2022constructive}.

In addition to unitary dynamics, local information can be leveraged to understand quantum information dynamics in a `hybrid' setting where the unitary dynamics is interspersed with local projective measurements. Hybrid circuits are known to host non-equilibrium phase transitions in the dynamics of quantum information, as probed by measurement-induced entanglement phase transitions~\cite{li2018quantum,skinner2019measurement,li2019measurement,Gullans2020Purification,nahum2021measurement}. The information lattice can be used to understand the effect of local measurements on information encoded at different length scales. Furthermore, given its utility in characterisation of quantum states and phases~\cite{artiaco2025universal}, it might serve as a diagnostic tool for aspects of measurement-induced criticality, beyond what is revealed by the scaling behaviour of bipartite entanglement. Specifically, stabiliser length distributions in hybrid Clifford circuits were shown to have a peak at $L/2$ in the volume-law phase, to be exponentially decaying in the area-law phase, and most importantly, to exhibit a power-law tail at the critical point~\cite{li2019measurement}.
In Sec.~\ref{sec:stab-pheno} we showed that stabiliser lengths correspond to the length scale of local information. This means that the information per scale $\mathcal I(l)$ is just the stabilizer length distribution. The information per scale can thus be treated as a generalization of the stabiliser length distribution for non-stabiliser states. Additionally, a power law decay in this quantity hints at scale invariance on the information lattice, similar to what is exhibited by critical quantum states\cite{artiaco2025universal}. We shall address these ideas in our upcoming work~\cite{thakur-il-mipt}.

\begin{acknowledgments}
We would like to thank S. Biswas for useful discussions on path-finding algorithms in the context of minimal cuts and A. Kundu for useful discussions about hydrodynamic theories. This work was supported by the Department of Atomic Energy, Government of India, under project nos. RTI4019 and RTI4013. S.R. acknowledges support from SERB-DST (India) under Grant No. SRG/2023/000858, from ANRF (India) under Grant No. ANRF/ARG/2025/004045/PS, and from a Max Planck Partner Group grant between ICTS-TIFR, Bengaluru and MPIPKS, Dresden.
\end{acknowledgments}


\appendix
\section{Details of the computation of local information for $q\rightarrow \infty$ Haar-random circuits}
\label{appendix:loc-info-q-inf}

In this appendix, we provide the details of the computation of local information and their dynamical rules for Haar-random circuits in the $q\to\infty$ limit. We first discuss the local information can be described in terms of the bigrams and then elaborate on the stochastic process for updating the bigram endpoints.

\subsection{Local information and bigrams}
We start by writing Eq.~\ref{eqn:von-neumann-cut} in the form
\begin{equation}
    E_n^l = \left[\left(p - \gamma_{p}\right) -\left( q + \gamma_{q}\right)\right]_+\,,\label{eq:Enl-pq}
\end{equation}
where $p = n+l/2+1$ and $q = n-l/2$. Using the identity
\begin{equation}
    \left[x + y\right]_+ = \sum_{\alpha\in \mathbb Z}^\infty \Theta(x - \alpha)\Theta(y+\alpha - 1)\,,
\end{equation}
back in Eq.~\ref{eq:Enl-pq}, we obtain
\begin{equation}
    E_n^l = \sum_{\alpha=0}^{L-1} \Theta\left(p - \gamma_{p} - \alpha - 1\right)\Theta\left(\alpha - q - \gamma_{q}\right)\,.\label{eq:Enl-step-func}
\end{equation}
Because of the constraints on $\gamma_i$, we have $p-\gamma_p-1 \leq L-1$ and $q+\gamma_q\geq 0$ which imply that the only non-zero terms in the sum are those which have $0\leq\alpha \leq L-1$. This explains the limits of the sum. 
Now consider the argument of the first step function in Eq.~\ref{eq:Enl-step-func},
\eq{
\begin{split}
p - \gamma_p - \alpha  - 1\geq 0&\implies \alpha< p - \gamma_p\\
&\implies \sum_{i=0}^{L-1}\Theta(\alpha - (i-\gamma_i)) \leq p\,,
\end{split}
\label{eqn-appendix-1}
}
where the last inequality holds because $i-\gamma_i$ is non-decreasing, so there can be a maximum of $p$ contributions to the sum. Similarly,
\begin{equation}
    p-\gamma_p - \alpha - 1 < 0 \implies \sum_{i=0}^{L-1}\Theta(\alpha - (i-\gamma_i))> p\,,
    \label{eqn-appendix-2}
\end{equation}
because there are at least $p+1$ terms that contribute to the sum. 
Using the expression for $e^+_\alpha$ from Eq.~\ref{eq:ealpha-pm} in Eq.~\ref{eqn-appendix-1} and Eq.~\ref{eqn-appendix-2}, we see that
\eq{
\begin{split}
    p - \gamma_p-\alpha-1 \geq 0 &\implies p-1 -e^+_\alpha \geq 0 \\
    p-\gamma_p - \alpha -1 <0 &\implies p-1-e_\alpha^+ < 0
\end{split}\,.
}
This leads to the following equality
\begin{equation}
\Theta\left(p-\gamma_p - \alpha - 1 \right) = \Theta\left(p - 1 - e^+_\alpha\right).
\end{equation}
For the second step function in Eq.~\ref{eq:Enl-step-func}, we proceed in the same manner to obtain 
\eq{
\begin{split}
    \alpha - q -\gamma_q \geq 0  &\implies \sum_{i=0}^{L-1}\Theta (\alpha - i-\gamma_i) \geq q +1\\
    \alpha-q-\gamma_q < 0 &\implies \sum_{i=0}^{L-1}\Theta(\alpha - i-\gamma_i) < q + 1
\end{split}
}
which are equivalent to,
\eq{
\begin{split}
    \alpha - q -\gamma_q \geq 0  &\implies e^-_\alpha - q \geq 0\\
    \alpha-q-\gamma_q < 0 &\implies e^-_\alpha - q < 0
\end{split}\,.
}
This leads to,
\begin{equation}
    \Theta\left(\alpha - q-\gamma_q \right) = \Theta\left(e^-_\alpha - q\right)\,.
\end{equation}
Finally, after substituting back the expressions for $p$ and $q$, we obtain
\begin{equation}
    E_n^l = \sum_{\alpha = 0}^{L-1} \Theta(n+l/2 - e^+_\alpha)\Theta(e^-_\alpha - n+l/2).
\end{equation}
The total von Neumann information in $\mathcal C_n^l$ is therefore the number of bigrams such that both if their endpoints lie inside $\mathcal C_n^l$  (inclusive of the boundaries). It then follows from the definition in Eq.~\ref{eqn:local-info} that $I_n^l$ is equal to the number of bigrams which have their left endpoint at $n-l/2$ and right endpoint at $n+l/2$, which is just the expression in Eq.~\ref{eqn:loc-info-q-inf}.

\subsection{Dynamical rules}

From the diagrammatic interpretation for the connection between $e^\pm_\alpha$ and $\gamma_i$, it follows that $\rho^\pm_i$ (the densities of the left and right end points, defined in Eq.~\ref{eq:rho-def}) can be written in terms of $\gamma_i$ as 
\begin{gather}
    \rho^-_i = \sum_{\alpha=0}^{L-1}\delta(e^-_\alpha,i) = \gamma_{i+1}- \gamma_i + 1,\\
    \rho^+_i = \sum_{\alpha=0}^{L-1}\delta(e^+_\alpha,i) = \gamma_{i}- \gamma_{i+1} + 1.
\end{gather}
Consider the action of a unitary gate on bond $j$. This gives $\gamma_i \rightarrow \gamma_i + \Delta_i$, where $\Delta_i = 0$ for $i\neq j$ and $\Delta_j = \Delta$. Using Eq.~\ref{eqn:cut-evolution}
\begin{equation}
    \Delta = \text{min}\left\{\rho_{j-1}^+,\rho^-_j\right\}\,.
    \label{eqn:appendix-delta}
\end{equation}
Using the expressions for $e^\pm_\alpha$, we see that under this change
\begin{multline}
    e^\pm_\alpha \rightarrow \sum_i \Theta(\alpha - i \pm \gamma_i \pm \Delta_i) - 1 \\= \sum_{i\neq j} \Theta(\alpha - i \pm \gamma_i) - 1+ \Theta(\alpha-j\pm\gamma_j \pm \Delta)\,.
\end{multline}
Employing the identities
\eq{
\begin{split}
    \Theta(x + n) &= \Theta(x) + \sum_{k=1}^n\delta_{x,-k}\\
    \Theta(x - n) &= \Theta(x) - \sum_{k=1}^n\delta_{x,k-1}
\end{split}\,,
}
which are true for $x \in \mathbb Z$ and $n\in \mathbb Z^+$, we obtain
\eq{
\begin{split}
    \epsilon_\alpha^+ \rightarrow \epsilon_\alpha^+ + \sum_{k=1}^\Delta \delta\left(j -  \gamma_j-k,\;\alpha\right)\,,\\
    \epsilon_\alpha^- \rightarrow \epsilon_\alpha^- - \sum_{k=1}^\Delta \delta\left(j +  \gamma_j+k-1,\;\alpha\right)\,.
\end{split}
}
We now show that these formal expressions for the change in functions $e^\pm_\alpha$ due to the operations $U_{j-1,j}$ are equivalent to to the dynamical rules (i)-(iv) of Sec.~\ref{sec:haar-q-inf-classical}. We treat all four cases separately:
\begin{enumerate}[label=(\roman*)]
    \item If $\rho^+_{j-1}= 0$ or $\rho^-_j =0$, then eqn. \ref{eqn:appendix-delta} gives $\Delta = 0$ and hence no change occurs in the bigram configuration under the action of a random unitary on sites $j-1,j$.
    \item If $\rho^+_{j-1}= 2$ and $\rho^-_j =1$, then $\Delta = 1$. So, the position of the right endpoint $e^+_{\alpha_1}$ with $\alpha_1 = j - \gamma_j- 1$ is exchanged with the the position of left endpoint $e^-_{\alpha_2}$ with $\alpha_2 = j+\gamma_i$. Using the expression for $e^+_\alpha$, given that $\gamma_{j-1} = \gamma_j + 1$ (which follows from $\rho^+_{j-1} = 2$), we find that the two endpoints with $e^+_\alpha = j-1$ have $\alpha = \alpha_1$ and $\alpha = \alpha_1-1$. So, the right endpoint which is exchanged is one of the two endpoints located at $j-1$ with the larger index. Analogously, $e^-_{\alpha_2} = j$ which is the only endpoint on site $j$. The rules for the configuration $\rho^+_{j-1}= 1$ and $\rho^-_j =2$ follow in the same manner.
    \item If $\rho^+_{j-1}= 1$ and $\rho^-_j =1$, then $\Delta = 1$. This again leads to an exchange in the positions of the right endpoint $e^+_{\alpha_1} = j-1$ with $\alpha_1 = j - \gamma_j- 1$ and the left endpoint $e^-_{\alpha_2} = j$ with $\alpha_2 = j+\gamma_i$.
    \item Finally, if $\rho^+_{j-1}= 2$ and $\rho^-_j =2-$, then $\Delta = 2$, meaning that the positions of right endpoints $e^+_{\alpha_1},e^+_{\alpha_1-1}$ with $\alpha_1 = j - \gamma_j -1$ are exchanged with those of left endpoints endpoints $e^-_{\alpha_2+1},e^-_{\alpha_2}$ with $\alpha_2 = j + \gamma_j$.
\end{enumerate}
These four cases above are in one-to-one correspondence with the rules (i)-(iv) of Sec.~\ref{sec:haar-q-inf-classical} and the panels (i)-(iv) in Fig.~\ref{fig:q-inf-rules}.

\section{Details for numerical computation of local information in Clifford circuits} \label{appendix:cliff-num}
In this appendix, we present the details of the computation of the local information in Clifford circuits within the stabiliser formalism.
For an $L$-qubit stabiliser state, identified by the stabiliser group $\mathcal S$, consider a minimal generating set $\{\mathcal Q_1,\mathcal Q_2,...,\mathcal Q_{L-1}\}$ of $\mathcal S$. We can write
\begin{equation}
    \mathcal Q_\alpha = (-1)^{\theta_\alpha}\bigotimes_{i=0}^{L-1} X^{x_{i,\alpha}}Z^{z_{i,\alpha}}\,,
\end{equation}
where $\theta_\alpha, x_{i,\alpha},z_{i,\alpha} \in \{0,1\}$. For computing bipartite entanglement, the overall phases can be ignored, leaving us with a representation of $\mathcal S$ in terms of an $L$-dimensional vector space over the field $GF(2)$ with $v_\alpha =\left(x_{i,\alpha},z_{i,\alpha}\right)$ as the basis vectors. For numerical simulations, $\mathcal S$, is represented by these vectors stored in a $L\times 2L$ dimensional matrix $M$ of binary numbers, where each row $(x_{0,\alpha},\; z_{0,\alpha},\;..., x_{L-1,\alpha},z_{L-1,\alpha})$ corresponds to the generator $\mathcal Q_\alpha$. Single-site Clifford operations on site $i$ then reduce to symplectic transformations\footnote{Symplectic transformation preserve the symplectic form $\Lambda = \begin{pmatrix}0&1\\1&0\end{pmatrix}$ such that, $\mathcal A\Lambda \mathcal A^T = \Lambda$. The requirement of $\mathcal A$ being symplectic stems from the fact that the the commutation relations amongst the generators must be preserved under Clifford evolution. In general, transformations corresponding to $n-$qubit Clifford operations preserve $\bigoplus_n \Lambda$.}.
\begin{equation}
    M_{\alpha,2i:2i+1} \rightarrow M_{\alpha,2i:2i+1}\mathcal A \;\;\;\;\forall\alpha,
\end{equation}
where $M_{\alpha,2i:2i+1} = \left(x_{\alpha,i}\;, z_{\alpha,i}\right)$ is a row vector.
The Hadamard and the Phase gates are represented by the symplectic transformations $\mathcal A_H$ and $\mathcal A_P$ respectively, where,
\begin{equation}
    \mathcal A_H = \begin{pmatrix}  0&1\\1&0
    \end{pmatrix}\;\;,\; 
    \mathcal A_P = \begin{pmatrix}
        1&1\\0&1
    \end{pmatrix}.
\end{equation}
Likewise, two site Clifford unitaries acting on sites $i,i+1$ are represented by $4\times 4$ symplectic matrices $\mathcal B$ which act on row vectors $M_{\alpha,2i:2i+3} = \left(x_{\alpha,i}\;,z_{\alpha,i}\;,x_{\alpha,i+1}\;,z_{\alpha,i+1}\right)$. In particular, the $\text{CNOT}_{L,R}$ gates are represented by
\begin{equation*}
    \mathcal B_{\text{CNOT}_L} = \begin{pmatrix}1&0&1&0\\0&1&0&0\\0&0&1&0\\0&1&0&1\end{pmatrix}\,,~~ \mathcal B_{\text{CNOT}_R} = \begin{pmatrix}
        1&0&0&0\\
        0&1&0&1\\
        1&0&1&0\\
        0&0&0&1
    \end{pmatrix}.
\end{equation*}
This achieves the intended Clifford evolution of $M$.

Bipartite entanglement across any bipartition in a stabiliser state can be determined directly in terms of the stabiliser group. For a region $A$ containing $L_A$ sites, we define the subgroup $\mathcal S_A \subset \mathcal S$ which contains all elements of $\mathcal S$ supported only in $A$. Then, a standard result proved in Ref.~\cite{fattal2004entanglement} relates the generator rank of $\mathcal S_A$ ({\it i.e.}, the minimal number of generators required to generate $\mathcal S_A$) to the von Neumann entropy of $A$. We state the result in terms of the von Neumann information, which is
\begin{equation}
    E_A = \vert \mathcal S_A\vert\,,
    \label{eqn:ent-cliff-formula}
\end{equation}
where $\vert \mathcal S_A\vert$ denotes the generator rank.
\par
For numerical computations, we utilise a related formula derived in Ref.~\cite{nahum2017quantum}, which gives
\begin{equation}
    E_A = 2L_A - {\rm rank}_{GF(2)}[M_A]\,,
    \label{eqn:ent-cliff-rank-formula}
\end{equation}
where ${\rm rank}_{GF(2)}[M_A]$ denotes the $GF(2)$ rank of a $L\times 2L_A$ matrix $M_A$ composed of the columns corresponding to the sites in $A$. From the set $\{E_A\}$ for all contiguous subsystems, the local information profiel can be readily computed. 

\section{Clipped gauge and entanglement}
\label{appendix:clipped-gauge}

The clipped gauge, which we defined in Section \ref{sec:stab-pheno}, was originally proposed in Ref.~\cite{nahum2017quantum}, offers valuable insights into the entanglement structure of one-dimensional stabiliser states. Here, we only provide a brief review of some key aspects leading up to Eq.~\ref{eqn:von-Neumann-info-stab-state} for completeness and to set up notation. We refer the reader to Ref.~\cite{li2019measurement} and the appendices therein for a detailed discussion and proofs. 

\paragraph*{The clipping procedure:} Starting with any generating set for $\mathcal S$, represented in the form of a $GF(2)$ matrix $M$, we perform Gaussian elimination on the matrix $M$ to reduce it to row echelon form. This ensures that every site has at most two left endpoints, and if a site has two right endpoints, their Pauli content must be different. Once this condition is satisfied, the sites with exactly two left endpoints cannot contain any right endpoints. This follows because the presence of a right endpoint will necessarily require that the corresponding generator anticommutes with at least one of the generators corresponding to the left endpoints, which we know have different Pauli content. This is not allowed since all generators of $\mathcal S$ are required to commute with each other. Commutation also ensures that if there is a single left endpoint on a site, presence of more than one right endpoint requires their Pauli content to be the same. This means that one of the endpoints can be eliminated by the other. In doing so, the right endpoint of the longer generator must be eliminated by the endpoint of a shorter generator. This ensures that the positions of the left endpoints do not change. Now, if a site contains no left endpoints and more than two right endpoints, it is always possible to eliminate one of them without a change in the left endpoints. This is possible even if all the three endpoints have different Pauli content, since multiplication of all three Pauli operators in any order always produces $\mathbb I$ up to an overall phase. With this, if elimination is now performed on the right endpoints, starting from the rightmost site and moving to the left in order to ensure that there are no more than two right endpoints on a site, the procedure will terminate with every site containing exactly two endpoints and the set of clipped generators satisfying the following properties:
\begin{enumerate}[label=(\roman*)]
    \item If both endpoints on a site are of the same type, the Pauli content of the endpoints is necessarily different.
    \item If both endpoints are of different type, then their Pauli content is same. This is required because the generators must mutually commute.
\end{enumerate}

\paragraph*{Bipartite entanglement of contiguous regions:} Consider a contiguous region $A$. Recall that the stabiliser subgroup $\mathcal S_A$ contains all elements of $\mathcal S$ which are supported within $A$. Since the generators $\left\{\mathcal Q_\alpha\right\}$ mutually commute and $\left(\mathcal Q_\alpha\right)^2 = \mathbb I$, any element $\mathcal Q \in\mathcal S_A$ can be written as
\begin{equation}
    \mathcal Q  = \prod_{\alpha=0}^{L-1} \left(\mathcal Q_\alpha\right)^{{p_\alpha}}\,,
    \label{eqn:subgroup-elem}
\end{equation}
where $p_\alpha \in \left\{0,1\right\}$. If the generators are in the clipped gauge, then it follows from property (i)\footnote{Since endpoints of the same kind present on the same site necessarily have different Pauli content, they cannot eliminate each other. Therefore, the region of support of of $\mathcal Q$ in Equation \ref{eqn:subgroup-elem} is determined by the leftmost left endpoint and the rightmost right endpoint of the generators appearing in the product in the right hand side of Eq.~\ref{eqn:subgroup-elem}. So, if there are one or more generators in the product which have support outside $A$, then $Q$ necessarily has support outside $A$ meaning $Q \notin S_A$.} that the only generators which can appear in this product are those which are supported only inside $A$. It follows that all elements in $S_A$ can be generated by products of generators of $\mathcal S$ which are supported only inside $A$, which gives 
\begin{equation}
    \vert \mathcal S_A\vert = \text{card}\left\{\mathcal Q_\alpha: e^+_\alpha \text{ and } e^-_\alpha\text{ are inside } A \right\},
\end{equation}
where $\text{card}\{.\}$ represents the cardinality (number of elements) of a set. Using this in Eq.~\ref{eqn:ent-cliff-formula} we find that the von Neumann information of $A$ is just the number of clipped bigrams contained entirely inside $A$ (inclusive of the boundaries) which gives Eq.~\ref{eqn:von-Neumann-info-stab-state}.

\paragraph*{A stabiliser bigram configuration is unique:} The clipping procedure applied to a different ordering of generators can produce a different set of clipped generators. Moreover, there can be gauge transformations which preserve the clipped gauge. For instance, if a generator $\mathcal Q_\alpha$ is replaced by $\mathcal Q_\alpha \mathcal Q_\beta$ where both endpoints of $\mathcal Q_\beta$ are inside the region of support of $\mathcal Q_\alpha$, the clipped gauge constraint is maintained. However, once the Pauli content of the generators is neglected and only the information about endpoints is retained, we obtain the set $\mathcal B(S)$ which is unique for a given state (up to a re-ordering of indices). This is true because the local information is independent of the choice of gauge, and given the direct relation between the bigrams and local information, non-uniqueness of the bigrams would lead to a contradiction.

\section{Additional analysis of the phenomenological model}
\label{appendix:pheno}

\begin{figure}[!t]
    \centering
    \includegraphics[width=\linewidth]{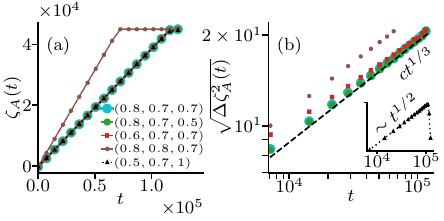}
    \caption{Results for $\zeta_A(t)$ and $\sqrt{\Delta\zeta^2_A(t)}$ obtained for alternate choices of parameters $(p_1,p_2,p_3)$ for the phenomenological model with $L=90000$. $\sqrt{\Delta \zeta^2_A(t)}$ takes significantly larger values for $(p_1,p_2,p_3) = (0.5,0.7,1)$ than for the other choices, so it plotted separately in the inset of panel (b). }
    \label{fig:pheno-2}
\end{figure}

In this appendix, we elaborate on the parameter-dependence, or lack thereof, of the universal features of the
phenomenological model proposed in Sec.~\ref{sec:stab-pheno} for local information dynamics of Clifford circuits.
Note that the phenomenology necessarily requires $p_2>0.5$ to obtain a positive growth rate of bigram lengths and $p_1>0.5$
to make sure that the shorter bigrams are more likely to grow than longer ones. This still leaves us with a sufficient freedom in the choice of these parameters. We provide numerical results for three of such choices in Fig.~\ref{fig:pheno-2}. The behaviour of $\zeta_A(t)$ indicates that the entanglement velocity is largely insensitive to $p_{1,3}$, but does depend on $p_2$. Validity of $t^{1/3}$ scaling at long times as seen in the results for $\Delta\zeta_A^2(t)$ for all three choices show that this behaviour is robust. The choice of parameters sets the time scale over which this universal scaling sets in. Additionally, we show that if we violate the requirement that information at shorter scales must be favoured for scrambling by setting $p_1=0.5$ and $p_3= 1$, we obtain a $t^{1/2}$ growth in $\sqrt{\Delta\zeta_A^2(t)}$ at long times. This highlights that this mechanism plays a crucial role in furnishing the expected universal behaviour.

We conclude that as long as the microscopic mechanisms which were deemed necessary are not violated, we obtain $\zeta_A(t) \sim 2v_Et$ and $\sqrt{\Delta\zeta_A^2(t)} \sim ct^{1/3}$ where the scaling exponents with $t$
are universal. 
Of course, $v_E$ and $c$ are non-universal constants which depend on the parameter choice. This also holds true for the dynamics generated by true quantum evolution-- both the entanglement velocity and the scale of fluctuation of the expected length scale can depend on the choice of circuit geometry and the gates. For example, within Clifford dynamics, one obtains a different entanglement velocity of $v_E = 3/10$ if the gates are drawn uniformly from the full Clifford group. The freedom in the choice of parameters can then be exploited to match this non-universal input from true quantum evolution which lead us to the choice $(p_1,p_2,p_3) = (0.8,0.7,0.7)$ in Sec.~\ref{sec:res-cliff}.

\bibliography{refs}

\begin{thebibliography}{69}%
\makeatletter
\providecommand \@ifxundefined [1]{%
 \@ifx{#1\undefined}
}%
\providecommand \@ifnum [1]{%
 \ifnum #1\expandafter \@firstoftwo
 \else \expandafter \@secondoftwo
 \fi
}%
\providecommand \@ifx [1]{%
 \ifx #1\expandafter \@firstoftwo
 \else \expandafter \@secondoftwo
 \fi
}%
\providecommand \natexlab [1]{#1}%
\providecommand \enquote  [1]{``#1''}%
\providecommand \bibnamefont  [1]{#1}%
\providecommand \bibfnamefont [1]{#1}%
\providecommand \citenamefont [1]{#1}%
\providecommand \href@noop [0]{\@secondoftwo}%
\providecommand \href [0]{\begingroup \@sanitize@url \@href}%
\providecommand \@href[1]{\@@startlink{#1}\@@href}%
\providecommand \@@href[1]{\endgroup#1\@@endlink}%
\providecommand \@sanitize@url [0]{\catcode `\\12\catcode `\$12\catcode
  `\&12\catcode `\#12\catcode `\^12\catcode `\_12\catcode `\%12\relax}%
\providecommand \@@startlink[1]{}%
\providecommand \@@endlink[0]{}%
\providecommand \url  [0]{\begingroup\@sanitize@url \@url }%
\providecommand \@url [1]{\endgroup\@href {#1}{\urlprefix }}%
\providecommand \urlprefix  [0]{URL }%
\providecommand \Eprint [0]{\href }%
\providecommand \doibase [0]{https://doi.org/}%
\providecommand \selectlanguage [0]{\@gobble}%
\providecommand \bibinfo  [0]{\@secondoftwo}%
\providecommand \bibfield  [0]{\@secondoftwo}%
\providecommand \translation [1]{[#1]}%
\providecommand \BibitemOpen [0]{}%
\providecommand \bibitemStop [0]{}%
\providecommand \bibitemNoStop [0]{.\EOS\space}%
\providecommand \EOS [0]{\spacefactor3000\relax}%
\providecommand \BibitemShut  [1]{\csname bibitem#1\endcsname}%
\let\auto@bib@innerbib\@empty
\bibitem [{\citenamefont {Deutsch}(1991)}]{deutsch1991quantum}%
  \BibitemOpen
  \bibfield  {author} {\bibinfo {author} {\bibfnamefont {J.~M.}\ \bibnamefont
  {Deutsch}},\ }\bibfield  {title} {\bibinfo {title} {Quantum statistical
  mechanics in a closed system},\ }\href
  {https://doi.org/10.1103/PhysRevA.43.2046} {\bibfield  {journal} {\bibinfo
  {journal} {Phys. Rev. A}\ }\textbf {\bibinfo {volume} {43}},\ \bibinfo
  {pages} {2046} (\bibinfo {year} {1991})}\BibitemShut {NoStop}%
\bibitem [{\citenamefont {Srednicki}(1994)}]{srednicki1994chaos}%
  \BibitemOpen
  \bibfield  {author} {\bibinfo {author} {\bibfnamefont {M.}~\bibnamefont
  {Srednicki}},\ }\bibfield  {title} {\bibinfo {title} {Chaos and quantum
  thermalization},\ }\href {https://doi.org/10.1103/PhysRevE.50.888} {\bibfield
   {journal} {\bibinfo  {journal} {Phys. Rev. E}\ }\textbf {\bibinfo {volume}
  {50}},\ \bibinfo {pages} {888} (\bibinfo {year} {1994})}\BibitemShut
  {NoStop}%
\bibitem [{\citenamefont {Rigol}\ \emph {et~al.}(2008)\citenamefont {Rigol},
  \citenamefont {Dunjko},\ and\ \citenamefont
  {Olshanii}}]{rigol2008thermalisation}%
  \BibitemOpen
  \bibfield  {author} {\bibinfo {author} {\bibfnamefont {M.}~\bibnamefont
  {Rigol}}, \bibinfo {author} {\bibfnamefont {V.}~\bibnamefont {Dunjko}},\ and\
  \bibinfo {author} {\bibfnamefont {M.}~\bibnamefont {Olshanii}},\ }\bibfield
  {title} {\bibinfo {title} {Thermalization and its mechanism for generic
  isolated quantum systems},\ }\href {https://doi.org/10.1038/nature06838}
  {\bibfield  {journal} {\bibinfo  {journal} {Nature}\ }\textbf {\bibinfo
  {volume} {452}},\ \bibinfo {pages} {854–858} (\bibinfo {year}
  {2008})}\BibitemShut {NoStop}%
\bibitem [{\citenamefont {D’Alessio}\ \emph {et~al.}(2016)\citenamefont
  {D’Alessio}, \citenamefont {Kafri}, \citenamefont {Polkovnikov},\ and\
  \citenamefont {Rigol}}]{dalessio2016from}%
  \BibitemOpen
  \bibfield  {author} {\bibinfo {author} {\bibfnamefont {L.}~\bibnamefont
  {D’Alessio}}, \bibinfo {author} {\bibfnamefont {Y.}~\bibnamefont {Kafri}},
  \bibinfo {author} {\bibfnamefont {A.}~\bibnamefont {Polkovnikov}},\ and\
  \bibinfo {author} {\bibfnamefont {M.}~\bibnamefont {Rigol}},\ }\bibfield
  {title} {\bibinfo {title} {From quantum chaos and eigenstate thermalization
  to statistical mechanics and thermodynamics},\ }\href
  {https://doi.org/10.1080/00018732.2016.1198134} {\bibfield  {journal}
  {\bibinfo  {journal} {Advances in Physics}\ }\textbf {\bibinfo {volume}
  {65}},\ \bibinfo {pages} {239–362} (\bibinfo {year} {2016})}\BibitemShut
  {NoStop}%
\bibitem [{\citenamefont {Deutsch}(2018)}]{deutsch2018eigenstate}%
  \BibitemOpen
  \bibfield  {author} {\bibinfo {author} {\bibfnamefont {J.~M.}\ \bibnamefont
  {Deutsch}},\ }\bibfield  {title} {\bibinfo {title} {Eigenstate thermalization
  hypothesis},\ }\href
  {http://iopscience.iop.org/article/10.1088/1361-6633/aac9f1/meta} {\bibfield
  {journal} {\bibinfo  {journal} {Rep. Prog. Phys.}\ }\textbf {\bibinfo
  {volume} {81}},\ \bibinfo {pages} {082001} (\bibinfo {year}
  {2018})}\BibitemShut {NoStop}%
\bibitem [{\citenamefont {Jaynes}(1957{\natexlab{a}})}]{jaynes1957information}%
  \BibitemOpen
  \bibfield  {author} {\bibinfo {author} {\bibfnamefont {E.~T.}\ \bibnamefont
  {Jaynes}},\ }\bibfield  {title} {\bibinfo {title} {{Information Theory and
  Statistical Mechanics}},\ }\href {https://doi.org/10.1103/PhysRev.106.620}
  {\bibfield  {journal} {\bibinfo  {journal} {Phys. Rev.}\ }\textbf {\bibinfo
  {volume} {106}},\ \bibinfo {pages} {620} (\bibinfo {year}
  {1957}{\natexlab{a}})}\BibitemShut {NoStop}%
\bibitem [{\citenamefont
  {Jaynes}(1957{\natexlab{b}})}]{jaynes1957information2}%
  \BibitemOpen
  \bibfield  {author} {\bibinfo {author} {\bibfnamefont {E.~T.}\ \bibnamefont
  {Jaynes}},\ }\bibfield  {title} {\bibinfo {title} {{Information Theory and
  Statistical Mechanics. II}},\ }\href
  {https://doi.org/10.1103/PhysRev.108.171} {\bibfield  {journal} {\bibinfo
  {journal} {Phys. Rev.}\ }\textbf {\bibinfo {volume} {108}},\ \bibinfo {pages}
  {171} (\bibinfo {year} {1957}{\natexlab{b}})}\BibitemShut {NoStop}%
\bibitem [{\citenamefont {Popescu}\ \emph {et~al.}(2006)\citenamefont
  {Popescu}, \citenamefont {Short},\ and\ \citenamefont
  {Winter}}]{popescu2006entanglement}%
  \BibitemOpen
  \bibfield  {author} {\bibinfo {author} {\bibfnamefont {S.}~\bibnamefont
  {Popescu}}, \bibinfo {author} {\bibfnamefont {A.~J.}\ \bibnamefont {Short}},\
  and\ \bibinfo {author} {\bibfnamefont {A.}~\bibnamefont {Winter}},\
  }\bibfield  {title} {\bibinfo {title} {Entanglement and the foundations of
  statistical mechanics},\ }\href {https://doi.org/10.1038/nphys444} {\bibfield
   {journal} {\bibinfo  {journal} {Nature Physics}\ }\textbf {\bibinfo {volume}
  {2}},\ \bibinfo {pages} {754–758} (\bibinfo {year} {2006})}\BibitemShut
  {NoStop}%
\bibitem [{\citenamefont {Rigol}\ \emph {et~al.}(2007)\citenamefont {Rigol},
  \citenamefont {Dunjko}, \citenamefont {Yurovsky},\ and\ \citenamefont
  {Olshanii}}]{rigol2007relaxation}%
  \BibitemOpen
  \bibfield  {author} {\bibinfo {author} {\bibfnamefont {M.}~\bibnamefont
  {Rigol}}, \bibinfo {author} {\bibfnamefont {V.}~\bibnamefont {Dunjko}},
  \bibinfo {author} {\bibfnamefont {V.}~\bibnamefont {Yurovsky}},\ and\
  \bibinfo {author} {\bibfnamefont {M.}~\bibnamefont {Olshanii}},\ }\bibfield
  {title} {\bibinfo {title} {{Relaxation in a Completely Integrable Many-Body
  Quantum System: An Ab Initio Study of the Dynamics of the Highly Excited
  States of 1D Lattice Hard-Core Bosons}},\ }\href
  {https://doi.org/10.1103/PhysRevLett.98.050405} {\bibfield  {journal}
  {\bibinfo  {journal} {Phys. Rev. Lett.}\ }\textbf {\bibinfo {volume} {98}},\
  \bibinfo {pages} {050405} (\bibinfo {year} {2007})}\BibitemShut {NoStop}%
\bibitem [{\citenamefont {Hayden}\ and\ \citenamefont
  {Preskill}(2007)}]{hayden2007black}%
  \BibitemOpen
  \bibfield  {author} {\bibinfo {author} {\bibfnamefont {P.}~\bibnamefont
  {Hayden}}\ and\ \bibinfo {author} {\bibfnamefont {J.}~\bibnamefont
  {Preskill}},\ }\bibfield  {title} {\bibinfo {title} {Black holes as mirrors:
  quantum information in random subsystems},\ }\href
  {https://doi.org/10.1088/1126-6708/2007/09/120} {\bibfield  {journal}
  {\bibinfo  {journal} {Journal of High Energy Physics}\ }\textbf {\bibinfo
  {volume} {2007}},\ \bibinfo {pages} {120–120} (\bibinfo {year}
  {2007})}\BibitemShut {NoStop}%
\bibitem [{\citenamefont {Sekino}\ and\ \citenamefont
  {Susskind}(2008)}]{sekino2008fast}%
  \BibitemOpen
  \bibfield  {author} {\bibinfo {author} {\bibfnamefont {Y.}~\bibnamefont
  {Sekino}}\ and\ \bibinfo {author} {\bibfnamefont {L.}~\bibnamefont
  {Susskind}},\ }\bibfield  {title} {\bibinfo {title} {Fast scramblers},\
  }\href {https://doi.org/10.1088/1126-6708/2008/10/065} {\bibfield  {journal}
  {\bibinfo  {journal} {Journal of High Energy Physics}\ }\textbf {\bibinfo
  {volume} {2008}},\ \bibinfo {pages} {065} (\bibinfo {year}
  {2008})}\BibitemShut {NoStop}%
\bibitem [{\citenamefont {Shenker}\ and\ \citenamefont
  {Stanford}(2014)}]{shenker2014black}%
  \BibitemOpen
  \bibfield  {author} {\bibinfo {author} {\bibfnamefont {S.~H.}\ \bibnamefont
  {Shenker}}\ and\ \bibinfo {author} {\bibfnamefont {D.}~\bibnamefont
  {Stanford}},\ }\bibfield  {title} {\bibinfo {title} {Black holes and the
  butterfly effect},\ }\href {http://dx.doi.org/10.1007/jhep03(2014)067}
  {\bibfield  {journal} {\bibinfo  {journal} {Journal of High Energy Physics}\
  }\textbf {\bibinfo {volume} {2014}},\ \bibinfo {pages} {067} (\bibinfo {year}
  {2014})}\BibitemShut {NoStop}%
\bibitem [{\citenamefont {Maldacena}\ \emph {et~al.}(2016)\citenamefont
  {Maldacena}, \citenamefont {Shenker},\ and\ \citenamefont
  {Stanford}}]{maldacena2016bound}%
  \BibitemOpen
  \bibfield  {author} {\bibinfo {author} {\bibfnamefont {J.}~\bibnamefont
  {Maldacena}}, \bibinfo {author} {\bibfnamefont {S.~H.}\ \bibnamefont
  {Shenker}},\ and\ \bibinfo {author} {\bibfnamefont {D.}~\bibnamefont
  {Stanford}},\ }\bibfield  {title} {\bibinfo {title} {A bound on chaos},\
  }\href {http://dx.doi.org/10.1007/JHEP08(2016)106} {\bibfield  {journal}
  {\bibinfo  {journal} {Journal of High Energy Physics}\ }\textbf {\bibinfo
  {volume} {2016}},\ \bibinfo {pages} {106} (\bibinfo {year}
  {2016})}\BibitemShut {NoStop}%
\bibitem [{\citenamefont {Harrow}\ and\ \citenamefont
  {Montanaro}(2017)}]{harrow2017nature}%
  \BibitemOpen
  \bibfield  {author} {\bibinfo {author} {\bibfnamefont {A.~W.}\ \bibnamefont
  {Harrow}}\ and\ \bibinfo {author} {\bibfnamefont {A.}~\bibnamefont
  {Montanaro}},\ }\bibfield  {title} {\bibinfo {title} {Quantum computational
  supremacy},\ }\href {https://doi.org/10.1038/nature23458} {\bibfield
  {journal} {\bibinfo  {journal} {Nature}\ }\textbf {\bibinfo {volume} {549}},\
  \bibinfo {pages} {203–209} (\bibinfo {year} {2017})}\BibitemShut {NoStop}%
\bibitem [{\citenamefont {Boixo}\ \emph {et~al.}(2018)\citenamefont {Boixo},
  \citenamefont {Isakov}, \citenamefont {Smelyanskiy}, \citenamefont {Babbush},
  \citenamefont {Ding}, \citenamefont {Jiang}, \citenamefont {Bremner},
  \citenamefont {Martinis},\ and\ \citenamefont
  {Neven}}]{boixo2018characterising}%
  \BibitemOpen
  \bibfield  {author} {\bibinfo {author} {\bibfnamefont {S.}~\bibnamefont
  {Boixo}}, \bibinfo {author} {\bibfnamefont {S.~V.}\ \bibnamefont {Isakov}},
  \bibinfo {author} {\bibfnamefont {V.~N.}\ \bibnamefont {Smelyanskiy}},
  \bibinfo {author} {\bibfnamefont {R.}~\bibnamefont {Babbush}}, \bibinfo
  {author} {\bibfnamefont {N.}~\bibnamefont {Ding}}, \bibinfo {author}
  {\bibfnamefont {Z.}~\bibnamefont {Jiang}}, \bibinfo {author} {\bibfnamefont
  {M.~J.}\ \bibnamefont {Bremner}}, \bibinfo {author} {\bibfnamefont {J.~M.}\
  \bibnamefont {Martinis}},\ and\ \bibinfo {author} {\bibfnamefont
  {H.}~\bibnamefont {Neven}},\ }\bibfield  {title} {\bibinfo {title}
  {Characterizing quantum supremacy in near-term devices},\ }\href
  {https://doi.org/10.1038/s41567-018-0124-x} {\bibfield  {journal} {\bibinfo
  {journal} {Nature Physics}\ }\textbf {\bibinfo {volume} {14}},\ \bibinfo
  {pages} {595–600} (\bibinfo {year} {2018})}\BibitemShut {NoStop}%
\bibitem [{\citenamefont {Arute}\ \emph {et~al.}(2019)\citenamefont {Arute},
  \citenamefont {Arya},\ and\ \citenamefont {Babbush~{\it et
  al.}}}]{arute2019quantum}%
  \BibitemOpen
  \bibfield  {author} {\bibinfo {author} {\bibfnamefont {F.}~\bibnamefont
  {Arute}}, \bibinfo {author} {\bibfnamefont {K.}~\bibnamefont {Arya}},\ and\
  \bibinfo {author} {\bibfnamefont {R.}~\bibnamefont {Babbush~{\it et al.}}},\
  }\bibfield  {title} {\bibinfo {title} {Quantum supremacy using a programmable
  superconducting processor},\ }\href
  {https://doi.org/10.1038/s41586-019-1666-5} {\bibfield  {journal} {\bibinfo
  {journal} {Nature}\ }\textbf {\bibinfo {volume} {574}},\ \bibinfo {pages}
  {505–510} (\bibinfo {year} {2019})}\BibitemShut {NoStop}%
\bibitem [{\citenamefont {Mi}\ \emph {et~al.}(2021)\citenamefont {Mi},
  \citenamefont {Roushan}, \citenamefont {Quintana}, \citenamefont {Mandrà},
  \citenamefont {Marshall}, \citenamefont {Neill}, \citenamefont {Arute},
  \citenamefont {Arya}, \citenamefont {Atalaya},\ and\ \citenamefont
  {Babbush~{\it et al.}}}]{mi2021information}%
  \BibitemOpen
  \bibfield  {author} {\bibinfo {author} {\bibfnamefont {X.}~\bibnamefont
  {Mi}}, \bibinfo {author} {\bibfnamefont {P.}~\bibnamefont {Roushan}},
  \bibinfo {author} {\bibfnamefont {C.}~\bibnamefont {Quintana}}, \bibinfo
  {author} {\bibfnamefont {S.}~\bibnamefont {Mandrà}}, \bibinfo {author}
  {\bibfnamefont {J.}~\bibnamefont {Marshall}}, \bibinfo {author}
  {\bibfnamefont {C.}~\bibnamefont {Neill}}, \bibinfo {author} {\bibfnamefont
  {F.}~\bibnamefont {Arute}}, \bibinfo {author} {\bibfnamefont
  {K.}~\bibnamefont {Arya}}, \bibinfo {author} {\bibfnamefont {J.}~\bibnamefont
  {Atalaya}},\ and\ \bibinfo {author} {\bibfnamefont {R.}~\bibnamefont
  {Babbush~{\it et al.}}},\ }\bibfield  {title} {\bibinfo {title} {Information
  scrambling in quantum circuits},\ }\href
  {https://doi.org/10.1126/science.abg5029} {\bibfield  {journal} {\bibinfo
  {journal} {Science}\ }\textbf {\bibinfo {volume} {374}},\ \bibinfo {pages}
  {1479} (\bibinfo {year} {2021})}\BibitemShut {NoStop}%
\bibitem [{\citenamefont {Roberts}\ and\ \citenamefont
  {Stanford}(2015)}]{roberts2015diagnosing}%
  \BibitemOpen
  \bibfield  {author} {\bibinfo {author} {\bibfnamefont {D.~A.}\ \bibnamefont
  {Roberts}}\ and\ \bibinfo {author} {\bibfnamefont {D.}~\bibnamefont
  {Stanford}},\ }\bibfield  {title} {\bibinfo {title} {Diagnosing chaos using
  four-point functions in two-dimensional conformal field theory},\ }\href
  {https://doi.org/10.1103/PhysRevLett.115.131603} {\bibfield  {journal}
  {\bibinfo  {journal} {Phys. Rev. Lett.}\ }\textbf {\bibinfo {volume} {115}},\
  \bibinfo {pages} {131603} (\bibinfo {year} {2015})}\BibitemShut {NoStop}%
\bibitem [{\citenamefont {Hosur}\ \emph {et~al.}(2016)\citenamefont {Hosur},
  \citenamefont {Qi}, \citenamefont {Roberts},\ and\ \citenamefont
  {Yoshida}}]{hosur2016chaos}%
  \BibitemOpen
  \bibfield  {author} {\bibinfo {author} {\bibfnamefont {P.}~\bibnamefont
  {Hosur}}, \bibinfo {author} {\bibfnamefont {X.-L.}\ \bibnamefont {Qi}},
  \bibinfo {author} {\bibfnamefont {D.~A.}\ \bibnamefont {Roberts}},\ and\
  \bibinfo {author} {\bibfnamefont {B.}~\bibnamefont {Yoshida}},\ }\bibfield
  {title} {\bibinfo {title} {Chaos in quantum channels},\ }\href
  {https://doi.org/https://doi.org/10.1007/JHEP02(2016)004} {\bibfield
  {journal} {\bibinfo  {journal} {Journal of High Energy Physics}\ }\textbf
  {\bibinfo {volume} {2016}},\ \bibinfo {pages} {1} (\bibinfo {year}
  {2016})}\BibitemShut {NoStop}%
\bibitem [{\citenamefont {Aleiner}\ \emph {et~al.}(2016)\citenamefont
  {Aleiner}, \citenamefont {Faoro},\ and\ \citenamefont
  {Ioffe}}]{aleiner2016microscopic}%
  \BibitemOpen
  \bibfield  {author} {\bibinfo {author} {\bibfnamefont {I.~L.}\ \bibnamefont
  {Aleiner}}, \bibinfo {author} {\bibfnamefont {L.}~\bibnamefont {Faoro}},\
  and\ \bibinfo {author} {\bibfnamefont {L.~B.}\ \bibnamefont {Ioffe}},\
  }\bibfield  {title} {\bibinfo {title} {{Microscopic model of quantum
  butterfly effect: Out-of-time-order correlators and traveling combustion
  waves}},\ }\href {https://doi.org/10.1016/j.aop.2016.09.006} {\bibfield
  {journal} {\bibinfo  {journal} {Annals of Physics}\ }\textbf {\bibinfo
  {volume} {375}},\ \bibinfo {pages} {378–406} (\bibinfo {year}
  {2016})}\BibitemShut {NoStop}%
\bibitem [{\citenamefont {Bohrdt}\ \emph {et~al.}(2017)\citenamefont {Bohrdt},
  \citenamefont {Mendl}, \citenamefont {Endres},\ and\ \citenamefont
  {Knap}}]{bohrdt2017scrambling}%
  \BibitemOpen
  \bibfield  {author} {\bibinfo {author} {\bibfnamefont {A.}~\bibnamefont
  {Bohrdt}}, \bibinfo {author} {\bibfnamefont {C.~B.}\ \bibnamefont {Mendl}},
  \bibinfo {author} {\bibfnamefont {M.}~\bibnamefont {Endres}},\ and\ \bibinfo
  {author} {\bibfnamefont {M.}~\bibnamefont {Knap}},\ }\bibfield  {title}
  {\bibinfo {title} {Scrambling and thermalization in a diffusive quantum
  many-body system},\ }\href {https://doi.org/10.1088/1367-2630/aa719b}
  {\bibfield  {journal} {\bibinfo  {journal} {New J. Phys.}\ }\textbf {\bibinfo
  {volume} {19}},\ \bibinfo {pages} {063001} (\bibinfo {year}
  {2017})}\BibitemShut {NoStop}%
\bibitem [{\citenamefont {Luitz}\ and\ \citenamefont
  {Bar~Lev}(2017)}]{luitz2017information}%
  \BibitemOpen
  \bibfield  {author} {\bibinfo {author} {\bibfnamefont {D.~J.}\ \bibnamefont
  {Luitz}}\ and\ \bibinfo {author} {\bibfnamefont {Y.}~\bibnamefont
  {Bar~Lev}},\ }\bibfield  {title} {\bibinfo {title} {Information propagation
  in isolated quantum systems},\ }\href
  {https://doi.org/10.1103/PhysRevB.96.020406} {\bibfield  {journal} {\bibinfo
  {journal} {Phys. Rev. B}\ }\textbf {\bibinfo {volume} {96}},\ \bibinfo
  {pages} {020406} (\bibinfo {year} {2017})}\BibitemShut {NoStop}%
\bibitem [{\citenamefont {Zhou}\ and\ \citenamefont
  {Luitz}(2017)}]{zhou2017operator}%
  \BibitemOpen
  \bibfield  {author} {\bibinfo {author} {\bibfnamefont {T.}~\bibnamefont
  {Zhou}}\ and\ \bibinfo {author} {\bibfnamefont {D.~J.}\ \bibnamefont
  {Luitz}},\ }\bibfield  {title} {\bibinfo {title} {Operator entanglement
  entropy of the time evolution operator in chaotic systems},\ }\href
  {https://doi.org/10.1103/PhysRevB.95.094206} {\bibfield  {journal} {\bibinfo
  {journal} {Phys. Rev. B}\ }\textbf {\bibinfo {volume} {95}},\ \bibinfo
  {pages} {094206} (\bibinfo {year} {2017})}\BibitemShut {NoStop}%
\bibitem [{\citenamefont {Nahum}\ \emph
  {et~al.}(2018{\natexlab{a}})\citenamefont {Nahum}, \citenamefont {Vijay},\
  and\ \citenamefont {Haah}}]{nahum2018operator}%
  \BibitemOpen
  \bibfield  {author} {\bibinfo {author} {\bibfnamefont {A.}~\bibnamefont
  {Nahum}}, \bibinfo {author} {\bibfnamefont {S.}~\bibnamefont {Vijay}},\ and\
  \bibinfo {author} {\bibfnamefont {J.}~\bibnamefont {Haah}},\ }\bibfield
  {title} {\bibinfo {title} {Operator spreading in random unitary circuits},\
  }\href {https://doi.org/10.1103/PhysRevX.8.021014} {\bibfield  {journal}
  {\bibinfo  {journal} {Phys. Rev. X}\ }\textbf {\bibinfo {volume} {8}},\
  \bibinfo {pages} {021014} (\bibinfo {year} {2018}{\natexlab{a}})}\BibitemShut
  {NoStop}%
\bibitem [{\citenamefont {von Keyserlingk}\ \emph {et~al.}(2018)\citenamefont
  {von Keyserlingk}, \citenamefont {Rakovszky}, \citenamefont {Pollmann},\ and\
  \citenamefont {Sondhi}}]{keyserlingk2018operator}%
  \BibitemOpen
  \bibfield  {author} {\bibinfo {author} {\bibfnamefont {C.~W.}\ \bibnamefont
  {von Keyserlingk}}, \bibinfo {author} {\bibfnamefont {T.}~\bibnamefont
  {Rakovszky}}, \bibinfo {author} {\bibfnamefont {F.}~\bibnamefont
  {Pollmann}},\ and\ \bibinfo {author} {\bibfnamefont {S.~L.}\ \bibnamefont
  {Sondhi}},\ }\bibfield  {title} {\bibinfo {title} {Operator hydrodynamics,
  otocs, and entanglement growth in systems without conservation laws},\ }\href
  {https://doi.org/10.1103/PhysRevX.8.021013} {\bibfield  {journal} {\bibinfo
  {journal} {Phys. Rev. X}\ }\textbf {\bibinfo {volume} {8}},\ \bibinfo {pages}
  {021013} (\bibinfo {year} {2018})}\BibitemShut {NoStop}%
\bibitem [{\citenamefont {Rakovszky}\ \emph {et~al.}(2018)\citenamefont
  {Rakovszky}, \citenamefont {Pollmann},\ and\ \citenamefont {von
  Keyserlingk}}]{rakovszky2018diffusive}%
  \BibitemOpen
  \bibfield  {author} {\bibinfo {author} {\bibfnamefont {T.}~\bibnamefont
  {Rakovszky}}, \bibinfo {author} {\bibfnamefont {F.}~\bibnamefont
  {Pollmann}},\ and\ \bibinfo {author} {\bibfnamefont {C.~W.}\ \bibnamefont
  {von Keyserlingk}},\ }\bibfield  {title} {\bibinfo {title} {Diffusive
  hydrodynamics of out-of-time-ordered correlators with charge conservation},\
  }\href {https://doi.org/10.1103/PhysRevX.8.031058} {\bibfield  {journal}
  {\bibinfo  {journal} {Phys. Rev. X}\ }\textbf {\bibinfo {volume} {8}},\
  \bibinfo {pages} {031058} (\bibinfo {year} {2018})}\BibitemShut {NoStop}%
\bibitem [{\citenamefont {Khemani}\ \emph {et~al.}(2018)\citenamefont
  {Khemani}, \citenamefont {Vishwanath},\ and\ \citenamefont
  {Huse}}]{khemani2018operator}%
  \BibitemOpen
  \bibfield  {author} {\bibinfo {author} {\bibfnamefont {V.}~\bibnamefont
  {Khemani}}, \bibinfo {author} {\bibfnamefont {A.}~\bibnamefont
  {Vishwanath}},\ and\ \bibinfo {author} {\bibfnamefont {D.~A.}\ \bibnamefont
  {Huse}},\ }\bibfield  {title} {\bibinfo {title} {{Operator Spreading and the
  Emergence of Dissipative Hydrodynamics under Unitary Evolution with
  Conservation Laws}},\ }\href {https://doi.org/10.1103/PhysRevX.8.031057}
  {\bibfield  {journal} {\bibinfo  {journal} {Phys. Rev. X}\ }\textbf {\bibinfo
  {volume} {8}},\ \bibinfo {pages} {031057} (\bibinfo {year}
  {2018})}\BibitemShut {NoStop}%
\bibitem [{\citenamefont {Hahn}\ \emph {et~al.}(2024)\citenamefont {Hahn},
  \citenamefont {Luitz},\ and\ \citenamefont {Chalker}}]{hahn2024eigenstate}%
  \BibitemOpen
  \bibfield  {author} {\bibinfo {author} {\bibfnamefont {D.}~\bibnamefont
  {Hahn}}, \bibinfo {author} {\bibfnamefont {D.~J.}\ \bibnamefont {Luitz}},\
  and\ \bibinfo {author} {\bibfnamefont {J.~T.}\ \bibnamefont {Chalker}},\
  }\bibfield  {title} {\bibinfo {title} {{Eigenstate Correlations, the
  Eigenstate Thermalization Hypothesis, and Quantum Information Dynamics in
  Chaotic Many-Body Quantum Systems}},\ }\href
  {https://doi.org/10.1103/PhysRevX.14.031029} {\bibfield  {journal} {\bibinfo
  {journal} {Phys. Rev. X}\ }\textbf {\bibinfo {volume} {14}},\ \bibinfo
  {pages} {031029} (\bibinfo {year} {2024})}\BibitemShut {NoStop}%
\bibitem [{\citenamefont {Pain}\ \emph {et~al.}(2026)\citenamefont {Pain},
  \citenamefont {Thakur},\ and\ \citenamefont {Roy}}]{pain2026imprints}%
  \BibitemOpen
  \bibfield  {author} {\bibinfo {author} {\bibfnamefont {B.}~\bibnamefont
  {Pain}}, \bibinfo {author} {\bibfnamefont {R.}~\bibnamefont {Thakur}},\ and\
  \bibinfo {author} {\bibfnamefont {S.}~\bibnamefont {Roy}},\ }\bibfield
  {title} {\bibinfo {title} {Imprints of information scrambling on eigenstates
  of a quantum chaotic system},\ }\href {https://doi.org/10.1103/r1gj-g99c}
  {\bibfield  {journal} {\bibinfo  {journal} {Phys. Rev. B}\ } (\bibinfo {year}
  {2026})}\BibitemShut {NoStop}%
\bibitem [{\citenamefont {Calabrese}\ and\ \citenamefont
  {Cardy}(2005)}]{calabrese2005evolution}%
  \BibitemOpen
  \bibfield  {author} {\bibinfo {author} {\bibfnamefont {P.}~\bibnamefont
  {Calabrese}}\ and\ \bibinfo {author} {\bibfnamefont {J.}~\bibnamefont
  {Cardy}},\ }\bibfield  {title} {\bibinfo {title} {Evolution of entanglement
  entropy in one-dimensional systems},\ }\href
  {https://doi.org/10.1088/1742-5468/2005/04/P04010} {\bibfield  {journal}
  {\bibinfo  {journal} {J. Stat. Mech.}\ }\textbf {\bibinfo {volume} {2005}},\
  \bibinfo {pages} {P04010} (\bibinfo {year} {2005})}\BibitemShut {NoStop}%
\bibitem [{\citenamefont {Kim}\ and\ \citenamefont
  {Huse}(2013)}]{kim2013ballistic}%
  \BibitemOpen
  \bibfield  {author} {\bibinfo {author} {\bibfnamefont {H.}~\bibnamefont
  {Kim}}\ and\ \bibinfo {author} {\bibfnamefont {D.~A.}\ \bibnamefont {Huse}},\
  }\bibfield  {title} {\bibinfo {title} {Ballistic spreading of entanglement in
  a diffusive nonintegrable system},\ }\href
  {https://doi.org/10.1103/PhysRevLett.111.127205} {\bibfield  {journal}
  {\bibinfo  {journal} {Phys. Rev. Lett.}\ }\textbf {\bibinfo {volume} {111}},\
  \bibinfo {pages} {127205} (\bibinfo {year} {2013})}\BibitemShut {NoStop}%
\bibitem [{\citenamefont {Liu}\ and\ \citenamefont
  {Suh}(2014)}]{liu2014entanglement}%
  \BibitemOpen
  \bibfield  {author} {\bibinfo {author} {\bibfnamefont {H.}~\bibnamefont
  {Liu}}\ and\ \bibinfo {author} {\bibfnamefont {S.~J.}\ \bibnamefont {Suh}},\
  }\bibfield  {title} {\bibinfo {title} {Entanglement tsunami: Universal
  scaling in holographic thermalization},\ }\href
  {https://doi.org/10.1103/PhysRevLett.112.011601} {\bibfield  {journal}
  {\bibinfo  {journal} {Phys. Rev. Lett.}\ }\textbf {\bibinfo {volume} {112}},\
  \bibinfo {pages} {011601} (\bibinfo {year} {2014})}\BibitemShut {NoStop}%
\bibitem [{\citenamefont {Zhang}\ \emph {et~al.}(2015)\citenamefont {Zhang},
  \citenamefont {Kim},\ and\ \citenamefont {Huse}}]{zhang2015thermalisation}%
  \BibitemOpen
  \bibfield  {author} {\bibinfo {author} {\bibfnamefont {L.}~\bibnamefont
  {Zhang}}, \bibinfo {author} {\bibfnamefont {H.}~\bibnamefont {Kim}},\ and\
  \bibinfo {author} {\bibfnamefont {D.~A.}\ \bibnamefont {Huse}},\ }\bibfield
  {title} {\bibinfo {title} {Thermalization of entanglement},\ }\href
  {https://doi.org/10.1103/PhysRevE.91.062128} {\bibfield  {journal} {\bibinfo
  {journal} {Phys. Rev. E}\ }\textbf {\bibinfo {volume} {91}},\ \bibinfo
  {pages} {062128} (\bibinfo {year} {2015})}\BibitemShut {NoStop}%
\bibitem [{\citenamefont {Nahum}\ \emph {et~al.}(2017)\citenamefont {Nahum},
  \citenamefont {Ruhman}, \citenamefont {Vijay},\ and\ \citenamefont
  {Haah}}]{nahum2017quantum}%
  \BibitemOpen
  \bibfield  {author} {\bibinfo {author} {\bibfnamefont {A.}~\bibnamefont
  {Nahum}}, \bibinfo {author} {\bibfnamefont {J.}~\bibnamefont {Ruhman}},
  \bibinfo {author} {\bibfnamefont {S.}~\bibnamefont {Vijay}},\ and\ \bibinfo
  {author} {\bibfnamefont {J.}~\bibnamefont {Haah}},\ }\bibfield  {title}
  {\bibinfo {title} {Quantum entanglement growth under random unitary
  dynamics},\ }\href {https://doi.org/10.1103/PhysRevX.7.031016} {\bibfield
  {journal} {\bibinfo  {journal} {Phys. Rev. X}\ }\textbf {\bibinfo {volume}
  {7}},\ \bibinfo {pages} {031016} (\bibinfo {year} {2017})}\BibitemShut
  {NoStop}%
\bibitem [{\citenamefont {Jonay}\ \emph {et~al.}(2018)\citenamefont {Jonay},
  \citenamefont {Huse},\ and\ \citenamefont {Nahum}}]{jonay2018coarse}%
  \BibitemOpen
  \bibfield  {author} {\bibinfo {author} {\bibfnamefont {C.}~\bibnamefont
  {Jonay}}, \bibinfo {author} {\bibfnamefont {D.~A.}\ \bibnamefont {Huse}},\
  and\ \bibinfo {author} {\bibfnamefont {A.}~\bibnamefont {Nahum}},\ }\href
  {https://arxiv.org/abs/1803.00089} {\bibinfo {title} {Coarse-grained dynamics
  of operator and state entanglement}} (\bibinfo {year} {2018}),\ \Eprint
  {https://arxiv.org/abs/1803.00089} {arXiv:1803.00089 [cond-mat.stat-mech]}
  \BibitemShut {NoStop}%
\bibitem [{\citenamefont {Calabrese}\ and\ \citenamefont
  {Cardy}(2007)}]{calabrese2007entanglement}%
  \BibitemOpen
  \bibfield  {author} {\bibinfo {author} {\bibfnamefont {P.}~\bibnamefont
  {Calabrese}}\ and\ \bibinfo {author} {\bibfnamefont {J.}~\bibnamefont
  {Cardy}},\ }\bibfield  {title} {\bibinfo {title} {Entanglement and
  correlation functions following a local quench: a conformal field theory
  approach},\ }\href {https://doi.org/10.1088/1742-5468/2007/10/p10004}
  {\bibfield  {journal} {\bibinfo  {journal} {J. Stat. Mech.}\ }\textbf
  {\bibinfo {volume} {2007}},\ \bibinfo {pages} {P10004} (\bibinfo {year}
  {2007})}\BibitemShut {NoStop}%
\bibitem [{\citenamefont {Zhou}\ and\ \citenamefont
  {Nahum}(2019)}]{zhou2019emergent}%
  \BibitemOpen
  \bibfield  {author} {\bibinfo {author} {\bibfnamefont {T.}~\bibnamefont
  {Zhou}}\ and\ \bibinfo {author} {\bibfnamefont {A.}~\bibnamefont {Nahum}},\
  }\bibfield  {title} {\bibinfo {title} {Emergent statistical mechanics of
  entanglement in random unitary circuits},\ }\href
  {https://doi.org/10.1103/PhysRevB.99.174205} {\bibfield  {journal} {\bibinfo
  {journal} {Phys. Rev. B}\ }\textbf {\bibinfo {volume} {99}},\ \bibinfo
  {pages} {174205} (\bibinfo {year} {2019})}\BibitemShut {NoStop}%
\bibitem [{\citenamefont {Zhou}\ and\ \citenamefont
  {Nahum}(2020)}]{zhou2020entanglement}%
  \BibitemOpen
  \bibfield  {author} {\bibinfo {author} {\bibfnamefont {T.}~\bibnamefont
  {Zhou}}\ and\ \bibinfo {author} {\bibfnamefont {A.}~\bibnamefont {Nahum}},\
  }\bibfield  {title} {\bibinfo {title} {{Entanglement Membrane in Chaotic
  Many-Body Systems}},\ }\href {https://doi.org/10.1103/PhysRevX.10.031066}
  {\bibfield  {journal} {\bibinfo  {journal} {Phys. Rev. X}\ }\textbf {\bibinfo
  {volume} {10}},\ \bibinfo {pages} {031066} (\bibinfo {year}
  {2020})}\BibitemShut {NoStop}%
\bibitem [{\citenamefont {Klein~Kvorning}\ \emph {et~al.}(2022)\citenamefont
  {Klein~Kvorning}, \citenamefont {Herviou},\ and\ \citenamefont
  {Bardarson}}]{kvorning2022time}%
  \BibitemOpen
  \bibfield  {author} {\bibinfo {author} {\bibfnamefont {T.}~\bibnamefont
  {Klein~Kvorning}}, \bibinfo {author} {\bibfnamefont {L.}~\bibnamefont
  {Herviou}},\ and\ \bibinfo {author} {\bibfnamefont {J.~H.}\ \bibnamefont
  {Bardarson}},\ }\bibfield  {title} {\bibinfo {title} {{Time-evolution of
  local information: thermalization dynamics of local observables}},\ }\href
  {https://doi.org/10.21468/SciPostPhys.13.4.080} {\bibfield  {journal}
  {\bibinfo  {journal} {SciPost Phys.}\ }\textbf {\bibinfo {volume} {13}},\
  \bibinfo {pages} {080} (\bibinfo {year} {2022})}\BibitemShut {NoStop}%
\bibitem [{\citenamefont {Artiaco}\ \emph {et~al.}(2024)\citenamefont
  {Artiaco}, \citenamefont {Fleckenstein}, \citenamefont {Aceituno~Ch\'avez},
  \citenamefont {Klein~Kvorning},\ and\ \citenamefont
  {Bardarson}}]{artiaco2024efficient}%
  \BibitemOpen
  \bibfield  {author} {\bibinfo {author} {\bibfnamefont {C.}~\bibnamefont
  {Artiaco}}, \bibinfo {author} {\bibfnamefont {C.}~\bibnamefont
  {Fleckenstein}}, \bibinfo {author} {\bibfnamefont {D.}~\bibnamefont
  {Aceituno~Ch\'avez}}, \bibinfo {author} {\bibfnamefont {T.}~\bibnamefont
  {Klein~Kvorning}},\ and\ \bibinfo {author} {\bibfnamefont {J.~H.}\
  \bibnamefont {Bardarson}},\ }\bibfield  {title} {\bibinfo {title} {Efficient
  large-scale many-body quantum dynamics via local-information time
  evolution},\ }\href {https://doi.org/10.1103/PRXQuantum.5.020352} {\bibfield
  {journal} {\bibinfo  {journal} {PRX Quantum}\ }\textbf {\bibinfo {volume}
  {5}},\ \bibinfo {pages} {020352} (\bibinfo {year} {2024})}\BibitemShut
  {NoStop}%
\bibitem [{\citenamefont {Artiaco}\ \emph {et~al.}(2025)\citenamefont
  {Artiaco}, \citenamefont {Klein~Kvorning}, \citenamefont {Aceituno~Ch\'avez},
  \citenamefont {Herviou},\ and\ \citenamefont
  {Bardarson}}]{artiaco2025universal}%
  \BibitemOpen
  \bibfield  {author} {\bibinfo {author} {\bibfnamefont {C.}~\bibnamefont
  {Artiaco}}, \bibinfo {author} {\bibfnamefont {T.}~\bibnamefont
  {Klein~Kvorning}}, \bibinfo {author} {\bibfnamefont {D.}~\bibnamefont
  {Aceituno~Ch\'avez}}, \bibinfo {author} {\bibfnamefont {L.}~\bibnamefont
  {Herviou}},\ and\ \bibinfo {author} {\bibfnamefont {J.~H.}\ \bibnamefont
  {Bardarson}},\ }\bibfield  {title} {\bibinfo {title} {{Universal
  Characterization of Quantum Many-Body States through Local Information}},\
  }\href {https://doi.org/10.1103/PhysRevLett.134.190401} {\bibfield  {journal}
  {\bibinfo  {journal} {Phys. Rev. Lett.}\ }\textbf {\bibinfo {volume} {134}},\
  \bibinfo {pages} {190401} (\bibinfo {year} {2025})}\BibitemShut {NoStop}%
\bibitem [{\citenamefont {Bauer}\ \emph {et~al.}(2025)\citenamefont {Bauer},
  \citenamefont {Trauzettel}, \citenamefont {Klein~Kvorning}, \citenamefont
  {Bardarson},\ and\ \citenamefont {Artiaco}}]{bauer2025local}%
  \BibitemOpen
  \bibfield  {author} {\bibinfo {author} {\bibfnamefont {N.~P.}\ \bibnamefont
  {Bauer}}, \bibinfo {author} {\bibfnamefont {B.}~\bibnamefont {Trauzettel}},
  \bibinfo {author} {\bibfnamefont {T.}~\bibnamefont {Klein~Kvorning}},
  \bibinfo {author} {\bibfnamefont {J.~H.}\ \bibnamefont {Bardarson}},\ and\
  \bibinfo {author} {\bibfnamefont {C.}~\bibnamefont {Artiaco}},\ }\bibfield
  {title} {\bibinfo {title} {Local information flow in quantum quench
  dynamics},\ }\href {https://doi.org/10.1103/v7gb-5gq8} {\bibfield  {journal}
  {\bibinfo  {journal} {Phys. Rev. A}\ }\textbf {\bibinfo {volume} {112}},\
  \bibinfo {pages} {022221} (\bibinfo {year} {2025})}\BibitemShut {NoStop}%
\bibitem [{\citenamefont {Flór}\ \emph {et~al.}(2025)\citenamefont {Flór},
  \citenamefont {Artiaco}, \citenamefont {Kvorning},\ and\ \citenamefont
  {Bardarson}}]{flor2025higher}%
  \BibitemOpen
  \bibfield  {author} {\bibinfo {author} {\bibfnamefont {I.~M.}\ \bibnamefont
  {Flór}}, \bibinfo {author} {\bibfnamefont {C.}~\bibnamefont {Artiaco}},
  \bibinfo {author} {\bibfnamefont {T.~K.}\ \bibnamefont {Kvorning}},\ and\
  \bibinfo {author} {\bibfnamefont {J.~H.}\ \bibnamefont {Bardarson}},\ }\href
  {https://arxiv.org/abs/2512.20793} {\bibinfo {title} {Higher-dimensional
  information lattice: Quantum state characterization through
  inclusion-exclusion local information}} (\bibinfo {year} {2025}),\ \Eprint
  {https://arxiv.org/abs/2512.20793} {arXiv:2512.20793 [quant-ph]} \BibitemShut
  {NoStop}%
\bibitem [{\citenamefont {Chan}\ \emph {et~al.}(2018)\citenamefont {Chan},
  \citenamefont {De~Luca},\ and\ \citenamefont {Chalker}}]{chan2018solution}%
  \BibitemOpen
  \bibfield  {author} {\bibinfo {author} {\bibfnamefont {A.}~\bibnamefont
  {Chan}}, \bibinfo {author} {\bibfnamefont {A.}~\bibnamefont {De~Luca}},\ and\
  \bibinfo {author} {\bibfnamefont {J.~T.}\ \bibnamefont {Chalker}},\
  }\bibfield  {title} {\bibinfo {title} {Solution of a minimal model for
  many-body quantum chaos},\ }\href {https://doi.org/10.1103/PhysRevX.8.041019}
  {\bibfield  {journal} {\bibinfo  {journal} {Phys. Rev. X}\ }\textbf {\bibinfo
  {volume} {8}},\ \bibinfo {pages} {041019} (\bibinfo {year}
  {2018})}\BibitemShut {NoStop}%
\bibitem [{\citenamefont {Fisher}\ \emph {et~al.}(2023)\citenamefont {Fisher},
  \citenamefont {Khemani}, \citenamefont {Nahum},\ and\ \citenamefont
  {Vijay}}]{fisher2023random}%
  \BibitemOpen
  \bibfield  {author} {\bibinfo {author} {\bibfnamefont {M.~P.~A.}\
  \bibnamefont {Fisher}}, \bibinfo {author} {\bibfnamefont {V.}~\bibnamefont
  {Khemani}}, \bibinfo {author} {\bibfnamefont {A.}~\bibnamefont {Nahum}},\
  and\ \bibinfo {author} {\bibfnamefont {S.}~\bibnamefont {Vijay}},\ }\bibfield
   {title} {\bibinfo {title} {Random quantum circuits},\ }\href
  {https://doi.org/https://doi.org/10.1146/annurev-conmatphys-031720-030658}
  {\bibfield  {journal} {\bibinfo  {journal} {Annual Review of Condensed Matter
  Physics}\ }\textbf {\bibinfo {volume} {14}},\ \bibinfo {pages} {335}
  (\bibinfo {year} {2023})}\BibitemShut {NoStop}%
\bibitem [{\citenamefont {Gottesman}(1997)}]{gottesman1997stabilizer}%
  \BibitemOpen
  \bibfield  {author} {\bibinfo {author} {\bibfnamefont {D.}~\bibnamefont
  {Gottesman}},\ }\href {https://arxiv.org/abs/quant-ph/9705052} {\bibinfo
  {title} {Stabilizer codes and quantum error correction}} (\bibinfo {year}
  {1997}),\ \Eprint {https://arxiv.org/abs/quant-ph/9705052}
  {arXiv:quant-ph/9705052 [quant-ph]} \BibitemShut {NoStop}%
\bibitem [{\citenamefont {Gottesman}(1998)}]{gottesman1998heisenberg}%
  \BibitemOpen
  \bibfield  {author} {\bibinfo {author} {\bibfnamefont {D.}~\bibnamefont
  {Gottesman}},\ }\href {https://arxiv.org/abs/quant-ph/9807006} {\bibinfo
  {title} {The heisenberg representation of quantum computers}} (\bibinfo
  {year} {1998}),\ \Eprint {https://arxiv.org/abs/quant-ph/9807006}
  {arXiv:quant-ph/9807006 [quant-ph]} \BibitemShut {NoStop}%
\bibitem [{\citenamefont {Aaronson}\ and\ \citenamefont
  {Gottesman}(2004)}]{aaronson2004improved}%
  \BibitemOpen
  \bibfield  {author} {\bibinfo {author} {\bibfnamefont {S.}~\bibnamefont
  {Aaronson}}\ and\ \bibinfo {author} {\bibfnamefont {D.}~\bibnamefont
  {Gottesman}},\ }\bibfield  {title} {\bibinfo {title} {Improved simulation of
  stabilizer circuits},\ }\href {https://doi.org/10.1103/PhysRevA.70.052328}
  {\bibfield  {journal} {\bibinfo  {journal} {Phys. Rev. A}\ }\textbf {\bibinfo
  {volume} {70}},\ \bibinfo {pages} {052328} (\bibinfo {year}
  {2004})}\BibitemShut {NoStop}%
\bibitem [{\citenamefont {Fawzi}\ and\ \citenamefont
  {Renner}(2015)}]{fawzi2015quantum}%
  \BibitemOpen
  \bibfield  {author} {\bibinfo {author} {\bibfnamefont {O.}~\bibnamefont
  {Fawzi}}\ and\ \bibinfo {author} {\bibfnamefont {R.}~\bibnamefont {Renner}},\
  }\bibfield  {title} {\bibinfo {title} {{Quantum Conditional Mutual
  Information and Approximate Markov Chains}},\ }\href
  {https://doi.org/10.1007/s00220-015-2466-x} {\bibfield  {journal} {\bibinfo
  {journal} {Communications in Mathematical Physics}\ }\textbf {\bibinfo
  {volume} {340}},\ \bibinfo {pages} {575–611} (\bibinfo {year}
  {2015})}\BibitemShut {NoStop}%
\bibitem [{\citenamefont {Lieb}\ and\ \citenamefont
  {Ruskai}(1973)}]{lieb1973proof}%
  \BibitemOpen
  \bibfield  {author} {\bibinfo {author} {\bibfnamefont {E.~H.}\ \bibnamefont
  {Lieb}}\ and\ \bibinfo {author} {\bibfnamefont {M.~B.}\ \bibnamefont
  {Ruskai}},\ }\bibfield  {title} {\bibinfo {title} {Proof of the strong
  subadditivity of quantum‐mechanical entropy},\ }\href
  {https://doi.org/10.1063/1.1666274} {\bibfield  {journal} {\bibinfo
  {journal} {Journal of Mathematical Physics}\ }\textbf {\bibinfo {volume}
  {14}},\ \bibinfo {pages} {1938} (\bibinfo {year} {1973})}\BibitemShut
  {NoStop}%
\bibitem [{\citenamefont {Page}(1993)}]{page1993average}%
  \BibitemOpen
  \bibfield  {author} {\bibinfo {author} {\bibfnamefont {D.~N.}\ \bibnamefont
  {Page}},\ }\bibfield  {title} {\bibinfo {title} {Average entropy of a
  subsystem},\ }\href {https://doi.org/10.1103/PhysRevLett.71.1291} {\bibfield
  {journal} {\bibinfo  {journal} {Phys. Rev. Lett.}\ }\textbf {\bibinfo
  {volume} {71}},\ \bibinfo {pages} {1291} (\bibinfo {year}
  {1993})}\BibitemShut {NoStop}%
\bibitem [{\citenamefont {Weingarten}(1978)}]{weingarten1978asymptotic}%
  \BibitemOpen
  \bibfield  {author} {\bibinfo {author} {\bibfnamefont {D.}~\bibnamefont
  {Weingarten}},\ }\bibfield  {title} {\bibinfo {title} {{Asymptotic behavior
  of group integrals in the limit of infinite rank}},\ }\href
  {https://doi.org/10.1063/1.523807} {\bibfield  {journal} {\bibinfo  {journal}
  {Journal of Mathematical Physics}\ }\textbf {\bibinfo {volume} {19}},\
  \bibinfo {pages} {999} (\bibinfo {year} {1978})}\BibitemShut {NoStop}%
\bibitem [{\citenamefont {Kardar}\ \emph {et~al.}(1986)\citenamefont {Kardar},
  \citenamefont {Parisi},\ and\ \citenamefont {Zhang}}]{kardar1986dynamic}%
  \BibitemOpen
  \bibfield  {author} {\bibinfo {author} {\bibfnamefont {M.}~\bibnamefont
  {Kardar}}, \bibinfo {author} {\bibfnamefont {G.}~\bibnamefont {Parisi}},\
  and\ \bibinfo {author} {\bibfnamefont {Y.-C.}\ \bibnamefont {Zhang}},\
  }\bibfield  {title} {\bibinfo {title} {Dynamic scaling of growing
  interfaces},\ }\href {https://doi.org/10.1103/PhysRevLett.56.889} {\bibfield
  {journal} {\bibinfo  {journal} {Phys. Rev. Lett.}\ }\textbf {\bibinfo
  {volume} {56}},\ \bibinfo {pages} {889} (\bibinfo {year} {1986})}\BibitemShut
  {NoStop}%
\bibitem [{\citenamefont {Li}\ \emph {et~al.}(2019)\citenamefont {Li},
  \citenamefont {Chen},\ and\ \citenamefont {Fisher}}]{li2019measurement}%
  \BibitemOpen
  \bibfield  {author} {\bibinfo {author} {\bibfnamefont {Y.}~\bibnamefont
  {Li}}, \bibinfo {author} {\bibfnamefont {X.}~\bibnamefont {Chen}},\ and\
  \bibinfo {author} {\bibfnamefont {M.~P.~A.}\ \bibnamefont {Fisher}},\
  }\bibfield  {title} {\bibinfo {title} {Measurement-driven entanglement
  transition in hybrid quantum circuits},\ }\href
  {https://doi.org/10.1103/PhysRevB.100.134306} {\bibfield  {journal} {\bibinfo
   {journal} {Phys. Rev. B}\ }\textbf {\bibinfo {volume} {100}},\ \bibinfo
  {pages} {134306} (\bibinfo {year} {2019})}\BibitemShut {NoStop}%
\bibitem [{\citenamefont {Nahum}\ \emph
  {et~al.}(2018{\natexlab{b}})\citenamefont {Nahum}, \citenamefont {Ruhman},\
  and\ \citenamefont {Huse}}]{nahum2018dynamics}%
  \BibitemOpen
  \bibfield  {author} {\bibinfo {author} {\bibfnamefont {A.}~\bibnamefont
  {Nahum}}, \bibinfo {author} {\bibfnamefont {J.}~\bibnamefont {Ruhman}},\ and\
  \bibinfo {author} {\bibfnamefont {D.~A.}\ \bibnamefont {Huse}},\ }\bibfield
  {title} {\bibinfo {title} {Dynamics of entanglement and transport in
  one-dimensional systems with quenched randomness},\ }\href
  {https://doi.org/10.1103/PhysRevB.98.035118} {\bibfield  {journal} {\bibinfo
  {journal} {Phys. Rev. B}\ }\textbf {\bibinfo {volume} {98}},\ \bibinfo
  {pages} {035118} (\bibinfo {year} {2018}{\natexlab{b}})}\BibitemShut
  {NoStop}%
\bibitem [{\citenamefont {Nandkishore}\ and\ \citenamefont
  {Huse}(2015)}]{nandkishore2015many}%
  \BibitemOpen
  \bibfield  {author} {\bibinfo {author} {\bibfnamefont {R.}~\bibnamefont
  {Nandkishore}}\ and\ \bibinfo {author} {\bibfnamefont {D.~A.}\ \bibnamefont
  {Huse}},\ }\bibfield  {title} {\bibinfo {title} {Many-body localization and
  thermalization in quantum statistical mechanics},\ }\href
  {https://doi.org/10.1146/annurev-conmatphys-031214-014726} {\bibfield
  {journal} {\bibinfo  {journal} {Annu. Rev. Condens. Matter Phys.}\ }\textbf
  {\bibinfo {volume} {6}},\ \bibinfo {pages} {15} (\bibinfo {year}
  {2015})}\BibitemShut {NoStop}%
\bibitem [{\citenamefont {Abanin}\ \emph {et~al.}(2019)\citenamefont {Abanin},
  \citenamefont {Altman}, \citenamefont {Bloch},\ and\ \citenamefont
  {Serbyn}}]{abanin2019colloquium}%
  \BibitemOpen
  \bibfield  {author} {\bibinfo {author} {\bibfnamefont {D.~A.}\ \bibnamefont
  {Abanin}}, \bibinfo {author} {\bibfnamefont {E.}~\bibnamefont {Altman}},
  \bibinfo {author} {\bibfnamefont {I.}~\bibnamefont {Bloch}},\ and\ \bibinfo
  {author} {\bibfnamefont {M.}~\bibnamefont {Serbyn}},\ }\bibfield  {title}
  {\bibinfo {title} {Colloquium: Many-body localization, thermalization, and
  entanglement},\ }\href {https://doi.org/10.1103/RevModPhys.91.021001}
  {\bibfield  {journal} {\bibinfo  {journal} {Rev. Mod. Phys.}\ }\textbf
  {\bibinfo {volume} {91}},\ \bibinfo {pages} {021001} (\bibinfo {year}
  {2019})}\BibitemShut {NoStop}%
\bibitem [{\citenamefont {Sierant}\ \emph {et~al.}(2024)\citenamefont
  {Sierant}, \citenamefont {Lewenstein}, \citenamefont {Scardicchio},
  \citenamefont {Vidmar},\ and\ \citenamefont
  {Zakrzewski}}]{sierant2024manybody}%
  \BibitemOpen
  \bibfield  {author} {\bibinfo {author} {\bibfnamefont {P.}~\bibnamefont
  {Sierant}}, \bibinfo {author} {\bibfnamefont {M.}~\bibnamefont {Lewenstein}},
  \bibinfo {author} {\bibfnamefont {A.}~\bibnamefont {Scardicchio}}, \bibinfo
  {author} {\bibfnamefont {L.}~\bibnamefont {Vidmar}},\ and\ \bibinfo {author}
  {\bibfnamefont {J.}~\bibnamefont {Zakrzewski}},\ }\href@noop {} {\bibinfo
  {title} {Many-body localization in the age of classical computing}} (\bibinfo
  {year} {2024}),\ \Eprint {https://arxiv.org/abs/2403.07111} {arXiv:2403.07111
  [cond-mat.dis-nn]} \BibitemShut {NoStop}%
\bibitem [{\citenamefont {Herviou}\ \emph {et~al.}(2019)\citenamefont
  {Herviou}, \citenamefont {Bera},\ and\ \citenamefont
  {Bardarson}}]{herviou2019multiscale}%
  \BibitemOpen
  \bibfield  {author} {\bibinfo {author} {\bibfnamefont {L.}~\bibnamefont
  {Herviou}}, \bibinfo {author} {\bibfnamefont {S.}~\bibnamefont {Bera}},\ and\
  \bibinfo {author} {\bibfnamefont {J.~H.}\ \bibnamefont {Bardarson}},\
  }\bibfield  {title} {\bibinfo {title} {Multiscale entanglement clusters at
  the many-body localization phase transition},\ }\href
  {https://doi.org/10.1103/PhysRevB.99.134205} {\bibfield  {journal} {\bibinfo
  {journal} {Phys. Rev. B}\ }\textbf {\bibinfo {volume} {99}},\ \bibinfo
  {pages} {134205} (\bibinfo {year} {2019})}\BibitemShut {NoStop}%
\bibitem [{\citenamefont {Garratt}\ \emph {et~al.}(2021)\citenamefont
  {Garratt}, \citenamefont {Roy},\ and\ \citenamefont
  {Chalker}}]{garratt2021resonances}%
  \BibitemOpen
  \bibfield  {author} {\bibinfo {author} {\bibfnamefont {S.~J.}\ \bibnamefont
  {Garratt}}, \bibinfo {author} {\bibfnamefont {S.}~\bibnamefont {Roy}},\ and\
  \bibinfo {author} {\bibfnamefont {J.~T.}\ \bibnamefont {Chalker}},\
  }\bibfield  {title} {\bibinfo {title} {Local resonances and parametric level
  dynamics in the many-body localized phase},\ }\href
  {https://doi.org/10.1103/PhysRevB.104.184203} {\bibfield  {journal} {\bibinfo
   {journal} {Phys. Rev. B}\ }\textbf {\bibinfo {volume} {104}},\ \bibinfo
  {pages} {184203} (\bibinfo {year} {2021})}\BibitemShut {NoStop}%
\bibitem [{\citenamefont {Morningstar}\ \emph {et~al.}(2022)\citenamefont
  {Morningstar}, \citenamefont {Colmenarez}, \citenamefont {Khemani},
  \citenamefont {Luitz},\ and\ \citenamefont
  {Huse}}]{morningstar2021avalanches}%
  \BibitemOpen
  \bibfield  {author} {\bibinfo {author} {\bibfnamefont {A.}~\bibnamefont
  {Morningstar}}, \bibinfo {author} {\bibfnamefont {L.}~\bibnamefont
  {Colmenarez}}, \bibinfo {author} {\bibfnamefont {V.}~\bibnamefont {Khemani}},
  \bibinfo {author} {\bibfnamefont {D.~J.}\ \bibnamefont {Luitz}},\ and\
  \bibinfo {author} {\bibfnamefont {D.~A.}\ \bibnamefont {Huse}},\ }\bibfield
  {title} {\bibinfo {title} {Avalanches and many-body resonances in many-body
  localized systems},\ }\href {https://doi.org/10.1103/PhysRevB.105.174205}
  {\bibfield  {journal} {\bibinfo  {journal} {Phys. Rev. B}\ }\textbf {\bibinfo
  {volume} {105}},\ \bibinfo {pages} {174205} (\bibinfo {year}
  {2022})}\BibitemShut {NoStop}%
\bibitem [{\citenamefont {Garratt}\ and\ \citenamefont
  {Roy}(2022)}]{garratt2022resonant}%
  \BibitemOpen
  \bibfield  {author} {\bibinfo {author} {\bibfnamefont {S.~J.}\ \bibnamefont
  {Garratt}}\ and\ \bibinfo {author} {\bibfnamefont {S.}~\bibnamefont {Roy}},\
  }\bibfield  {title} {\bibinfo {title} {Resonant energy scales and local
  observables in the many-body localized phase},\ }\href
  {https://doi.org/10.1103/PhysRevB.106.054309} {\bibfield  {journal} {\bibinfo
   {journal} {Phys. Rev. B}\ }\textbf {\bibinfo {volume} {106}},\ \bibinfo
  {pages} {054309} (\bibinfo {year} {2022})}\BibitemShut {NoStop}%
\bibitem [{\citenamefont {Crowley}\ and\ \citenamefont
  {Chandran}(2022)}]{crowley2022constructive}%
  \BibitemOpen
  \bibfield  {author} {\bibinfo {author} {\bibfnamefont {P.~J.~D.}\
  \bibnamefont {Crowley}}\ and\ \bibinfo {author} {\bibfnamefont
  {A.}~\bibnamefont {Chandran}},\ }\bibfield  {title} {\bibinfo {title} {{A
  constructive theory of the numerically accessible many-body localized to
  thermal crossover}},\ }\href {https://doi.org/10.21468/SciPostPhys.12.6.201}
  {\bibfield  {journal} {\bibinfo  {journal} {SciPost Phys.}\ }\textbf
  {\bibinfo {volume} {12}},\ \bibinfo {pages} {201} (\bibinfo {year}
  {2022})}\BibitemShut {NoStop}%
\bibitem [{\citenamefont {Li}\ \emph {et~al.}(2018)\citenamefont {Li},
  \citenamefont {Chen},\ and\ \citenamefont {Fisher}}]{li2018quantum}%
  \BibitemOpen
  \bibfield  {author} {\bibinfo {author} {\bibfnamefont {Y.}~\bibnamefont
  {Li}}, \bibinfo {author} {\bibfnamefont {X.}~\bibnamefont {Chen}},\ and\
  \bibinfo {author} {\bibfnamefont {M.~P.~A.}\ \bibnamefont {Fisher}},\
  }\bibfield  {title} {\bibinfo {title} {Quantum zeno effect and the many-body
  entanglement transition},\ }\href
  {https://doi.org/10.1103/PhysRevB.98.205136} {\bibfield  {journal} {\bibinfo
  {journal} {Phys. Rev. B}\ }\textbf {\bibinfo {volume} {98}},\ \bibinfo
  {pages} {205136} (\bibinfo {year} {2018})}\BibitemShut {NoStop}%
\bibitem [{\citenamefont {Skinner}\ \emph {et~al.}(2019)\citenamefont
  {Skinner}, \citenamefont {Ruhman},\ and\ \citenamefont
  {Nahum}}]{skinner2019measurement}%
  \BibitemOpen
  \bibfield  {author} {\bibinfo {author} {\bibfnamefont {B.}~\bibnamefont
  {Skinner}}, \bibinfo {author} {\bibfnamefont {J.}~\bibnamefont {Ruhman}},\
  and\ \bibinfo {author} {\bibfnamefont {A.}~\bibnamefont {Nahum}},\ }\bibfield
   {title} {\bibinfo {title} {Measurement-induced phase transitions in the
  dynamics of entanglement},\ }\href
  {https://doi.org/10.1103/PhysRevX.9.031009} {\bibfield  {journal} {\bibinfo
  {journal} {Phys. Rev. X}\ }\textbf {\bibinfo {volume} {9}},\ \bibinfo {pages}
  {031009} (\bibinfo {year} {2019})}\BibitemShut {NoStop}%
\bibitem [{\citenamefont {Gullans}\ and\ \citenamefont
  {Huse}(2020)}]{Gullans2020Purification}%
  \BibitemOpen
  \bibfield  {author} {\bibinfo {author} {\bibfnamefont {M.~J.}\ \bibnamefont
  {Gullans}}\ and\ \bibinfo {author} {\bibfnamefont {D.~A.}\ \bibnamefont
  {Huse}},\ }\bibfield  {title} {\bibinfo {title} {Dynamical purification phase
  transition induced by quantum measurements},\ }\href
  {https://doi.org/10.1103/PhysRevX.10.041020} {\bibfield  {journal} {\bibinfo
  {journal} {Phys. Rev. X}\ }\textbf {\bibinfo {volume} {10}},\ \bibinfo
  {pages} {041020} (\bibinfo {year} {2020})}\BibitemShut {NoStop}%
\bibitem [{\citenamefont {Nahum}\ \emph {et~al.}(2021)\citenamefont {Nahum},
  \citenamefont {Roy}, \citenamefont {Skinner},\ and\ \citenamefont
  {Ruhman}}]{nahum2021measurement}%
  \BibitemOpen
  \bibfield  {author} {\bibinfo {author} {\bibfnamefont {A.}~\bibnamefont
  {Nahum}}, \bibinfo {author} {\bibfnamefont {S.}~\bibnamefont {Roy}}, \bibinfo
  {author} {\bibfnamefont {B.}~\bibnamefont {Skinner}},\ and\ \bibinfo {author}
  {\bibfnamefont {J.}~\bibnamefont {Ruhman}},\ }\bibfield  {title} {\bibinfo
  {title} {Measurement and entanglement phase transitions in all-to-all quantum
  circuits, on quantum trees, and in {L}andau-{G}insburg theory},\ }\href
  {https://doi.org/10.1103/PRXQuantum.2.010352} {\bibfield  {journal} {\bibinfo
   {journal} {PRX Quantum}\ }\textbf {\bibinfo {volume} {2}},\ \bibinfo {pages}
  {010352} (\bibinfo {year} {2021})}\BibitemShut {NoStop}%
\bibitem [{\citenamefont {Thakur}\ and\ \citenamefont
  {Roy}()}]{thakur-il-mipt}%
  \BibitemOpen
  \bibfield  {author} {\bibinfo {author} {\bibfnamefont {R.}~\bibnamefont
  {Thakur}}\ and\ \bibinfo {author} {\bibfnamefont {S.}~\bibnamefont {Roy}},\
  }\href@noop {} {\bibinfo {title} {Information length scales across
  measurement-induced entanglement transitions}},\ \bibinfo {note} {{\it in
  preparation}}\BibitemShut {NoStop}%
\bibitem [{\citenamefont {Fattal}\ \emph {et~al.}(2004)\citenamefont {Fattal},
  \citenamefont {Cubitt}, \citenamefont {Yamamoto}, \citenamefont {Bravyi},\
  and\ \citenamefont {Chuang}}]{fattal2004entanglement}%
  \BibitemOpen
  \bibfield  {author} {\bibinfo {author} {\bibfnamefont {D.}~\bibnamefont
  {Fattal}}, \bibinfo {author} {\bibfnamefont {T.~S.}\ \bibnamefont {Cubitt}},
  \bibinfo {author} {\bibfnamefont {Y.}~\bibnamefont {Yamamoto}}, \bibinfo
  {author} {\bibfnamefont {S.}~\bibnamefont {Bravyi}},\ and\ \bibinfo {author}
  {\bibfnamefont {I.~L.}\ \bibnamefont {Chuang}},\ }\href
  {https://arxiv.org/abs/quant-ph/0406168} {\bibinfo {title} {Entanglement in
  the stabilizer formalism}} (\bibinfo {year} {2004}),\ \Eprint
  {https://arxiv.org/abs/quant-ph/0406168} {arXiv:quant-ph/0406168 [quant-ph]}
  \BibitemShut {NoStop}%
\end{thebibliography}%
\end{document}